\documentclass[a4paper,fleqn,usenatbib]{mnras}

\usepackage[T1]{fontenc}
\usepackage{ae,aecompl}

\usepackage{graphicx}	
\usepackage{amsmath}	
\usepackage{amssymb}	
\usepackage{natbib}

\usepackage{txfonts}

\usepackage{subfig}
\usepackage{rotfloat}

\title[MOCCA -- V. Initial globular cluster conditions influence on blue stragglers]{MOCCA code for star cluster
simulations -- V. Initial globular cluster conditions influence on blue stragglers}

\author[Arkadiusz Hypki and Mirek Giersz]{
Arkadiusz Hypki,$^{1,2}$\thanks{E-mail: ahypki@strw.leidenuniv.nl} and  
Mirek Giersz$^{2}$\\
\\
$^{1}$Leiden Observatory, Leiden University, PO Box 9513, NL-2300 RA
Leiden, the Netherlands\\
$^{2}$Nicolaus Copernicus Astronomical Center, Bartycka 18, 00--716 Warsaw,
Poland
}

\date{Accepted XXX. Received YYY; in original form ZZZ}

\pubyear{2016}

\begin{document}
\label{firstpage}
\pagerange{\pageref{firstpage}--\pageref{lastpage}}
\maketitle

\begin{abstract}

The paper presents an analysis of properties of populations of blue stragglers
(BSs) in evolving globular clusters, based on numerical simulations done with
the \textsc{mocca} code for various initial globular clusters conditions.

We find that various populations of BSs strongly depend on the initial
semi-major axes distributions.
With a significant number of compact binaries, the number of evolutionary BSs
can be also significant.
In turn, for semi-major axes distributions preferring binaries with wider
orbits, dynamical BSs are the dominant ones.
Their formation scenario is very distinct: for wide binaries the number of
dynamical interactions is significantly larger.
Most interactions are weak and increase only slightly the eccentricities.
However, due to a large number of such interactions, the eccentricities of a
number of binaries finally get so large that the stars collide.

We study how larger initial clusters' concentrations influence the BSs.
Besides the expected increase of the number of dynamically created BSs (for
denser GCs the probabilities of strong dynamical interactions and collisions are
higher), we find that the number
of the evolutionary BSs is not affected even by very high initial
concentrations. This has a very important implication on observations -- it
supports the theory that the evolutionary BSs are the result of the unperturbed
evolution of the primordial binaries.

In addition, the paper presents the evolution of the ratio between
the number of BSs in binaries and as single stars ($R_{B/S}$). For a
vast diversity of models, the ratio $R_{B/S}$ approaches the value $\sim 0.4$.
Additionally, we identified two subgroups which differ in the
initial semi-major axes distributions. The first group starts with a high ratio
$R_{B/S}$, it decreases with time and settles around 0.4. The second group
starts with lower values of the ratio $R_{B/S}$ and increases to about the
same level 0.4. The first group is dominated by the
evolutionary BSs originating from the semi-major axes distribution which create
some number of compact binaries. In turn, the second group is dominated by the
dynamical ones with the initial semi-major axes distribution preferring the wider
binaries.

We find also that the initial eccentricity
distributions seems to have a small or no influence on the population of BSs.

\end{abstract}

\begin{keywords}
stellar dynamics - methods: numerical - globular clusters: evolution - stars:
blue stragglers
\end{keywords}



\section{Introduction}
\label{sec:Intro}

The subject of this paper concerns properties of blue straggler stars.
They are particularly interesting today, because by studying these type of
objects, one can get important constraints on the link between the stellar and
dynamical evolution of star clusters. Star clusters are very efficient
environments for creating such exotic objects. By studying them, one can
reveal e.g. the dynamical history of a cluster and the role of dynamics on the
stellar evolution. BSs properties can also provide some constraints for initial
binary properties.

BSs are defined as stars that are brighter and bluer (hotter) than the
main-sequence turn-off point (more than 2 mag above the turn-off point).
These stars lie along an extension of the main-sequence (MS) in the Color
Magnitude Diagram (CMD) and appear to be a rejuvenated stellar population. BSs
are on the place in the CMD where they should already evolve away from the MS.
Their mass is larger that the turn-off mass and is of the order of $M = 1.0 -
1.7 M_{\odot}$ \citep{DeMarco2005ApJ...632..894D}, which suggests some stellar
merger or a mass transfer scenario for their creation.
They were first discovered by \citet{1953AJ.....58...61S} in M3 and later
observations showed that BSs are present essentially in all star clusters.
\citep{2004ApJ...604L.109P} observed 3000 BSs in 56 different size clusters.
BSs were discovered also in open clusters, e.g.
\citet{Mathieu2009Natur.462.1032M} and dwarf galaxies e.g.
\citet{Mateo1995AJ....110.2166M}, \citet{Mapelli2007MNRAS.380.1127M} or
\citet{Monelli2012ApJ...744..157M}.

\subsection{Channels of formation of blue stragglers}
\label{sec:Intro:BSS:Channels}

Currently, there are two main scenarios considered as possible formation
mechanisms for BSs. The first scenario is a mass transfer between binary
companions which can possibly lead to the coalescence of the binary system
\citep{1964MNRAS.128..147M, 1976ApJ...209..734Z, Mateo1990AJ....100..469M,
Pritchet1991ApJ...373..105P, Knigge2009Natur.457..288K}.
The second leading scenario for creating BSs is a physical collision between
stars \citep{1976ApL....17...87H}.
Channels of formation combine together dynamical interactions between stars
(collisions) and stellar evolution (mas transfer).
However, the exact nature of channels of formation of these objects and their
relative importance is still unclear.
Moreover, there is still no observational mechanism able to distinguish BSs from
both channels (first steps were however already made
\citep{Ferraro2006ApJ...647L..53F}).

According to \citet{1992AJ....104.1831F}, different environments could be
responsible for different origins of BSs. In globular clusters which are not
dense, BSs could form as evolutionary mergers of primordial binaries, while in
high density GCs, BSs could form from dynamical interactions, particularly from
interactions involving binaries. Recently more evidence appeared suggesting that
these all scenarios are actually working simultaneously in the GCs
\citep{Ferraro1995A&A...294...80F, 1997A&A...324..915F,
Ferraro2009Natur.462.1028F}.

The relative efficiency of these two main formation channels is still unknown.
However, it is believed that they act with different efficiencies according to
the cluster structural parameters \citep{1992AJ....104.1831F} and additionally
they can work simultaneously in different radial parts of a star cluster
\citep{1997A&A...324..915F, 2006MNRAS.373..361M}. Particularly, the number of
BSs formed in the cluster does not correlate with the predicted collision rate
\citep{2004ApJ...604L.109P, Leigh2007ApJ...661..210L, 2008IAUS..246..331L}. This
is one of the reasons why it is believed that mass transfer mechanisms are more
important in the creation of BSs, instead of collisions between stars
\citep{Knigge2009Natur.457..288K}. Unfortunately, there is still no simple
observational distinction between BSs formation through mass transfer or
collisions between stars. One of the first attempts to clarify this issue is the
approach of \citet{2009RMxAC..37...62F}, who observed a significant depletion of
C and O suggesting mass transfer mechanisms for creating some BSs subpopulations
in 47~Tuc. According to \citet{Davies2004MNRAS.349..129D} primordial binaries
with BSs are vulnerable to exchange encounters in the crowded environments of
star clusters.
Low-mass components are replaced by more massive single stars. The authors claim
that these encounters tend to reduce the number of binaries containing primaries
with masses close to the present turn-off mass. Thus, the population of
primordial BSs is reduced in more massive star clusters.

\subsection{Masses of blue stragglers}
\label{sec:Intro:BSS:Masses}

First estimates of masses of BSs were performed by
\citet{Shara1997ApJ...489L..59S}. They performed direct measurements of BSs in
47~Tuc GC where they used spectroscopic analysis of HST data. They derived the
mass of $M = 1.7 M_{\odot}$, which is twice as large as the turn-off mass for
47~Tuc. Later, \citet{DeMarco2005ApJ...632..894D} calculated masses for 4 BSs
(1.27, 1.0.5, 0.99 and 0.99 $M_{\odot}$) but with slightly larger errors and
thus evolutionary tracks did not have a very good agreement with BSs masses. In
turn, very precise masses determinations were performed by using spectroscopic
and photometric analysis of eclipsing binaries,
\citet{Thompson2010AJ....139..329T} for 47~Tuc,
\citet[a]{Kaluzny2007AJ....133.2457K}, \citet[b]{Kaluzny2007AJ....134..541K},
\citet{Kaluzny2009AcA....59..371K} for 47~Tuc, $\omega$~Cen and NGC~6752. The
last papers showed a very good agreement between estimated BSs masses and
predicted masses from single-star evolutionary tracks. However, there are
examples of works which studied BSs in binaries and found that single-star
evolutionary models overestimate the dynamical masses. For example
\citet{Geller2012AJ....144...54G} gives values of overestimation of 15-30\% for
BSs in NGC~188 ($\sim 7$~Gyr old).

\citet{Lanzoni2007ApJ...663.1040L} determined masses for 34 BSs for NGC~1904
using theoretical isochrones and trying to fit them to the photometric data in V
and B-V colors. The metallicity was chosen to be $Z = 6 \times 10^{-4}$ (33
times smaller than solar metallicity) and the reddening $E(B-V) = 0.01$
\citep{Ferraro1999AJ....118.1738F}. The isochrone for the age of 12~Gyr
reproduced the main sequence nicely, while isochrones for BSs were calculated
for ages of 1-6~Gyr, with 0.5~Gyr step, which covered the whole BSs population
on CMD. The computed isochrones created a mesh of possible evolutionary tracks.
For all BSs colors and magnitudes of the closest evolutionary track were chosen,
and after a simple projection the masses for BSs were derived. They are in the
range range from $\sim 0.95$ to $\sim 1.6 M_{\odot}$, the mean and median BS mass is $1.2
M_{\odot}$ and the turn-off mass is estimated to be $M_{turn-off} = 0.8
M_{\odot}$.

Blue stragglers masses can be also calculated based on the pulsation properties
\citep{Fiorentino2014ApJ...783...34F}. They used HST images to study the
population of variable BSs in the central region of NGC~6541. NGC~6541 is an old
GC $13.25 \pm 1$~Gyr \citep{Dotter2010ApJ...708..698D}, metal poor $[Fe/H] =
-1.76 \pm -0.02$ \citep{Lee2002AJ....124.1511L}, 3~kpc from the center of the
Milky Way, and it is a dynamically old, classified as post core-collapse cluster
\citep{Harris1996AJ....112.1487H}. Among all BSs
\citet{Fiorentino2014ApJ...783...34F} discovered three W~UMa and nine
SX~Phoenicis stars (SXP). SXP stars cross the faint extension of the classical
instability strip (IS, see e.g.
\citet{Pych2001A&A...367..148P}).
IS is a place on the Hertzsprung-Russell diagram (HR) where pulsating,
constantly expanding and contracting, stars are located due to imbalance of
their thermal pressure and the gravitational force. SXP show a photometric
variability on very short scales with periods $P \lesssim 0.1$ [d] and can be
unstable for radial and nonradial pulsations.
However, these variables follow the classical pulsation equation, relating the
observed period to the intrinsic stellar parameters such as mass, luminosity,
and effective temperature $P(M, L, T_{eff})$, for any given pulsation mode and
chemical composition. Thus, this equation can be used to estimate SXP masses.
For SXP stars periods, mean magnitudes and pulsation modes were obtained and
masses were calculated using pulsation equations obtained from linear
nonadiabatic models \citep{Santolamazza2001ApJ...554.1124S}. The masses are in
the range $1.0 - 1.1 M_{\odot}$, which is significantly larger than the main
sequence turn-off mass ($\sim 0.75 M_{\odot}$). The computed masses turned out
to be in agreement with evolutionary tracks for single stars
\citep{Fiorentino2014ApJ...783...34F}.
The SXP stars are of the great importance because if they e.g. pulsate in the
fundamental modes they can be used as ``standard candles" for precise distance
calculations (e.g.
\citet{Otulakowska2011AcA....61..161O} for NGC~2155).

\subsection{Other blue straggler properties}
\label{sec:BSS:Other}

\citet{2008A&A...481..701S} tested 13 low-density GCs for correlations between
the specific frequency of BSs and cluster parameters like binary fraction, total
magnitude, age, central velocity dispersion, metallicity, cluster central
density, half-mass relaxation time, half-mass radius, stellar collision rate,
concentration, and cluster evaporation rate. The BSs specific frequency was
defined as the ratio between the estimated BSs number and MS number. MSs were
chosen, instead of horizontal branch (HB) or red giant branch (RGB) stars,
because of their abundance in all clusters and their completeness. They found
the strongest correlation between the number of BSs and the binary fraction. It
suggests that the primordial binary fraction is one of the most important factor
for producing BSs. Additionally, a noticeable correlation exists with the
absolute magnitude and an anticorrelation with the cluster age and central
velocity dispersion. The age estimates are uncertain and span a narrow range, so
one has to be careful while making any generalizations. However, if such
anticorrelation with the cluster ages is confirmed in the future, it could
suggest that binary disruptions in cores of GCs become more efficient with time,
which would in consequence reduce the fraction of binaries and also BSs in the
core.
\citet{2008A&A...481..701S} suggest that the strong correlation between the
number of BSs and the binary fraction is a result of the formation channel of
BSs in the form of unperturbed evolution of primordial binary systems. They
found no correlations for central density, concentration, stellar collision rate
and half-mass relaxation time. This indicates that the collisional channel of
BSs formation has a very small efficiency in low-density GCs.

Finding an observational mechanism to determine the channel of formation of a BS
is very important. It could provide valuable boundaries for the processes which
lead to their creation. It would also help to investigate how different channels
of formation of BSs depend on the properties of GCs.
\citet{Ferraro2006ApJ...647L..53F} gave the first results of the chemical
composition of BSs for some selected GCs. They examined 43 BSs in 47~Tuc and
found the first evidence that some subpopulations of these BSs have significant
depletion of C and O with respect to the normal cluster stars. They argue that
this is caused by CNO burning products on the BS surface, coming from the core
of a deeply peeled primary star.
This scenario is expected for the case of the mass transfer formation mechanism
and could be the first direct proof of this formation process. Later,
\citet{Fossati2010A&A...510A...8F} attempted to develop a formation scenario for
HD~73666, a known BS from the Praesepe cluster, and showed that the abundance of
CNO is consistent with a collisional formation.
However, they were unable to determine whether HD~73666 is a product of a
collision between two stars, components of a binary, or between binary systems.
Further studies of these phenomena could create some statistics on how efficient
this mechanism could be in producing BSs.

\citet{Ferraro2009Natur.462.1028F} reported two distinct sequences of BSs in GC
M30.
These two groups are clearly separated in the CMD and nearly parallel to each
other \citep[Fig. 1]{Ferraro2009Natur.462.1028F}. The first BSs sequence was
accurately reproduced by the collisional isochrones \citep[Fig. 4, blue
points]{Ferraro2009Natur.462.1028F}. The second BSs sequence
 corresponds well to the zero-age main sequence shifted by 0.75 mag, marking the
 position of the
low-luminosity boundary predicted for a population of mass-transfer binary
systems \citep[Fig. 4, red points]{Ferraro2009Natur.462.1028F}.

\citet{Knigge2009Natur.457..288K} focused on BSs in cores of star clusters,
because in these regions collisions between stars should be frequent.
They used existing data from a large set of HST-based CMDs and confirmed that
there is no global correlation between the observed core BSs number and the
collision rate (different core densities have different predicted collision
rates and it does not correlate with the number of BSs). However, there is a
significant correlation if one would restrict this relation to the clusters with
dense cores (see \citet{Knigge2009Natur.457..288K} black points in Fig. 1). The
second relation which was tested by this group concerns the binary fraction in
the core. If most of BSs were formed in binaries, the number of BSs should scale
with the binary fraction simply as $N_{BSS} \propto f_{bin} M_{core}$, where
$f_{bin}$ is the binary fraction in the core, and $M_{core}$ is the total
stellar mass contained in the core. Indeed, they found a clear correlation
between the number of BSs and core masses of the clusters, as it is expected for
the scenario where most BSs originate from binaries (see
\citet{Knigge2009Natur.457..288K} Fig.~2). They interpret this result as a
strong evidence that more BSs originates from binaries instead of collisions
between stars. They found that the dependence $N_{BSS} \propto
M_{core}^{\delta}$ can be estimated with $\delta \simeq 0.4-0.5$. Furthermore,
they estimated the power-law correlation $f_{bin} \propto M_{core}^{-0.35}$
based on the data from \citet{2008MmSAI..79..623M} who described global
parameters for 35 clusters spanning a wide range of density and other dynamical
star cluster parameters.
Those two estimates combined together shows that the number of BSs found in the
cores of GCs scales roughly as $N_{BSS} \propto f_{bin} M_{core}$, just as
expected if most core BSs are formed in binary systems
\citep{Knigge2009Natur.457..288K}.

BSs are being found in the halo and in the bulge of the Galaxy
\citep{Bragaglia2005IAUS..228..243B,Fuhrmann2011MNRAS.416..391F,Clarkson2011ApJ...735...37C}.
\citet{Tillich2010A&A...517A..36T} found a BS from the halo which has a radial
velocity of about $504.6 \pm 5$~km/s. With a Galactic rest-frame velocity of
about 467 km/s, it makes this BS one of the fastest moving BSs (but it is still
bound to the Galaxy).

Recently, \citet{Geller2011Natur.478..356G} reported that BSs in long-period
binaries in an old (7~Gyr) open cluster, NGC~188, have companions with masses of
about half of the solar mass, which is a surprisingly narrow mass distribution.
This rules out the collisional origin for these long-period BSs, because
otherwise, for the collision hypothesis, there would be significantly more
companions with higher masses. The data is consistent with a mass transfer origin for the
long-period blue straggler binaries in NGC 188, in which the companions would be
white dwarfs of about half of the solar mass.

This paper is organized as follows. In the Sect.~\ref{sec:NumericalSimulations}
there is shortly described the MOCCA code, summary of performed
numerical simulations and description of the data analysis methods.
Sect.~\ref{sec:PropBSS} contains the detailed analysis how various initial
conditions influence on the population of BSs. Additionally, we discuss how
the ratio of BSs in binaries and as single stars changes for various models. 
Finally, Sect.~\ref{sec:Summary} summarizes our findings and
presents discussion about channels of formation of BSS.

\section{Numerical simulations}
\label{sec:NumericalSimulations}

Numerical simulations were performed with the
\textsc{mocca}\footnote{\url{http://moccacode.net}} code
\citep{Hypki2013MNRAS.429.1221H}.

The \textsc{mocca} code is currently one of the most advanced codes which is
able to simulate real size GCs and at the same time, it allows to have a full
dynamical history of the evolution of all stars in the system. \textsc{mocca} is
an improved version of the Monte Carlo code, originally developed by
\citet{Henon1971Ap&SS..14..151H}, improved by
\citet{Stodolkiewicz1986AcA....36...19S}, and finally heavily developed by
Giersz and his collaborators \citep{Giersz1998MNRAS.298.1239G,
Giersz2001MNRAS.324..218G, Giersz2006MNRAS.371..484G,
Giersz2011arXiv1112.6246G,Giersz2013MNRAS.431.2184G}.
The \textsc{mocca} code combines together the old version of the code (Monte
Carlo method) with strong dynamical interactions performed with the
\textsc{fewbody} code \citep{Fregeau2007ApJ...658.1047F}. The stellar evolution
is done for both single and binary stars using SSE and BSE codes
\citep{Hurley2000MNRAS.315..543H, Hurley2002MNRAS.329..897H}.
All these codes together create one package, called the \textsc{mocca} code,
which stands for MOnte Carlo Cluster simulAtor (for detail description of the
\textsc{mocca} code see \citet{Hypki2013MNRAS.429.1221H}).

The speed of the \textsc{mocca} code is its greatest advantage in comparison to
N-body codes. During the same amount of time one can run multiple simulations
with the \textsc{mocca} code to cover a very wide range of initial cluster
parameters. Instead of having one simulation from N-body code, one can have
hundreds of simulations from the \textsc{mocca} code and one can perform
detailed statistical analysis of the results. Additionally, \textsc{mocca}
simulations give practically the same amount of information about the evolution
of star clusters as N-body codes, which makes it even more attractive.
There is already number of papers, which shows this agreement across all
previous version of the \textsc{mocca} code \citep{Giersz1994MNRAS.268..257G,
Giersz1994MNRAS.270..298G, Giersz1996MNRAS.279.1037G, Giersz1997MNRAS.286..709G,
Giersz1994MNRAS.269..241G}, and especially with the current version
\citet{Giersz2013MNRAS.431.2184G,Wang2016MNRAS.458.1450W}.

\subsection{Initial parameters for the \textsc{mocca} code simulations}
\label{sec:InitialParameters}

For the purpose of this paper a large number of simulations was computed. These
simulations vary in many aspects. They have different initial mass functions,
binary properties, different sizes, concentrations (thus different scales of the
dynamical evolution), and more. The purpose of computing many simulations was to
check how different properties of blue stragglers depend on the initial
conditions of GCs and distributions of initial binary properties.

The \textsc{mocca} code allows to define many different initial conditions.
However, only a subset of them was used. The chosen parameters are believed to
be the ones which could have the biggest influence on the population of the BSs.
The simulations, together with their initial conditions, are summarized in the
Tab.~\ref{tab:Sim:Ini1} and Tab.~\ref{tab:Sim:Ini2}.

A majority of the models have 20\% of the primordial binaries. Typically, GCs
contain at a present time $\sim 5 - 10\%$ of binaries. We decided to start the
models with a higher value to simply have a larger number of BSs, in order to
make some features easier to notice. All models use only the Plummer model as a
density distribution. We decided to use it, instead of the King models, because
it is simple and accurate enough for the initial mass distribution.

\begin{table*}
    \begin{tabular}{c|c|c|c|c|c|c|c|c|c|c|c}
\multicolumn{12}{c}{\textsc{initial mass function of the \textsc{mocca} simulations (Part~I)}}\\
Name            &$N$ &$f_b$&IM&$IMF_s$&$IMF_b$&q&a&e&z&$r_{tid}$&$r_h$\\
\hline
\textsc{mocca-1}&300k&0.1&P&K93&K91&U&UL&T&0.001&69&6.9\\
\textsc{mocca-2}&300k&0.2&P&K93&K91&U&UL&T&0.001&15&1.5\\
\textsc{mocca-3}&300k&0.2&P&K93&K91&U&UL&T&0.001&25&2.5\\
\textsc{mocca-4}&300k&0.2&P&K93&K91&U&UL&T&0.001&35&3.5\\
\textsc{mocca-5}&300k&0.2&P&K93&K91&U&UL&T&0.001&45&4.5\\
\textsc{mocca-6}&300k&0.2&P&K93&K91&U&UL&T&0.001&69&1.2\\ 
\textsc{mocca-7}&300k&0.2&P&K93&K91&U&UL&T&0.001&69&1.7\\ 
\textsc{mocca-8}&300k&0.2&P&K93&K91&U&UL&T&0.001&69&2.3\\ 
\textsc{mocca-9}&300k&0.2&P&K93&K91&U&UL&T&0.001&69&2.8\\ 
\textsc{mocca-10}&300k&0.2&P&K93&K91&U&UL&T&0.001&69&3.5\\ 
\textsc{mocca-11}&300k&0.2&P&K93&K91&U&UL&T&0.001&69&4.6\\ 
\textsc{mocca-12}&300k&0.2&P&K93&K91&U&UL&T&0.001&69&6.9\\ 
\textsc{mocca-13}&300k&0.2&P&K93&K91&U&UL&T&0.001&69&9.9\\ 
\textsc{mocca-14}&300k&0.2&P&K93&K91&U&UL&T&0.001&69&17.3\\ 
\textsc{mocca-15}&300k&0.2&P&K93&K91&U&UL&T&0.001&85&8.5\\
\textsc{mocca-16}&300k&0.2&P&K93&K91&U&UL&T&0.001&135&13.5\\
\textsc{mocca-17}&300k&0.2&P&K93&K91&U&UL&T&0.001&235&23.5\\
\textsc{mocca-18}&300k&0.2&P&K93&K91&U&UL&T&0.001&335&33.5\\
\textsc{mocca-19}&300k&0.3&P&K93&K91&U&UL&T&0.001&69&9.6\\
\textsc{mocca-20}&300k&0.5&P&K93&K91&U&UL&T&0.001&69&9.6\\

\textsc{mocca-21}&600k&0.05&P&K93&K91&U&UL&T&0.001&100&10.0\\ 
\textsc{mocca-22}&600k&0.1&P&K93&K91&U&UL&T&0.001&100&10.0\\ 
\textsc{mocca-23}&600k&0.2&P&K93&K91&U&UL&T&0.001&25&2.5\\ 
\textsc{mocca-24}&600k&0.2&P&K93&K91&U&UL&T&0.001&35&0.9\\ 
\textsc{mocca-25}&600k&0.2&P&K93&K91&U&UL&T&0.001&35&1.2\\ 
\textsc{mocca-26}&600k&0.2&P&K93&K91&U&UL&T&0.001&35&1.8\\ 
\textsc{mocca-27}&600k&0.2&P&K93&K91&U&UL&T&0.001&35&3.5\\ 
\textsc{mocca-28}&600k&0.2&P&K93&K91&U&UL&T&0.001&55&1.4\\ 
\textsc{mocca-29}&600k&0.2&P&K93&K91&U&UL&T&0.001&55&1.8\\ 
\textsc{mocca-30}&600k&0.2&P&K93&K91&U&UL&T&0.001&55&2.8\\ 
\textsc{mocca-31}&600k&0.2&P&K93&K91&U&UL&T&0.001&55&5.5\\ 
\textsc{mocca-32}&600k&0.2&P&K93&K91&U&UL&T&0.001&100&1.7\\ 
\textsc{mocca-33}&600k&0.2&P&K93&K91&U&UL&T&0.001&100&2.5\\ 
\textsc{mocca-34}&600k&0.2&P&K93&K91&U&UL&T&0.001&100&5.0\\ 
\textsc{mocca-35}&600k&0.2&P&K93&K91&U&UL&T&0.001&100&10.0\\ 
\textsc{mocca-36}&600k&0.2&P&K93&K91&U&UL&T&0.001&100&20.0\\ 
\textsc{mocca-37}&600k&0.2&P&K93&K91&U&UL&T&0.001&180&18.0\\ 
\textsc{mocca-38}&600k&0.2&P&K93&K91&U&UL&T&0.001&130&13.0\\ 
\textsc{mocca-39}&600k&0.2&P&K93&K91&U&UL&T&0.001&230&23.0\\ 
\textsc{mocca-40}&600k&0.2&P&K93&K91&U&UL&T&0.001&300&30.0\\ 
\textsc{mocca-41}&600k&0.2&P&K93&K91&U&UL&T&0.001&400&40.0\\ 
\textsc{mocca-42}&600k&0.4&P&K93&K91&U&UL&T&0.001&100&10.0\\ 
\textsc{mocca-43}&600k&0.5&P&K93&K91&U&UL&T&0.001&100&10.0\\ 

    \end{tabular}
\caption[Initial conditions of \textsc{mocca} simulations done for the purpose
of this paper -- Part I]{Initial conditions of \textsc{mocca} simulations done
for the purpose of this paper. Symbols have the following meaning: N -- initial
number of objects (single + binary stars), $f_b$ -- initial binary fraction,
$f_b = N_b / N$ ($N_b$ -- number of binaries), $IM$ -- initial model, P --
Plummer model, $IMF_s$ -- Initial Mass Function for single stars, K93 --
\citet{Kroupa1993MNRAS.262..545K} in the range $[0.1; 100] \mathrm{M_{\odot}}$,
$IMF_b$ -- Initial Mass Function for binary stars, K91 --
\citet[eq.~1]{Kroupa1991MNRAS.251..293K}, binary masses from 0.2 to
100~$\mathrm{M_{\odot}}$, q -- distribution of mass ratios between stars in
binaries, U -- uniform distribution of mass ratios, R -- random pairing of
masses for binary components, a -- semi-major axes distribution, UL -- uniform
distribution of semi-major axes in the logarithmic scale from $2(R_1+R_2)$ to
100~AU, L -- lognormal distribution of semi-major axes from $2(R_1+R_2)$ to
100~AU, K95 -- binary period distribution from
\citet{Kroupa1995aMNRAS.277.1491K}, K95E -- distribution of semi-major axes with
eigenevolution and feeding algorithm \citep{Kroupa1995aMNRAS.277.1491K}, K13 --
new eigenevolution and feeding algorithm \citep{Kroupa2013pss5.book..115K}, e --
eccentricity distribution, T -- thermal eccentricity distribution, TE -- thermal
eccentricity distribution with eigenevolution, z -- mettalicity (e.g.~0.001 =
1/20 of the solar metallicity 0.02), $r_{tid}$ -- tidal radius in pc, $r_h$ --
half-mass radius in pc. 
}
	\label{tab:Sim:Ini1}
\end{table*}

\begin{table*}
    \begin{tabular}{c|c|c|c|c|c|c|c|c|c|c|c}
\multicolumn{12}{c}{\textsc{initial mass function of the \textsc{mocca} simulations (Part~I)}}\\
Name            &$N$ &$f_b$&IM&$IMF_s$&$IMF_b$&q&a&e&z&$r_{tid}$&$r_h$\\
\hline
\textsc{mocca-44}&300k&0.2&P&K93&K91&R&UL  &T&0.001&69&6.9\\
\textsc{mocca-45}&300k&0.2&P&K93&K91&R&L   &T&0.001&69&6.9\\
\textsc{mocca-46}&300k&0.2&P&K93&K91&R&K95 &T&0.001&69&6.9\\
\textsc{mocca-47}&300k&0.2&P&K93&K91&R&K95E&T&0.001&69&6.9\\
\textsc{mocca-48}&300k&0.2&P&K93&K91&R&K13 &T&0.001&69&6.9\\

\textsc{mocca-49}&300k&0.2&P&K93&K91&R&UL  &TE&0.001&69&6.9\\
\textsc{mocca-50}&300k&0.2&P&K93&K91&R&L   &TE&0.001&69&6.9\\
\textsc{mocca-51}&300k&0.2&P&K93&K91&R&K95 &TE&0.001&69&6.9\\
\textsc{mocca-52}&300k&0.2&P&K93&K91&R&K95E&TE&0.001&69&6.9\\
\textsc{mocca-53}&300k&0.2&P&K93&K91&R&K13 &TE&0.001&69&6.9\\

\textsc{mocca-54}&300k&0.2&P&K93&K91&U&L   &T&0.001&69&6.9\\
\textsc{mocca-55}&300k&0.2&P&K93&K91&U&K95 &T&0.001&69&6.9\\
\textsc{mocca-56}&300k&0.2&P&K93&K91&U&K95E&T&0.001&69&6.9\\
\textsc{mocca-57}&300k&0.2&P&K93&K91&U&K13 &T&0.001&69&6.9\\

\textsc{mocca-58}&300k&0.2&P&K93&K91&U&UL  &TE&0.001&69&6.9\\
\textsc{mocca-59}&300k&0.2&P&K93&K91&U&L   &TE&0.001&69&6.9\\
\textsc{mocca-60}&300k&0.2&P&K93&K91&U&K95 &TE&0.001&69&6.9\\
\textsc{mocca-61}&300k&0.2&P&K93&K91&U&K95E&TE&0.001&69&6.9\\
\textsc{mocca-62}&300k&0.2&P&K93&K91&U&K13 &TE&0.001&69&6.9\\

\textsc{mocca-63}&600k&0.2&P&K93&K91&U&K13&TE&0.001&55&5.5\\ 

    \end{tabular}
\caption[Initial conditions of \textsc{mocca} simulations done for the purpose
of this paper -- Part II]{For description see Tab.~\ref{tab:Sim:Ini1}}
	\label{tab:Sim:Ini2}
\end{table*}

Tables Tab.~\ref{tab:Sim:Ini1} and Tab.~\ref{tab:Sim:Ini2} contain over 60
models. The models from Tab.~\ref{tab:Sim:Ini1} differ mainly in the values of
initial number of stars, tidal radii ($r_{tid}$), and concentrations ($c =
r_{tid} / r_{h}$). These are the parameters which define GCs with different
dynamical scales, from slowly evolving models, up to models with fast dynamical
evolution. Various dynamical scales of these models should have a different
influence on the spatial distribution of BSs in GCs, which is essential for the
studies of the formation of the bimodal spatial distribution observed in many
real GCs. Thus, the models from Tab.~\ref{tab:Sim:Ini1} (from \textsc{mocca-1}
up to \textsc{mocca-43}) were mainly used to study the spatial distributions of
BSs in evolving GCs (see Hypki (2016), \textit{MOCCA code for star cluster
simulations -- VI. Bimodal spatial distribution of blue stragglers}, submitted).
However, these models were also used to study the relation between the number of
BSs in binaries and as single stars (see Sect.~\ref{sec:PropBSs:Binaries}).

The models with identifiers larger than 43 (Tab.~\ref{tab:Sim:Ini2}) were mainly
used to study how different initial binary conditions influence the population
of BSs of different types. They have different mass ratios for components in
binaries, different distributions of semi-major axes and eccentricities but the
same initial number of stars and concentrations (except \textsc{mocca-63}). The
diversity of initial properties of binaries allows to study how the number of
BSs from different channels depends on the initial conditions.

All models from Tab.~\ref{tab:Sim:Ini1} and Tab.~\ref{tab:Sim:Ini2} were used to
study the ratio between BSs in binaries and as single stars (see
Sect.~\ref{sec:PropBSs:Binaries}). Different sizes and concentrations, as well
as different initial properties of binaries, are expected to have an influence
on this ratio.

The models from Tab.~\ref{tab:Sim:Ini1} and Tab.~\ref{tab:Sim:Ini2} are only a
small subset of the models actually computed. The total number of models was
much higher and concerned even broader range of initial conditions. However, the
models from these two tables compose a complete subset of models which is
sufficient enough to support conclusions stated in this paper.

From this point any core radius ($r_c$) refers to this calculated according to
\citet{1985ApJ...298...80C}, and relaxation time ($t_{rh}$) refers to the
half-mass relaxation time unless it is noted otherwise. BS is detected in the
\textsc{mocca} code if it exceeds the turn-off mass by at least 2\% (to be
consistent with the first results on BSs presented by
\citet{Hypki2013MNRAS.429.1221H}).

\subsection{Data analysis}
\label{sec:DataAnalysis}

Data analysis of the results of the \textsc{mocca} code is very challenging. The
output files are large. One simulation with 600k initial stars can easily exceed
a few GBs. When there are several dozens of simulations, the analysis of such
large data sets is not trivial.

Each \textsc{mocca} simulation contains almost 20 different files. Each file
stores different kind of data. For many cases querying the data is simple -- it
is just the extraction and visualization of a few columns. However, real life
queries are much more complicated. In the analysis there is often a need to read
data from many files simultaneously in order to prepare meaningful results. If
the same procedure has to be applied for many simulations, the overall
complexity of data analysis increases significantly. Thus, for the data analysis
of the results of the \textsc{mocca} code there were created many scripts which
simplify this process.

All scripts for data analysis are written in Java. 
They share the same core library, which means
that the process of building next scripts is significantly simplified. The
scripts are built with Object Oriented Programming (OOP) paradigm in mind. It
means that it consists of small Java classes responsible for small tasks. By
combining them into larger Java classes, one can create a modular code able to
solve complex tasks while still being easy to understand and change. OOP
programming is especially useful for the data analysis of the \textsc{mocca}
simulations because each entity from the \textsc{mocca} code, like star or
binary, can be expressed as one Java class. Each Java class can have an
arbitrary number of properties. In the case of \textsc{mocca}, they are for
instance: mass, radius, luminosity of stars, and semi-major axis and
eccentricity of binaries. In this way one can create very clean and fast scripts
to analyze many \textsc{mocca} simulations.

The output from the \textsc{mocca} code was split into a number of files. Each
file contains only one type of information. Some of them store information on
global parameters of GC, positions and velocities of stars (data on dynamics),
interactions between two binaries or binaries with single stars,  stellar
evolution etc. What is more, such output divided into separate files is much
easier to maintain and to understand by new users of the \textsc{mocca} code.

The scripts allowed to optimize the disk usage of the \textsc{mocca}
simulations. The largest file which is produced is a snapshot file, which
contains the full image of a GC with a number of parameters for each star (in
total 30 parameters per object). The snapshots are produced usually every 50 or
200 Myr. Thus, the output file becomes very large (even $> 20$~GBs for one
simulation). In order to save disk space, a more advanced solution was
implemented. The snapshot can be saved in a compact form with only 4 values: ID,
position, radial and tangential velocities -- the only values which are not
stored in other output files. All the other properties, like masses, radii,
semi-major axes, one can recreate from other output files (e.g. from files
storing data of stellar evolution or dynamical interactions). The script
automatically detects whether a snapshot is in the compact form or in a default
mode (with all columns). If the snapshot is compact, then the scripts
can automatically rebuild full snapshot. Additionally, all output files can be
compressed using gzip algorithm and thus saving even more disk space.
The scripts handle compressed data on-the-fly as well. All these efforts made
the need for disk space decrease a lot. It is especially useful for Big
Survey project, which goal is to produce and maintain thousands of
\textsc{mocca} simulations of real size GCs for the vast mesh of initial parameters. Easy estimations
indicate that a simulation for Big Survey will take 10-15 TBs. Thus, it is
crucial to simplify this process and make the data analysis as straightforward
as possible.

One of great advantages of the scripts developed for the \textsc{mocca}
simulations is that they are ready for High Performance Computing. They can be
executed on clusters of computers and analyze simulations in parallel if needed.
In this way one can start series of jobs simultaneously for many simulations and
get results much faster. The scripts have all of the dependencies built-in (they
work on any machine equipped with Java). This feature is also very important for the future
Big Survey project.

As a result of the extensive data analysis of the \textsc{mocca} simulations
many useful scripts were developed. A few of them are described here.

One of the most complicated scripts reads the data from the output of the
\textsc{mocca} simulations and prepares a detailed summary of properties of BSs.
It checks dozens of parameters like masses at the time when a star has been
recognized to be a BS and when it stopped to be the BS. It checks the time of
the last mass transfer or merger (the event which actually creates a BS), to
check whether a BS was created immediately or rather was dormant for some time.
The script saves positions of BSs at the time of the detection and when it stops
to be a BS. It checks if/when the BS escapes from GC, it stores information
about initial channel of formation of each BS (see
Sect.~\ref{sec:PropBSs:IniitialConditions}) and the changes of types due to e.g.
dynamical interactions. The script stores also many other parameters. All of
this provides a detail information on the history of formation and
changes of properties of BSs.

Another example of a complex and time-saving script is the one which follows the
complete history of a selected star. The script reads the whole output of the
\textsc{mocca} code and follows every possible event which concerns the selected
star. It gathers all information on stars' properties (masses, radii,
luminosity), all information on the dynamical interactions, stellar evolution
events, etc. It follows also any change in the radial distance or in velocities
available in the output. As a result, the script builds the complete history of
the star, so one can study the evolution of masses, positions or binary
properties. The script is very complicated since following the whole history of
a star is not an easy task. Such star can change its identifier due to a merger
event or it can change its binary companion. Thus, the properties of a given
star may be stored in different columns in the same file.
The script traces the history of the star starting from the end, it moves back in time and follows all
these events as well as the history of the stars' predecessors (before mergers),
until it reaches the first star at the time $T = 0$. In this way one can study
in detail the complete stellar and dynamical evolution of any star in the
system.

The last example which shows the power of the scripts concerns gathering data
from all available \textsc{mocca} simulations. The script traverse through all
selected directories. It looks for \textsc{mocca} simulations and extracts some
useful information from them. In this way one can study the properties of the
whole set of GCs together. This script was extensively used e.g. to study the
ratio between BSs in binaries and as single stars (see
Sect.~\ref{sec:PropBSs:Binaries}).

\section{Properties of blue stragglers}
\label{sec:PropBSS}

This section presents how initial conditions of GCs influence the
population of BSs of different types. It is a continuation of the work published
by \citet{Hypki2013MNRAS.429.1221H}, where the channels of formation were
presented and discussed in detail but only for a single test model.

The last section of this chapter presents in detail the ratio between BSs in
binaries and as single stars. Some hidden properties of different populations of
BSs might be revealed by this ratio.

Before the influence of various initial conditions on different populations of
BSs will be discussed, the channels of formations have to be introduced. Then,
rough estimates of the errors of numbers of blue stragglers will be presented.

\subsection{Channels of formation of BSs}
\label{sec:PropBSs:Channels}

The first type of formation of BSs is called \textit{Evolutionary Merger} (EM)
and represents the scenario when two stars from a binary merge into one star.
The merger is a result of the stellar evolution only, without involving other
stars through dynamical interactions.
The second channel is called \textit{Evolutionary Mass Transfer} (EMT). This
scenario creates BSs through a mass transfer in a binary, so that the mass of
one of the stars overcomes the turn-off mass. In this case the stellar evolution
does not have to lead immediately to a binary merger. A merger can occur later,
and then, if the star would be the main-sequence star, it would still be
considered as BSs.
The third channel is called \textit{Evolutionary Dissolution} (ED). It is the
scenario when the stellar evolution leads to a disruption of a binary (e.g. SN
explosion) with some mass accretion by the companion, which in consequence
becomes a BSs.
The EM, EMT and ED channels are connected to stellar evolution only.

Channels of formation of BSs which we include in the dynamical category are
connected strictly to dynamical interactions and are described by the following
cases. The channel of formation called \textit{Collision Single-Single} (CSS)
describes a physical collision between two single stars. This is the only
channel, both from evolution and dynamical categories, which involves only two
single stars. All other channels of formation involve at least one binary. The
second channel called \textit{Collision Binary-Single/Binary} (CBS, CBB)
describes the scenario when there is a collision between any two or more
stars in a binary-single (CBS) or binary-binary interaction (CBB).

The rest of the channels do not in fact create a new BSs but rather describe the
change of BSs type. \textit{Exchange Binary-Single/Binary}, corresponds to the
situation when BSs changes its companion in a binary, becomes a single star, or
goes into a binary. EXBS stands for an exchange event in a binary-single
dynamical interaction and EXBB means an exchange in a binary-binary interaction.
The last dynamical channel is called \textit{Dissolution Binary-Single/Binary}
and corresponds to the scenario when BSs was present in a binary, which was
disrupted by a binary-single dynamical interaction (DBS) or binary-binary
interaction (DBB).
The EXBS, EXBB, DBS and DBB cannot be the initial types of BSs. Initial BSs type
can be EM, EMT, ED, CSS, CBS or CBB, and only later BSs can change its type into
another one.

More details on the definitions of BSs and the physical processes of their
creation one can find in \citet[Sect.~4.1]{Hypki2013MNRAS.429.1221H}.

The fluctuations of the number of BSs were discussed in \citet[Fig.~2,
Fig.~3]{Hypki2013MNRAS.429.1221H}. The error $\sigma~\sim~5$~BSs is a standard
deviation of the mean number of BSs computed from five simulations with the same
initial conditions but with different initial seed values. Thus, it is safe to
assume that the fluctuations are $\pm~10$~BSs ($2 \sigma$). Every feature, in
terms of the number of BSs, which is of the order of 10 should not be
considered.

\subsection{Influence of initial conditions on populations of BSs}
\label{sec:PropBSs:IniitialConditions}

The initial conditions of simulations used in this section are summarized in
Tab.~\ref{tab:Sim:Ini1} and Tab.~\ref{tab:Sim:Ini2}. The simulations differ in
the initial properties of semi-major axes, eccentricities, initial mass
functions for single and binary stars, and different pairing of stars in
binaries.

\begin{figure*}
	\includegraphics[width=\columnwidth]{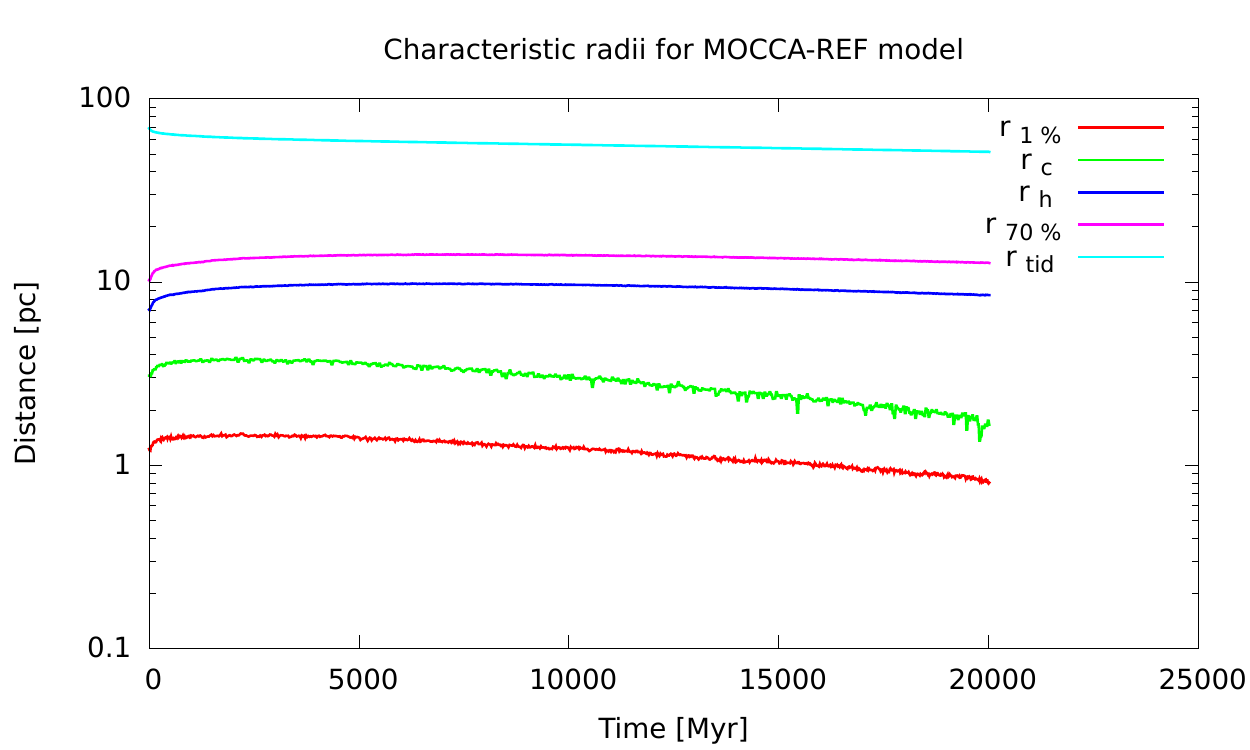}
	\includegraphics[width=\columnwidth]{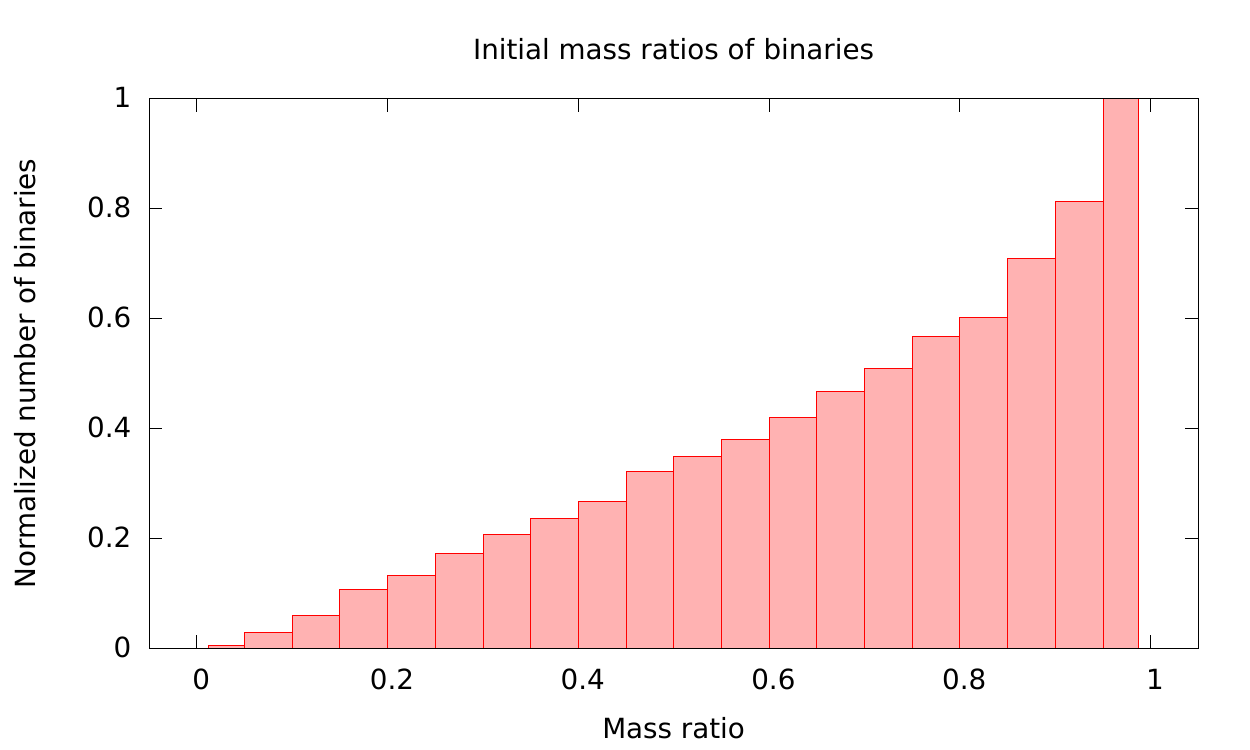}
	\includegraphics[width=\columnwidth]{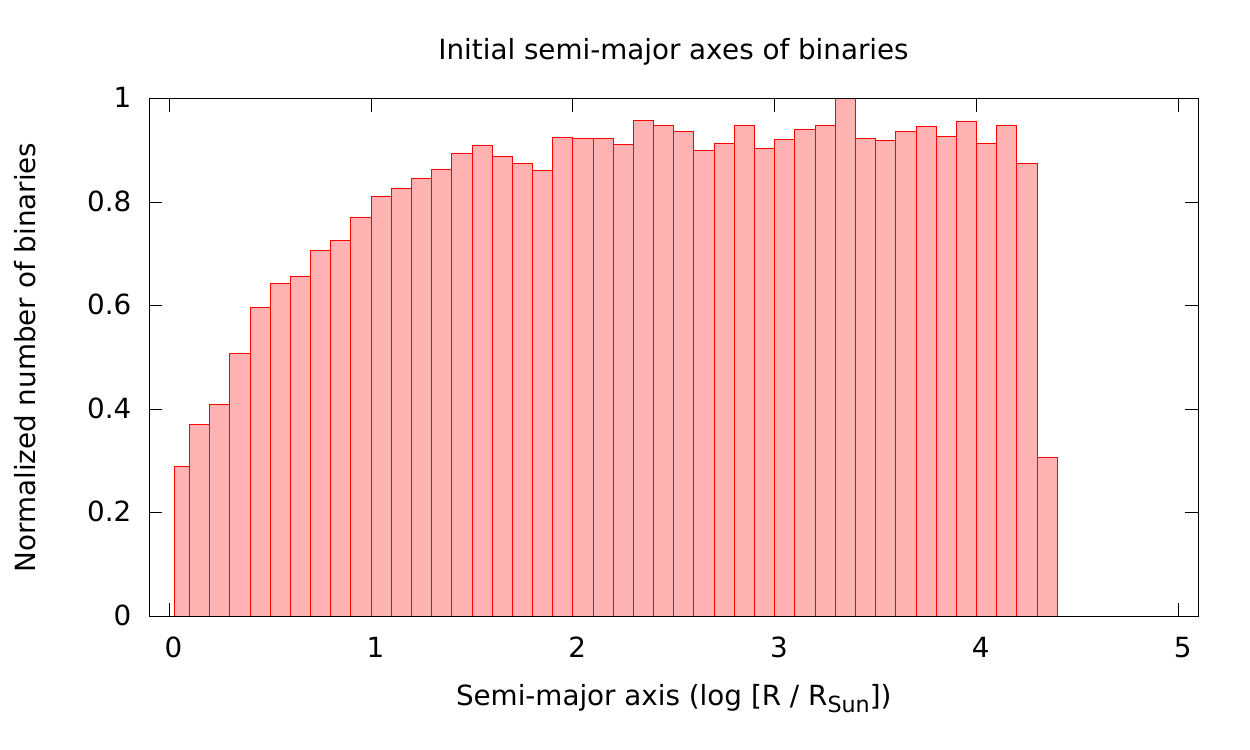}
	\includegraphics[width=\columnwidth]{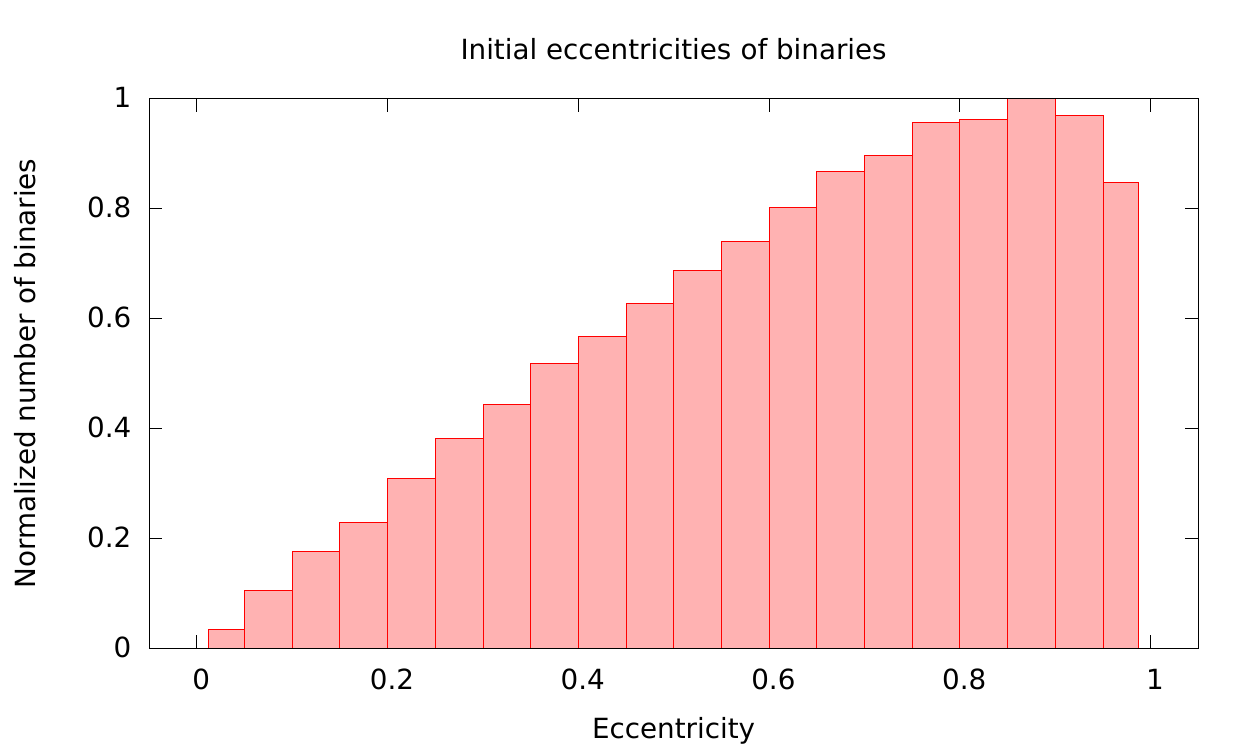}
	\includegraphics[width=\columnwidth]{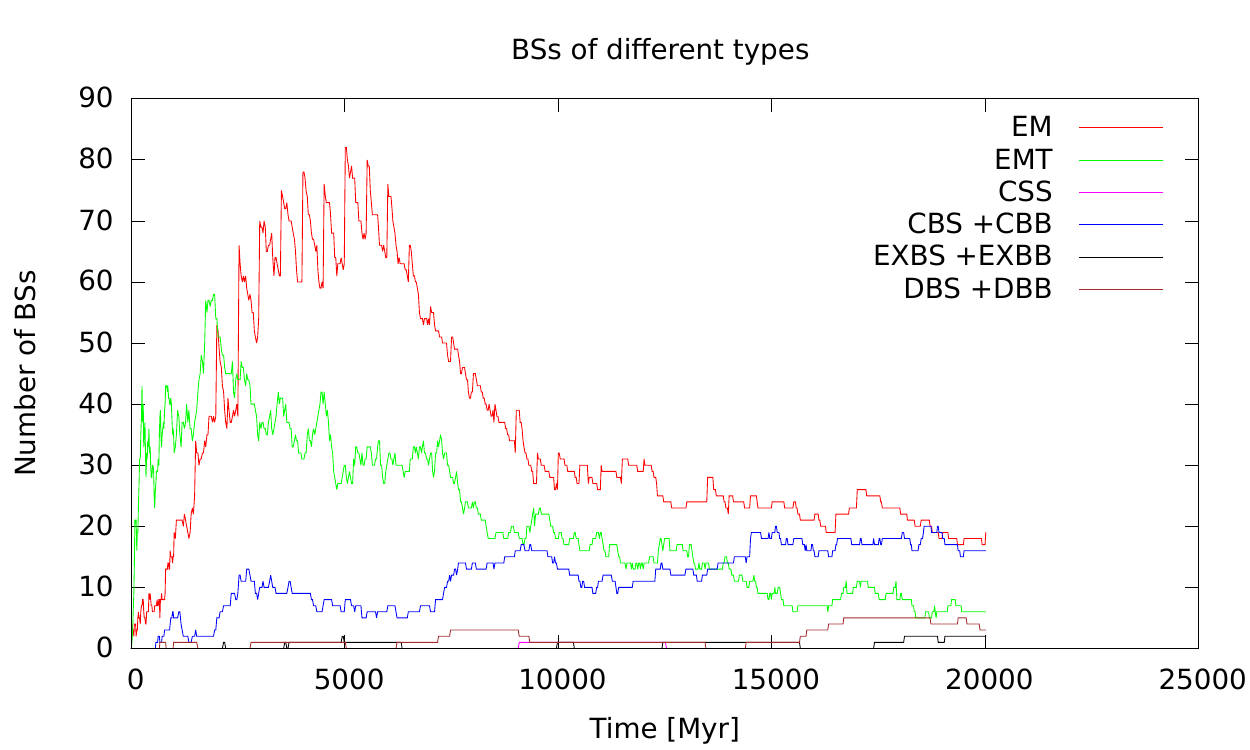}
	\includegraphics[width=\columnwidth]{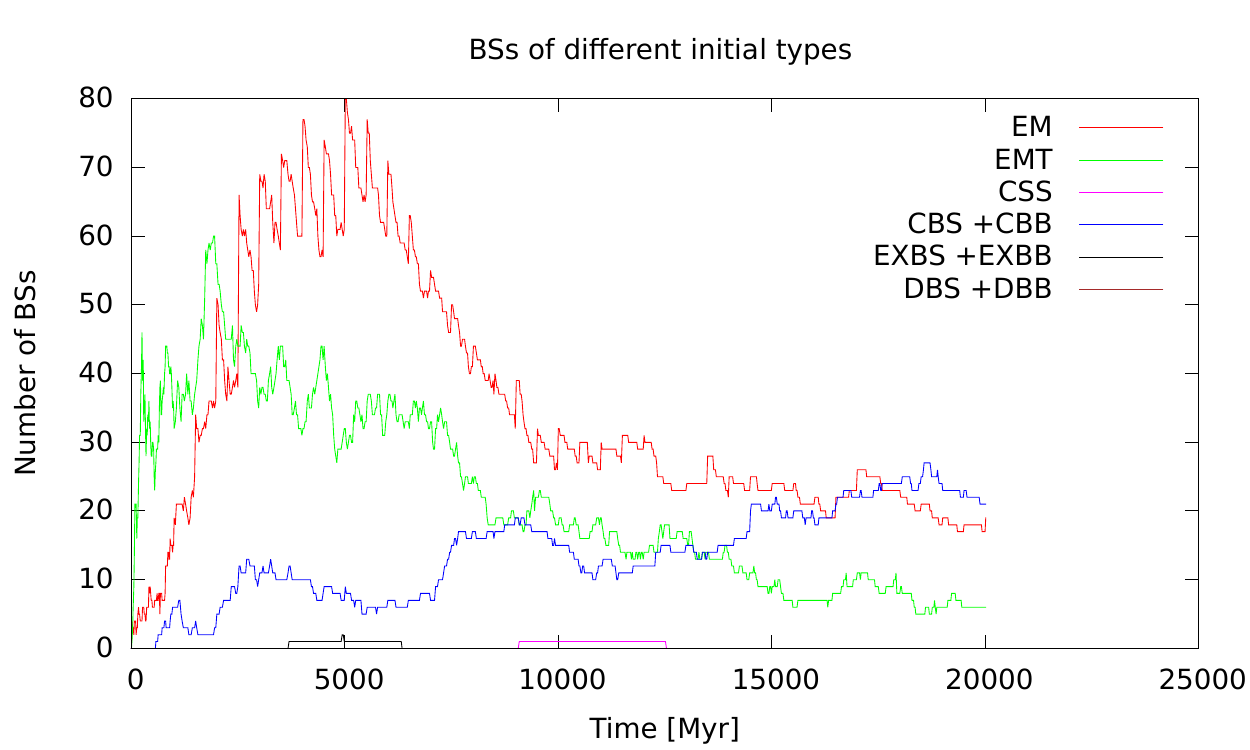}
\caption[Properties of the \textsc{mocca-ref} model used as a reference model to
study the influence of initial conditions on the populations of BSs]{Properties
of the \textsc{mocca-ref} model which is used in the
Sect.~\ref{sec:PropBSs:IniitialConditions} as a reference model to study how
different initial conditions influence the population of BSs of different types.
The first plot (top-left) presents several characteristic radii like $r_c$,
$r_h$ and $r_{tid}$. The second plot (top-right) shows initial distribution of
mass ratios for binaries (denoted as U in Tab.~\ref{tab:Sim:Ini1} and
Tab.~\ref{tab:Sim:Ini2}). The third plot (middle-left) shows the
initial distribution of semi-major axes of binaries. It is a uniform distribution in
logarithmic scale, between $4 R_n$ and 100~AU ($R_n$ is a star with the smallest
radii). It is denoted as UL in Tab.~\ref{tab:Sim:Ini1} and
Tab.~\ref{tab:Sim:Ini2}. The next plot (middle-right) presents the initial
distribution of eccentricities (thermal distribution -- T in
Tab.~\ref{tab:Sim:Ini1} and Tab.~\ref{tab:Sim:Ini2}). The last two plots show
the number of BSs of present types of BSs (bottom-left) and the initial types
of BSs (bottom-right). See the text for a description.
}
\label{fig:PropBSs:RefModel}
\end{figure*}

The changes of BSs populations are discussed only through one parameter at once
to avoid any additional complexity. As a reference simulation we chose the
\textsc{mocca-12} model (see Tab.~\ref{tab:Sim:Ini1}). This model contains 300k
initial stars, 20\% of primordial binaries. Its initial tidal radius is 69 pc
and the half-mass radius is 6.9 pc. The \textsc{mocca-12} model is called
\textsc{mocca-ref} in this section for clarity.

\subsubsection{Properties of \textsc{mocca-ref} reference simulation}
\label{sec:MOCCAREF}

The properties of the \textsc{mocca-ref} model are presented in
Fig.~\ref{fig:PropBSs:RefModel}. The meaning of the plots in this Figure,
starting from the top-left corner, is the following. The first plot presents a
few characteristic radii, like $r_c$, $r_h$ and $r_{tid}$. Please note, that the
$r_c$ is slowly getting smaller. It will be important in the discussion on the
influence of the initial concentration on the populations of BSs
(Sect.\ref{sec:PropBSs:concentrations}). The second plot shows the initial
distribution of mass ratios for binaries (denoted as U (uniform) in
Tab.~\ref{tab:Sim:Ini1} and Tab.~\ref{tab:Sim:Ini2}). The plot does not resemble
uniform (flat) distribution because not all initially drawn mass ratios can be
chosen -- they cannot exceed the maximum mass for a binary ($100 M_{\odot}$) and
the minimum mass of a star ($0.08 M_{\odot}$). The third plot presents the 
initial distribution of semi-major axes of binaries. It is a uniform
distribution in logarithmic scale, between $4 R_n$ and 100~AU ($R_n$ is a star
with the smallest radii). It is denoted as UL in Tab.~\ref{tab:Sim:Ini1} and
Tab.~\ref{tab:Sim:Ini2}. Here again, the distribution is not entirely flat.
There are some missing compact binaries in this distribution around ($\approx 0
[log(R/R_{Sun})]$) -- the semi-major axis cannot be too small because it would
create an immediate merger just in the first call of the stellar evolution. The
forth plot shows the initial distribution of eccentricities (thermal
distribution, denoted with T in Tab.~\ref{tab:Sim:Ini1} and
Tab.~\ref{tab:Sim:Ini2}, it is uniform in $e^2$). There are here some missing
binaries with high eccentricities too. For some of the binaries eccentricity
cannot be close to 1 because it would create an immediate merger too. The fifth
plot shows the number of BSs of different types at a given time. BSs can
change types (see Sect.~\ref{sec:PropBSs:Channels}), thus in the sixth plot
there is the number of BSs of different initial types (the types of BSs at the
time of their creation). Here the differences between the initial and present
populations of BSs are not significant. They are more important for initially
more concentrated clusters (see e.g. Sect.~\ref{sec:PropBSs:concentrations}).

The number of EM BSs increases within the first few Gyr (see
Fig.~\ref{fig:PropBSs:RefModel}) as a result of two formation scenarios. EM in
the first few Gyr are formed due to the Roche lobe overflow in compact binaries.
The semi-major axis of the binary decreases slightly, after some time the
heavier star leaves the main-sequence and its radius increases. The
semi-detached phase starts, which leads to a merger. The second scenario of EM
formation involves magnetic braking for slightly wider binaries and works for
stars with masses less than about 1.25~$M_{\odot}$. Around the time 3~Gyr, the
turn-off mass equals 1.25~$M_{\odot}$ and magnetic braking starts to work for
both components in the binary (if they are main-sequence stars). This causes
that the EM channel is most efficient around that time. The peak of EM channel
for the \textsc{mocca-ref} model (and many others) is around 5~Gyr (for more
details see \citet[Sect.~4.1.3]{Hypki2013MNRAS.429.1221H}).

The EMT channel is the most active in the model \textsc{mocca-ref} (and many
others) during the first few Gyr as a result of the initial binary properties
(see Fig.~\ref{fig:PropBSs:RefModel}). There are two scenarios of forming EMT
BSs. The first one creates EMT through the mass transfer in the Roche lobe
overflow in a compact binary. In the second scenario in a wide binary a mass is
transfered through stellar winds when a companion goes through the AGB phase.
Both scenarios of formation of EMT BSs are the most active during the first few
Gyr because the mass transfer concerns compact binaries and wide binaries
together \citep[Fig.~5]{Hypki2013MNRAS.429.1221H}. During the first few Gyr the
mass transfer is possible for the largest number of binaries. The significance
of EMT decreases with time because the mass transfer is less effective for less
massive stars (for more details see
\citet[Sect.~4.1.2]{Hypki2013MNRAS.429.1221H}).

The number of the dynamical BSs (DBS, DBB) increases steady with time. It is
caused by the increasing density in the GC (see radii in
Fig.~\ref{fig:PropBSs:RefModel}). Eventually, the dynamical BSs, at least for
the \textsc{mocca-ref} model, become more important than EMT and more or less as
numerous as from the EM channel.

\textsc{mocca-12} is a very standard model which slowly evolves toward the
core collapse. Its density in the core raises during the Hubble time so that the
number of BSs created due to dynamical interactions becomes important. It is a
very reasonable standard model of a real size GC. The only significant
difference as compared to real clusters is a slightly larger fraction of
primordial binaries. Usually, in GCs one can observe a fraction $\lesssim 10$\%,
whereas the \textsc{mocca-12} has 20\%. The larger number is chosen to have a
larger number of BSs and thus to highlight their features. The chosen $r_h$ is
also slightly larger than for a typical GC. Such value of $r_h$ was chosen to
give us a freedom in both increasing and decreasing its value for other models.

\subsubsection{Influence of semi-major axes distribution on BSs population}
\label{sec:PropBSs:isemi}

This subsection shows how different semi-major axes distributions change the
populations of BSs.

\begin{figure}
	\includegraphics[trim={0 2cm 0
	2cm},width=\columnwidth]{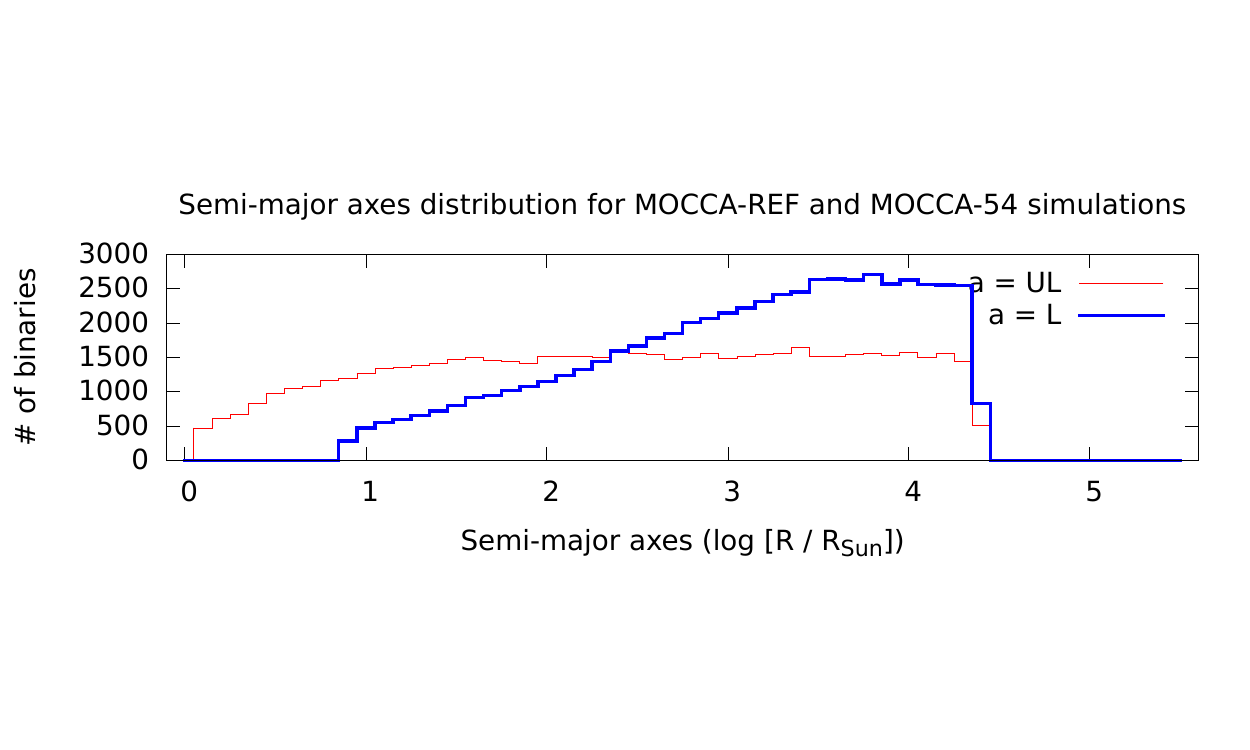}
	\includegraphics[width=\columnwidth]{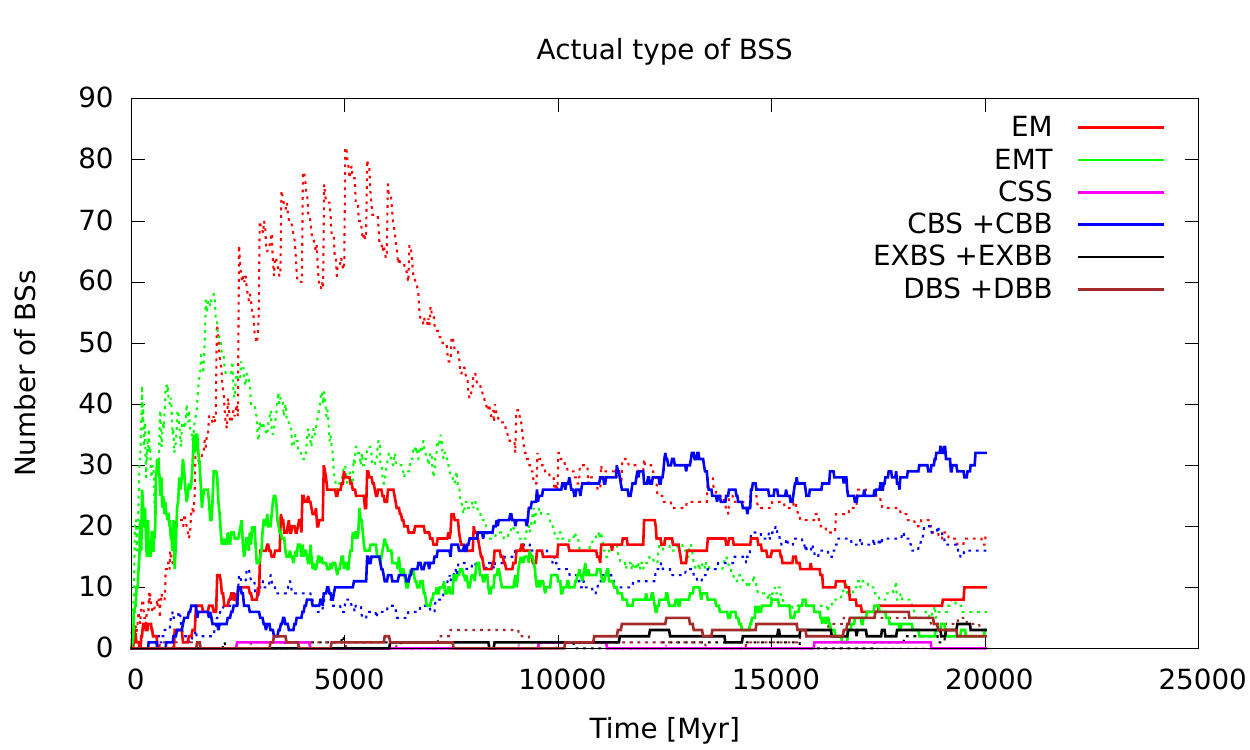}
\caption[Comparison of different populations of BSs for two models with
different initial semi-major axes distributions (UL, L)]{Comparison between the
reference model \textsc{mocca-ref} and the model \textsc{mocca-54}. The
difference between models concerns the initial semi-major axes distributions
(top). The reference model \textsc{mocca-ref} has a uniform distribution in log
scale (a = UL, red line, top plot), while the model \textsc{mocca-54} has
lognormal distribution (a = L, blue line, top plot). The result differences in
the number of BSs of different types is presented in the bottom plot. BSs from
the \textsc{mocca-ref} model are presented with dashed lines, whereas the BSs
from \textsc{mocca-54} models with solid lines. For definitions of BSs types see
Sect.~\ref{sec:PropBSs:Channels} and for details see the text.}
\label{fig:PropBSs:isemi0_vs_isemi1}
\end{figure}

Fig.~\ref{fig:PropBSs:isemi0_vs_isemi1} shows the differences between the
reference model \textsc{mocca-ref} and the model \textsc{model-54}. The
properties of the \textsc{mocca-ref} were discussed in the
Sect.~\ref{sec:MOCCAREF}. The difference between models
concerns only the initial semi-major axes distributions (top plot in
Fig.~\ref{fig:PropBSs:isemi0_vs_isemi1}). The reference model \textsc{mocca-ref}
has a uniform distribution in log scale (a = UL, red line, top plot) and the
model \textsc{mocca-54} has lognormal distribution (a = L, blue line, top plot).
The model \textsc{mocca-ref} has many more binaries with small and medium
semi-major axes ($< 10^{2.5} [R_{\odot}]$). In turn, \textsc{mocca-54} has more
binaries with longer semi-major axes ($> 10^{2.5} [R_{\odot}]$).

The differences in semi-major axes distributions between \textsc{mocca-ref} and
\textsc{mocca-54} models yield significant differences in the channels of
formation of BSs. The number of EMT and EM is significantly lower for \textsc{mocca-54}
model.
However, the number of dynamical BSs (CBS, CBB) is higher. The number of
exchanges and dissolution in BSs is not important for both models.

The lower number of EM and EMT BSs for \textsc{mocca-54} is a consequence of its
semi-major axes distribution for which there is less compact binaries. Small
semi-major axes are expected for EM BSs, because in order to have an
evolutionary merger, one has to have a compact binary.
Additionally, binaries in \textsc{mocca-54} model need more time to have
evolutionary mergers, thus, the number of EM BSs becomes significant after 5~Gyr
-- the number of EM for \textsc{mocca-ref} model is then already at its peak and
is a dominant channel of formation. For more details about the physical
processes of formation of EM see
\citet[Sect.~4.1.3]{Hypki2013MNRAS.429.1221H}.

A similar explanation applies also for the lower number of EMT BSs for
\textsc{mocca-54} model. There are two subgroups of EMT BSs. The first group
consists of harder binaries ($< 10$~days) for which there is some Roche lobe
overflow. The second group consists of wider binaries ($> 100$~days) for which
future BS gains some additional mass trough stellar winds, when the companion
goes through the AGB phase. In the second subgroup the eccentricities are
significantly larger then 0.1 (even 0.9) which makes the mass transfer a bit
easier. The second subgroup creates less BSs than the first one. In
\textsc{mocca-54} there is much less compact binaries (see top panel in
Fig.~\ref{fig:PropBSs:isemi0_vs_isemi1}) thus the number of EMT is smaller too.
For detail description of the physical
processes of formation of both subgroups of EMT channel see
\citet[Sect.~4.1.2]{Hypki2013MNRAS.429.1221H}.

Less intuitive explanation concerns the higher number of dynamical BSs (CBS,
CBB) for \textsc{mocca-54} model. The dynamical BS is created due to a physical
collision (or collisions) which occurs during a dynamical interaction. For
\textsc{mocca-54} there are in overall many more dynamical interactions because
it contains a larger number of wider binaries (see top panel in
Fig.~\ref{fig:PropBSs:isemi0_vs_isemi1}). There are 35k dynamical interactions
for \textsc{mocca-ref} model within 20~Gyr and 60k for \textsc{mocca-54}. It is
almost twice as many. As a result, there is also more physical collisions (140)
during these interactions for \textsc{mocca-54} model, whereas for
\textsc{mocca-ref} there are only 80 collisions. The models are identical,
except the semi-major axes distributions. The number of binaries and the
concentration are the same, the GCs for both models evolve very similarly, the
characteristic radii like $r_c$, or $r_h$ are very similar too. 

The only difference between \textsc{mocca-ref} and \textsc{mocca-54} is 
that for \textsc{mocca-54} the average semi-major axes for binaries are larger. Wider
binaries have larger probabilities of having dynamical interactions. Many of
them are in fact only distant fly-by interactions, which do not change
significantly semi-major axes. However, these interactions increase the
eccentricities. Larger eccentricities increase the probabilities of the
collisions further. At some point, the \textsc{fewbody} code detects a
collision, when the periastron distance gets smaller than the sum of the radii
of stars. As a result there are more dynamical BSs (CBS and CBB) for the models
which have initial semi-major axes distribution containing more wide binaries,
despite the fact that the initial concentrations for both models are the same
(e.g. \textsc{mocca-54}). This scenario of formation of the BSs is discussed
in details in Sect.~\ref{sec:PropBSs:MoccaRaising}.

\begin{figure}
	\includegraphics[trim={0 2cm 0
	2cm},width=\columnwidth]{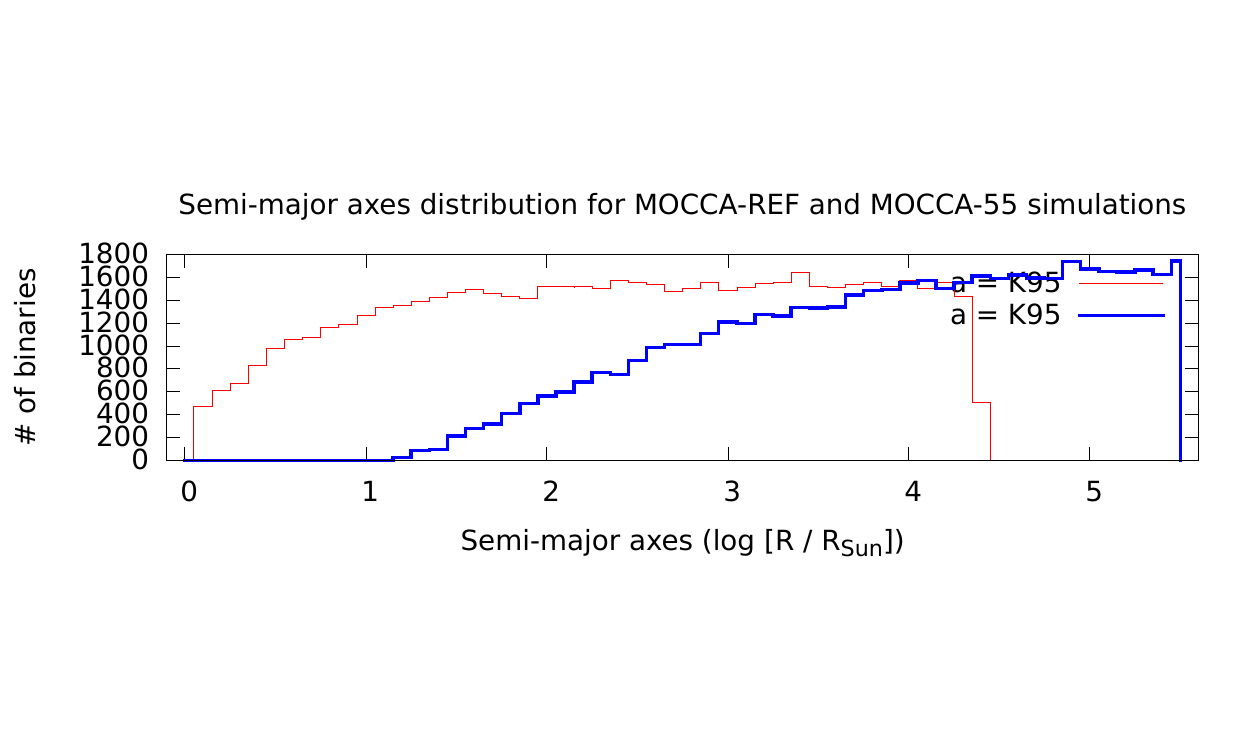}
	\includegraphics[width=\columnwidth]{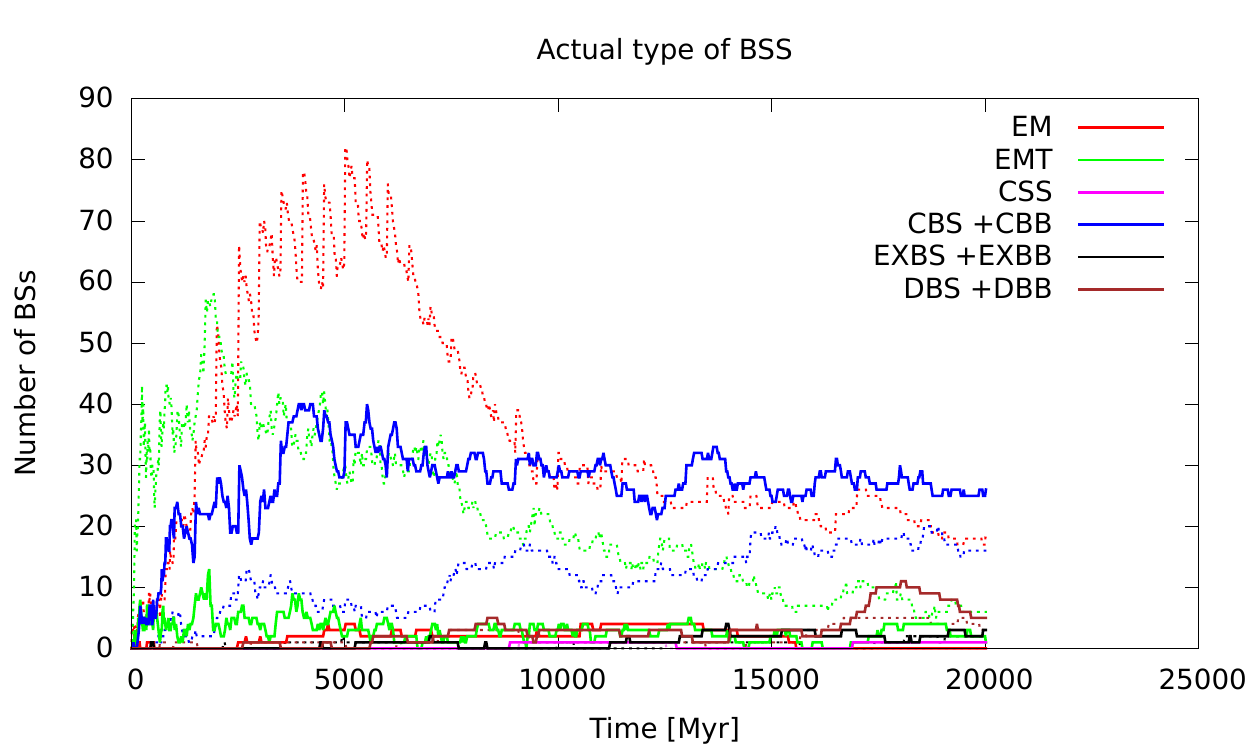}
\caption[Comparison of different populations of BSs for two models with
different initial semi-major axes distributions (UL, K95)]{Comparison between
the reference model \textsc{mocca-ref} and model \textsc{mocca-55}. The
difference between models concerns the initial semi-major axes distributions
(top). The reference model \textsc{mocca-ref} has a uniform distribution in log
scale (a = UL, red line, top plot) and the model \textsc{mocca-55} has the
distribution from \citet{Kroupa1995bMNRAS.277.1507K} (a = K95, blue line, top
plot). The result differences in the number of BSs of different types are
presented in the bottom plot. BSs from the \textsc{mocca-ref} model are
presented with dashed lines, whereas the BSs from \textsc{mocca-55} models with
solid lines. For definitions of BSs types see Sect.~\ref{sec:PropBSs:Channels}
and for details see the text.}
\label{fig:PropBSs:isemi0_vs_isemi2}
\end{figure}

Fig.~\ref{fig:PropBSs:isemi0_vs_isemi2} shows the differences between the
reference model \textsc{mocca-ref} and another model, \textsc{model-55}. The
difference concerns only the initial semi-major axes distributions (top plot in
Fig.~\ref{fig:PropBSs:isemi0_vs_isemi2}). The reference model \textsc{mocca-ref}
has a uniform distribution in log scale and the model \textsc{mocca-55} has the
distribution of \citet{Kroupa1995bMNRAS.277.1507K}, which results in the given
semi-major axes distribution (a = K95, blue line, top plot). The model
\textsc{mocca-ref} has many more binaries with small and medium semi-major axes
($< 10^{2.5} [R_{\odot}]$). In turn, \textsc{mocca-55} has binaries with larger
semi-major axes ($> 10^{4.5} [R_{\odot}]$).
Fig.~\ref{fig:PropBSs:isemi0_vs_isemi2} shows that the number of EMT, EM is very
low for \textsc{mocca-55} model. However, the number of dynamical BSs (CBS, CBB) is much higher.

When it comes to the differences between \textsc{mocca-55} and previously
discussed model \textsc{mocca-54} (see Fig.~\ref{fig:PropBSs:isemi0_vs_isemi1}),
the \textsc{mocca-55} has even less binaries with semi-major axes $< 10^{2.5}
[R_{\odot}]$ than model \textsc{mocca-54}. 
This has an influence on the number of BSs.
Previous \textsc{mocca-54} model has the average number of EM
around 20 and EMT around 15. Whereas the \textsc{mocca-55} models has only 
a few of EM and EMT BSs. The number of CBS and CBB BSs increase faster for
\textsc{mocca-55} model until reaching the average value around 30 -- the same
as for the previous \textsc{mocca-54} model.

The differences in the number of BSs for the \textsc{mocca-ref} and
\textsc{mocca-55} models confirm and make it even more explicit to notice the
previously stated conclusions. The EM and most of EMT BSs are created from
compact binaries. Model \textsc{mocca-55} has the lowest number of such binaries
and thus the number of BSs of these types is not significant at all. There are
on average only a few EM and EMT BSs for model \textsc{mocca-55}.

The number of CBS and CBB BSs is larger for model \textsc{mocca-55} than for
\textsc{mocca-ref} and previously
discussed model \textsc{mocca-54}. The explanation is the same as previously. For
\textsc{mocca-55} model there are more binaries with larger semi-major axes.
Thus, the probability of dynamical interactions is higher. The interactions
increase eccentricities to such values that binaries eventually merge. For
\textsc{mocca-ref} model for the first 5~Gyr there were around 17k dynamical
interactions, whereas for the \textsc{mocca-55} there were already 176k
interactions. Thus, the number of CBS+CBB BSs increases so fast during the first
5~Gyr. This is also the reason why the number of CBS and CBB BSs in
\textsc{mocca-55} increases faster than for  \textsc{mocca-54} model. There are
many more dynamical interactions which lead to collisional events.
Interestingly, the number of CBS+CBB raises to the same level of about 30 BSs
after the first 5~Gyr for both models. The reason why the number of CBS and CBB
stopped to increase for \textsc{mocca-55} model is caused by the number of
binaries destroyed by dynamical interactions. The number of destroyed binaries
for \textsc{mocca-55} after 10~Gyr is 28.3k binaries, whereas for
\textsc{mocca-54} it is only 3.5k. Most of the widest binaries were destroyed
and thus the number of CBS+CBB stopped to increase.

The last remark is needed to complete the discussion on the influence of the
initial semi-major axes distribution on the populations of BSs. The
\textsc{mocca} code allows to change the maximum semi-major axis for binaries.
By default it is 100~AU ($\sim 10^{4.3} [R_{\odot}]$). Of course, if one set
this value to a larger one, the number of BSs of different channel would change.
However, it is expected to see differences obeying the conclusions specified in
the previous paragraphs. It is expected to see more EM and EMT BSs if there are
initially more compact binaries. And when the number of wide binaries is larger,
then the number of dynamical BSs should increase. The maximum values of
semi-major axes for the \textsc{mocca-ref} and \textsc{mocca-54} models are the
same (100~AU). Only \textsc{mocca-55}, which generates semi-major axes according
to \citet{Kroupa1995bMNRAS.277.1507K}, sets up its own value for maximum
semi-major axis. Testing the influence of wider orbits than
100~AU on the population of BSs is planned for the future research.

\subsubsection{Influence of eccentricity distribution on BSs population}
\label{sec:PropBSs:ecc}

\begin{figure}
	\includegraphics[trim={0 2cm 0
	2cm},width=\columnwidth]{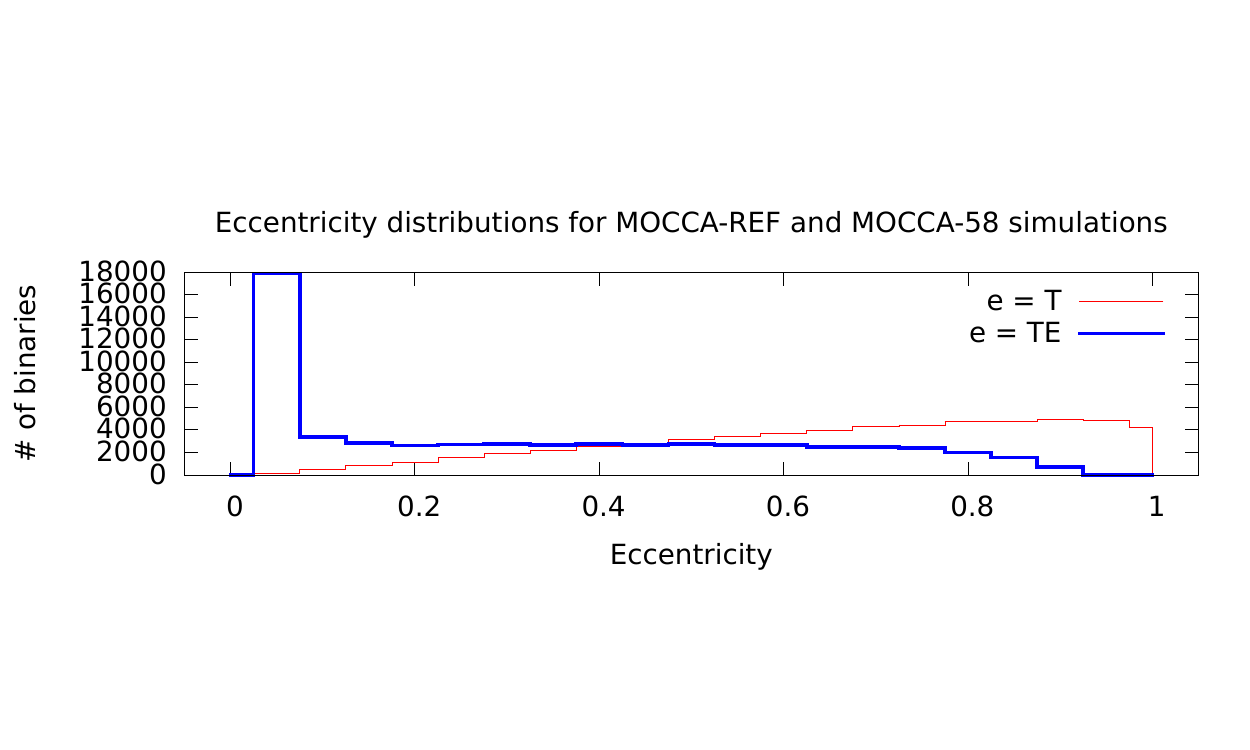}
	\includegraphics[width=\columnwidth]{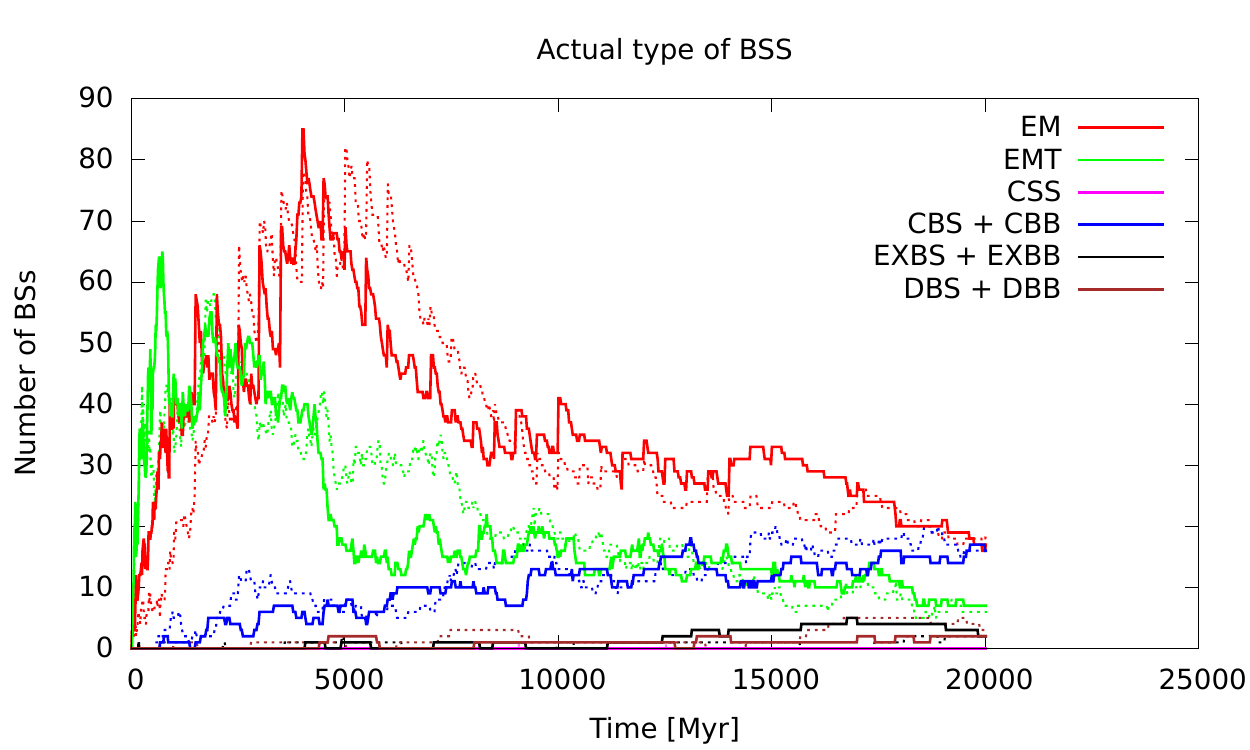}
\caption[Comparison of different populations of BSs for two models with
different initial eccentricity distributions (T, TE)]{Comparison between the
reference model \textsc{mocca-ref} and model \textsc{mocca-58}. The difference
between models concerns the initial eccentricity distributions (top). The
reference model \textsc{mocca-ref} has a thermal distribution (e = T, red line,
top plot) and the model \textsc{mocca-58} has a thermal distribution with
eigenevolution \citet{Kroupa1995bMNRAS.277.1507K} (e = TE, blue line, top plot).
The result differences in the number of BSs of different types are presented in
the bottom plot. BSs from the \textsc{mocca-ref} model are presented with dashed
lines, whereas the BSs from \textsc{mocca-58} model with solid lines. For
definitions of BSs types see Sect.~\ref{sec:PropBSs:Channels} and for details
see the text.}
\label{fig:PropBSs:ikroupa1}
\end{figure}

Fig.~\ref{fig:PropBSs:ikroupa1} shows how different initial eccentricity
distributions (top panel) influence the population of BSs of different types
(bottom panel). The model \textsc{mocca-ref} has a thermal distribution (e = T)
and the model \textsc{mocca-58} has a thermal distribution as well but with a
procedure called \textit{eigenevolution} applied to the eccentricity
distribution (not applied to orbital periods)
\citep{Kroupa1995bMNRAS.277.1507K}. The basic idea behind eigenevolution is to
modify a binary for small pericenter distances due to pre-main-sequence
evolution, when stellar radii are larger. In this procedure eccentricity is
sampled from a thermal distribution. A pre-main-sequence radius of a star of $5
m^{1/2} [M_{\odot}]$ is used to reflect the earlier contraction stage. Using the
tidal circularization theory \citep{Mardling2001MNRAS.321..398M}, modified
eccentricity $e_i$ for characteristic time interval of $10^5$ yr is computed.
Angular momentum conservation is used afterwards to compute semi-major axes.
Finally, any case of overlapping enlarged radii is defined as collision. It is
rejected and the procedure is repeated. As a result of the eigenevolution
procedure some fraction of compact binaries are circularized (see top plot in
Fig.~\ref{fig:PropBSs:ikroupa1}).

Fig.~\ref{fig:PropBSs:ikroupa1} shows that the population of evolutionary BSs is
not affected significantly by the changed eccentricity distribution. The overall
number of EM and EMT BSs is very similar for both models. The only noticeable
difference is that the EM channel raise a bit faster for \textsc{mocca-58}
model. The eigenevolution procedure circularize some of the compact binaries,
thus the EM channel seems to be raising in the beginning of the simulations more
quickly -- EM BSs do not need any additional time to circularize their orbits
and finally merge. However, the changes are within the fluctuations ($\pm 10$
BSs) thus cannot be recognised as certain.

The population of dynamical BSs is not affected by the different eccentricity
distribution. The eccentricities neither have influence on the probabilities of
the dynamical interactions, nor the probabilities of the collisions. Thus, the
populations of CBS and CBB for both models are well within the fluctuations.

\subsubsection{Influence of concentration on BSs population}
\label{sec:PropBSs:concentrations}

\begin{figure}
	\includegraphics[width=\columnwidth]{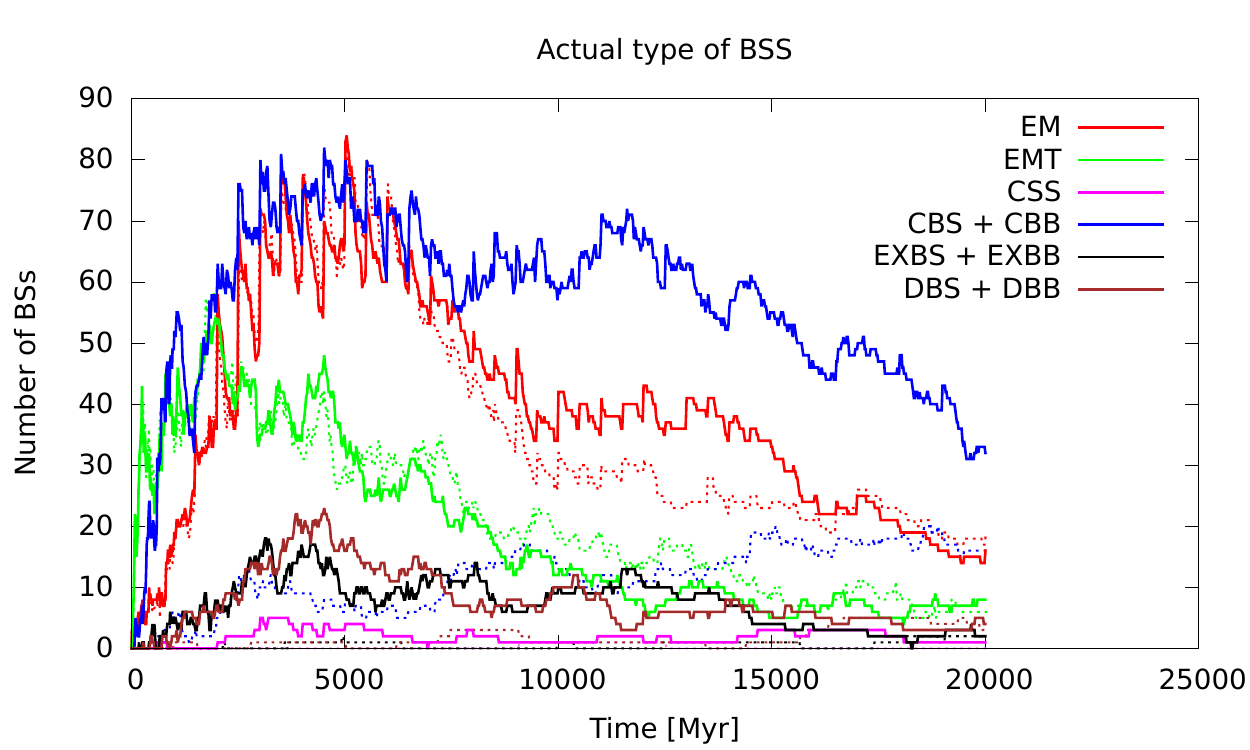}
\caption[Comparison of different populations of BSs for two models with
different initial concentrations ($c = 10, 30$)]{Comparison of the number of BSs
of different types between the reference model \textsc{mocca-ref} (dashed lines)
and model \textsc{mocca-8} (solid lines). The difference between models concerns
only the initial concentration $c = r_{tid}/r_h$. For the \textsc{mocca-ref} $c
= 10$, and for the \textsc{mocca-8} $c = 30$. The reference model
\textsc{mocca-ref} is described in details in
Sect.~\ref{sec:PropBSs:IniitialConditions}. For definitions of BSs types see
Sect.~\ref{sec:PropBSs:Channels} and for details see
Sect.~\ref{sec:PropBSs:concentrations}.}
\label{fig:PropBSs:rplum30}
\end{figure}

Fig.~\ref{fig:PropBSs:rplum30} shows how the population of BSs of different
types depends on the cluster concentration parameter. It shows differences
between \textsc{mocca-ref} model with concentration $c = r_{tid}/r_h = 10$, and
\textsc{mocca-8} model with $c = 30$. All other parameters between these models
are the same.

The number of evolutionary BSs (EM, EMT) is the same for both models. There are
some small differences but they fall well within the error ($\pm 10 BSs$, see
Sect.~\ref{sec:PropBSs:Channels}). The same number of EM and EMT BSs for the
models suggests that the concentration does not influence their population.
Much more important for them are the initial conditions -- especially the
semi-major axes (see Sect.~\ref{sec:PropBSs:isemi}). This suggest also that the
evolutionary BSs have been mostly created in unperturbed, primordial binaries.
The binaries, which later on created BSs, were not affected by close
interactions, even for the \textsc{mocca-8} model which has much greater
concentration.

The differences between \textsc{mocca-ref} and \textsc{mocca-8} models concerns
the number of dynamical BSs (CBS, CBB). For the more concentrated model,
\textsc{mocca-8}, their number increases fast and this channel of formation
becomes dominant just after $\sim~7$~Gyr. GCs with higher concentrations have
higher probabilities of interactions. Thus, the number of physical collisions
also increases. Interestingly, for higher concentrations the number of BSs of
types EXBS, EXBB, DBS, and DBB increases too (black and blue lines in
Fig.~\ref{sec:PropBSs:concentrations}).
It is caused by the numerous, strong dynamical interactions in which BSs change
their companions or are dissolved. It is the natural outcome of the highly
concentrated systems.

\begin{figure}
\includegraphics[width=\columnwidth]{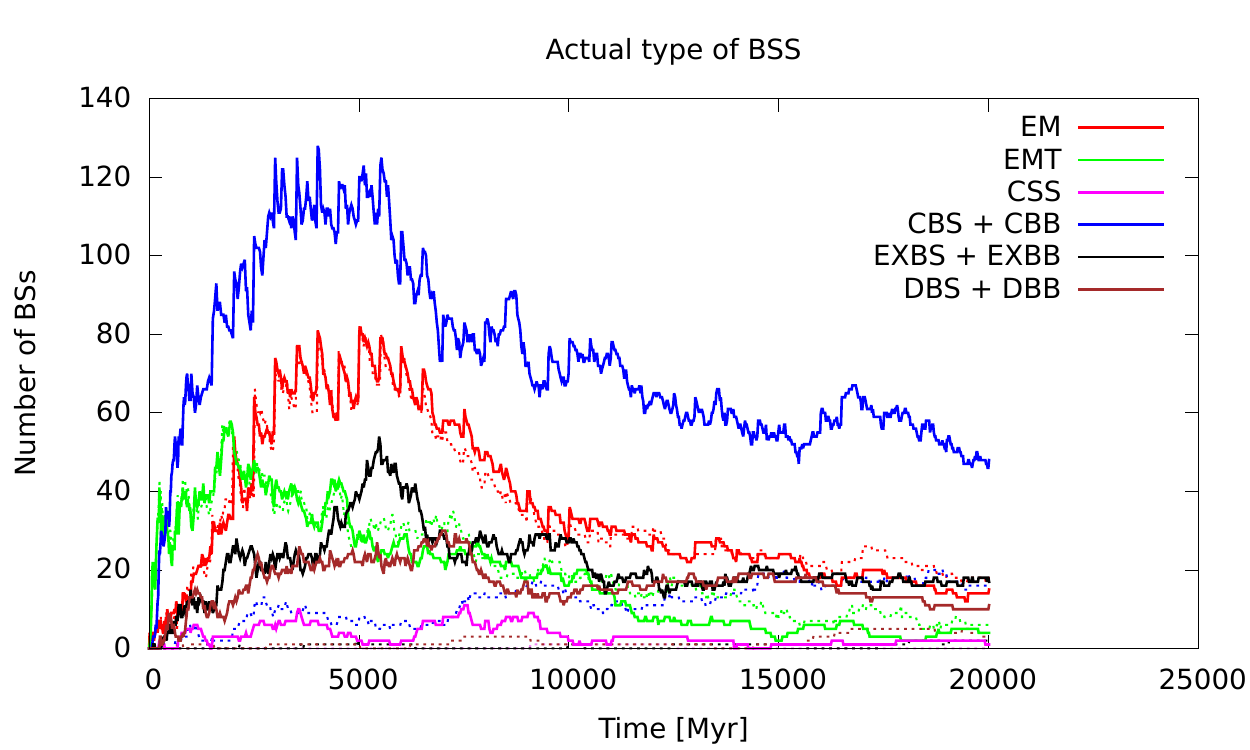}
\caption[Comparison of different populations of BSs for two models with
different initial concentrations ($c = 10, 40$)]{Comparison of the number of BSs
of different types between the reference model \textsc{mocca-ref} (dashed lines)
and model \textsc{mocca-7} (solid lines). The difference concerns
only the initial concentration $c = r_{tid}/r_h$. For the \textsc{mocca-ref} $c
= 10$, and for the \textsc{mocca-7} $c = 40$. The reference model
\textsc{mocca-ref} is described in details in
Sect.~\ref{sec:PropBSs:IniitialConditions}. For definitions of BSs types see
Sect.~\ref{sec:PropBSs:Channels} and for details see
Sect.~\ref{sec:PropBSs:concentrations}.}
\label{fig:PropBSs:rplum40}
\end{figure}

Even the larger concentration, $c = 40$ (\textsc{mocca-7}, see
Tab.~\ref{tab:Sim:Ini1}), does not change the population of EM and EMT BSs.
Their number is very much the same as for the model \textsc{mocca-ref} (see
Fig.~\ref{fig:PropBSs:rplum40}). The number of the dynamical BSs is naturally
larger. There are many more strong dynamical interactions leading to collisions.
There are more exchanges and dissolutions of binaries as well (EXBS, EXBB,
DBS, DBB).

The higher concentration ($c = 40$) puts even more confidence into the fact that
EM and EMT BSs seem to be formed in primordial, unperturbed binaries. In
Fig.~\ref{fig:PropBSs:rplum30} ($c = 30$) there is some difference in the number
of EM between $\sim 10$~Gyr and $\sim 15$~Gyr. However, these are most likely only some 
fluctuations, because for the higher concentration there is no such discrepancy ($c = 40$,
Fig.~\ref{fig:PropBSs:rplum40}).

\subsubsection{Relation between hard and soft binaries}

Different initial semi-major axes distributions and initial clusters'
concentrations have variuos influence on the populations of dynamical BSs because of (i) the
boundary between hard and soft binaries and (ii) the probabilities of
the dynamical interactions. The boundary between hard and soft binaries plays
a very important role in GCs. Hard binaries have binding energies larger
than the average kinetic energies of stars, whereas soft binaries inversely.
\citet{Heggie1975MNRAS.173..729H} showed that in general hard binaries, due
to dynamical evolution of GCs, get harder, and soft binaries get softer (wider)
or disrupt.

The division between the hard and soft binaries can be expressed as an
average binary semi-major axis of a GC.

The cluster binding energy is equal to $E_b = - \frac{0.4 G M^2}{r_h}$, where G
is the gravitational constant, M -- the total mass, and $r_h$ -- the half-mass
radius. The cluster kinetic energy equals $E_k = 0.5 M v_h^2$, where $v_h$ is
the average velocity in the system. Thus, using the virial theorem, the average
velocity can be expressed as $v_h^2 = \frac{0.8 G M}{r_h}$. The total mass can
be expressed as $M = \overline{m} N$, where $\overline{m}$ is the average mass
in the system and N -- the number of stars.

The boundary semi-major axis ($\overline{a}$) between soft and hard binaries can
be derived from the equality between average kinetic and binding
energy of a binary:

\[0.5 \overline{m} v_h^2 = \frac{G \overline{m}^2}{2 \overline{a}} \Rightarrow
\overline{a} = \frac{G \overline{m}^2}{\overline{m} v_h^2} \]

using expressions for $v_h^2$, and M we get:

\begin{equation} \label{eq:Hard:a}
\overline{a} = \frac{G M r_h}{N 0.8 G M} = \frac{r_h}{0.8 N}
\end{equation}

In a single time step, a probability of a strong dynamical interaction between a
binary and another object (single star or binary) is given by:

\begin{equation}
P = \sigma_h n_h u_h \Delta t
\end{equation}

where $\sigma_h = \pi \overline{a}^2 (\frac{2 G (m_b +
\overline{m})}{\overline{a} u_h} + 1)$ is the cross section for the interaction,
$u_h$ -- the relative velocity at the infinity, $n_h$ -- the local number density of
stars (single or binaries), $m_b$ -- the average mass of a binary, and $\Delta~t$
is the time step. It is assumed that the strong interaction occurs when an incoming
object approaches binary at the distance comparable to the binary semi-major
axis. The relative velocity can be expressed by the average velocity as $u_h^2
= 2 v_h^2$, the average mass of the binary as $m_b = 2 \overline{m}$, and taking the
expression for $v_h^2$, we get:

\[ \sigma_h = \pi \overline{a}^2 (\frac{2 G 3 \overline{m}}{\overline{a} 2
v_h^2} + 1) = \sigma_h = \pi \overline{a}^2 (\frac{3}{\overline{a}} \frac{r_h
\overline{m}}{0.8 M} + 1) = \pi \overline{a}^2 (\frac{3}{\overline{a}} \frac{r_h}{0.8 N} +
1)\]

using expression for $\overline{a}$, we get:

\begin{equation}
\sigma_h = 4 \pi \overline{a}^2
\end{equation}

The local number density (computed only inside $r_h$, thus only half of the
mass is taken) is:

\begin{equation}
\rho_h = \overline{m} n_h = \frac{3 M}{8 \pi r_h^3} \Rightarrow n_h = \frac{3 N}{8 \pi r_h^3}
\end{equation}

Using the above expressions and simplifying the probability for the strong dynamical
interactions, we get:

\begin{equation} \label{eq:Hard:P}
P = \frac{1.5 \sqrt{1.6 G}}{0.8^2} \frac{M^{1/2}}{ N r_h^2} \propto
\frac{\overline{m}}{M^{1/2} r_h^{3/2}} =
\frac{\overline{m}^{1/2}}{N^{1/2} r_h^{3/2}}
\end{equation}

From the Eq.~\ref{eq:Hard:a} we have $\overline{a} \propto \frac{r_h}{N}$ and
from the Eq.~\ref{eq:Hard:P}, $P \propto \frac{1}{N^{1/2} r_h^{3/2}}$. The
influence of the initial semi-major axes distributions and initial
concentrations of the GCs on the CBS and CBB BSs is a result of the interplay
between these two equations.

In the case of different semi-major axes distributions (but the same $r_h$
and N) the average semi-major axis $\overline{a} = 1275 [R_{\odot}] = 3.1 [log
R_{\odot}]$ is the same for models \textsc{mocca-ref}, \textsc{mocca-54} (see
Fig.~\ref{fig:PropBSs:isemi0_vs_isemi1}), and \textsc{mocca-55} (see
Fig.~\ref{fig:PropBSs:isemi0_vs_isemi2}). However, the boundary between hard and
soft binaries ($\overline{a}$) causes that for \textsc{mocca-54} there are more
binaries which are softer (wider) from the point of view of the GC. There are
even more softer binaries for the \textsc{mocca-55} model.
It causes that there will be many more dynamical interactions for the models
with a large number of soft binaries.
Mostly they are just fly-bys which continuously increases binaries'
eccentricities. Eventually, because of their high eccentricities, more binaries
collide.
Thus, for the semi-major axes distributions with a larger number of soft binaries,
the number of dynamical BSs is higher as well.

In the case of different initial clusters' concentrations (see
Fig.~\ref{fig:PropBSs:rplum30} and Fig.~\ref{fig:PropBSs:rplum40}), but the same
initial semi-major axes distributions, the values of $\overline{a}$ decrease for
the higher concentrations. For the \textsc{mocca-ref}  model ($r_h = 6.9$, $c =
10$, see Fig.~\ref{fig:PropBSs:RefModel}) the average semi-major axis is
$\overline{a} = 3.1 [log R_{\odot}]$, whereas for the model \textsc{mocca-7}
($r_h = 1.7$, $c = 40$) it decreases to $\overline{a} = 2.5 [log R_{\odot}]$.
Because of Eq.~\ref{eq:Hard:P} the probabilities of dynamical interactions
increase for higher concentrations (lower $r_h$, but the same N). For the
\textsc{mocca-ref} model $P \propto \frac{1}{r_h^{3/2}} \propto 0.05$. In turn,
for the \textsc{mocca-7} model $P \propto \frac{1}{r_h^{3/2}} \propto 0.45$ --
the probability is 9 times higher. For higher concentrated models (smaller
$\overline{a}$) there are more binaries with small semi-major axes which
simultaneously are soft from the point of view of the GC. This causes that there
are more collisions for them and the number of CBS and CBB increase with the
higher concentrations.

For higher initial concentrations (smaller $\overline{a}$) the number of EM and
EMT stays the same because binaries which created these BSs are hard 
for all models (\textsc{mocca-ref}, \textsc{mocca-7}, and \textsc{mocca-8}). The
value of $\overline{a}$ is not small enough to start to disrupt the binaries
which lead to the creation of the EM and EMT BSs. Only the population of EMT
which is created from wide binaries (through the stellar winds, see
Sect.~\ref{sec:PropBSs:Channels}) could be affected. However, their number is
low.

\subsection{Blue stragglers in binaries}
\label{sec:PropBSs:Binaries}

The ratio of the number of BSs in binaries and as single stars is a very interesting
subject to study.
It might reveal some hidden properties of BSs and lead to some methods of
narrowing down the initial distributions of semi-major axes of binaries.
This section presents an analysis of these ratios for all of the \textsc{mocca} simulations described in 
Sect.~\ref{sec:InitialParameters}.

\begin{figure*}
  \includegraphics[width=18cm]{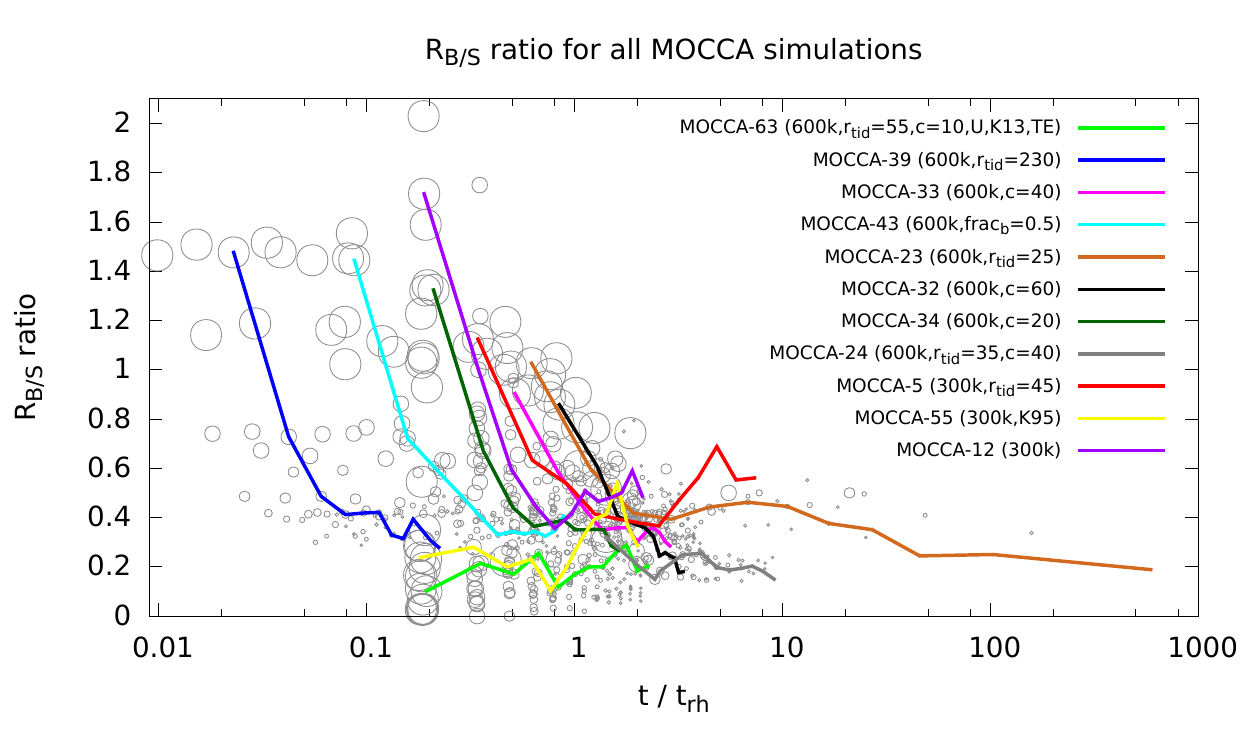}
  \caption[Evolving ratios between BSs in binaries to BSs as a single
  stars]{Evolving ratios between BSs in binaries and as single stars ($R_{B/S}$)
  for all \textsc{mocca} simulations (Tab.~\ref{tab:Sim:Ini1},
  Tab.~\ref{tab:Sim:Ini2}) as a function of time in the units of the present
  half-mass relaxation times ($t_{rh}$) but only up to 12~Gyr. The X axis shows
  the dynamical scales of the GCs -- higher values for dynamically older GCs.
  Each circle represents one ratio $R_{B/S}$ for one \textsc{mocca} simulation
  for one snapshot in time. The BSs from all channels are combined together into
  one circle. For each \textsc{mocca} simulation the ratio $R_{B/S}$ is computed
  every 1~Gyr (thus, max 12 circles for one simulation -- 1...12~Gyr). The sizes
  of circles decrease with time. For several selected simulations there are
  plotted lines to show the evolution of the ratio $R_{B/S}$. The values in
  brackets are copied from Tab.~\ref{tab:Sim:Ini1} and Tab.~\ref{tab:Sim:Ini2}
  for clarity. They specify only these parameters which are different from the
  parameters of models \textsc{mocca-35} and \textsc{mocca-12} (standard models
  for 600k and 300k respectively). For details see the text.
  }
  \label{fig:BinSin:All}
\end{figure*}

Fig.~\ref{fig:BinSin:All} shows the ratios between BSs in binaries to BSs as 
single stars ($R_{B/S}$) for the \textsc{mocca} simulations (Y axis). The time
on X axis is given in the units of the present half-mass relaxation times
($t_{rh}$), but only up to 12~Gyr. It shows the dynamical scales of the GCs. The
higher values are characteristic for dynamically older GCs. By using such a
scale one can easily compare GCs with very different initial conditions. Each
circle represents one ratio $R_{B/S}$ computed for one \textsc{mocca} simulation for one
snapshot in time. The BSs from all channels are combined together into one
circle. For each \textsc{mocca} simulation the ratio $R_{B/S}$ is computed every
1~Gyr (thus, max 12 circles for one simulation -- 1...12~Gyr). The sizes of
circles decrease with time. For several selected simulations there are plotted
lines to show overall evolution of the ratio $R_{B/S}$.

The lines with ratios $R_{B/S}$ in Fig.~\ref{fig:BinSin:All} for all
\textsc{mocca} simulations reveal two separate groups of models.

The first group of models, called \textsc{mocca-dropping}, starts with the ratio
$R_{B/S} \gtrsim 1.0$ (at $T = 1$~Gyr). Then, as the evolution of GCs proceeds,
the ratio drops continuously until it settles around $R_{B/S} \sim 0.3$. This
trend is well represented by models e.g. \textsc{mocca-39} (blue),
\textsc{mocca-33} (magenta), \textsc{mocca-34} (dark green). For a few 
models of this group there are observed some exceptions. The models
\textsc{mocca-5} (red), \textsc{mocca-12} (violet) raise after several Gyr to
values around $R_{B/S} \sim 0.5$. For a few other models, e.g.
\textsc{mocca-32}, \textsc{mocca-24}, the ratio drops to values $R_{B/S}
\sim 0.2$ and intersects with models of the second group.

The second group of models, called \textsc{mocca-raising}, starts with the
ratios $R_{B/S} \lesssim 0.3$. The ratios stay at the same level or raise
slightly with time (e.g. \textsc{mocca-63}, light green; \textsc{mocca-55},
yellow). The ratios from this group, in general, do not increase significantly
above the level $R_{B/S} \sim 0.3$.

The models \textsc{mocca-dropping} and \textsc{mocca-raising} have various
initial conditions (see Sect.~\ref{sec:InitialParameters}). There are models
which evolve slowly almost as isolated clusters, the other ones are very dense
and thus they evolve quickly. Some models have a large number of compact
binaries, whereas other have many wide binaries. Despite all these differences,
all models seem to evolve towards the values $R_{B/S} \sim 0.3$. The level
$R_{B/S} \sim 0.3$ seem to be a universal one.

\begin{figure*}
  \includegraphics[width=18cm]{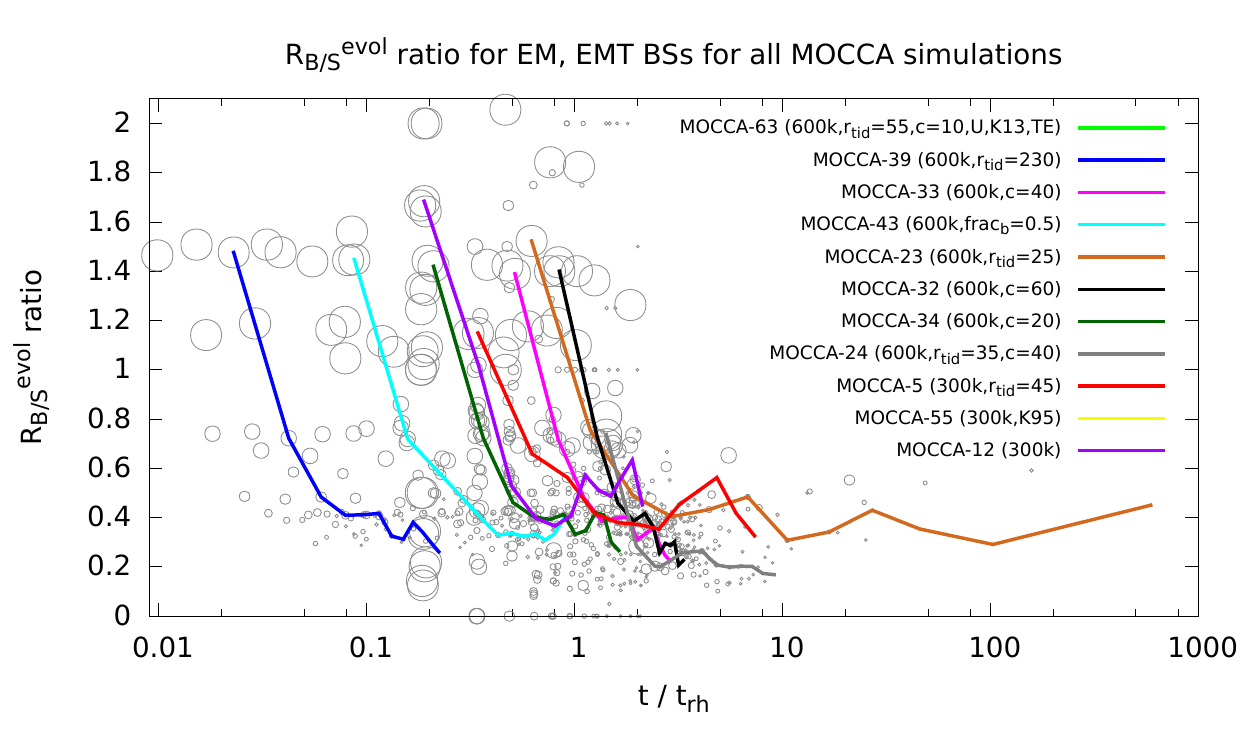}
  \includegraphics[width=18cm]{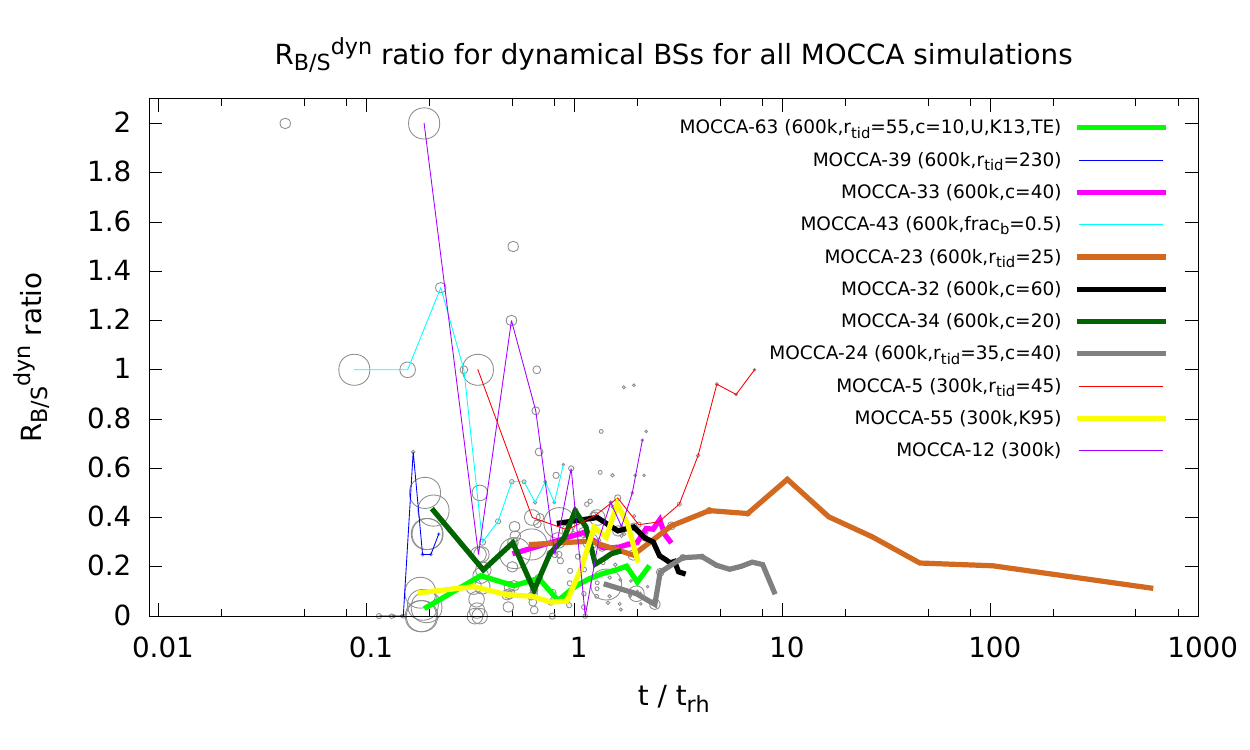}
  \caption[Evolving ratios between BSs in binaries to BSs as a single stars for
  evolutionary and dynamical BSs separately]{Evolving ratios between BSs in
  binaries and as single stars for all \textsc{mocca} simulations described in 
  Sect.~\ref{sec:InitialParameters} for evolutionary (top, $R_{B/S}^{evol}$) and
  dynamical (bottom, $R_{B/S}^{dyn}$) BSs separately. The meaning of symbols and
  lines is the same as for Fig.~\ref{fig:BinSin:All}. The colors of lines for
  selected models are consistent with Fig.~\ref{fig:BinSin:All} too. For the top
  plot two models (\textsc{mocca-63}, \textsc{mocca-55}) are not presented
  because of the low number of EM and EMT BSs (around 10) and thus a large
  scatter. For  dynamical BSs (bottom panel) not all simulations are plotted
  (thus, a lower number of gray circles). Thin lines in the bottom plot indicate
  the models with a very low number of dynamical BSs ($\sim 10$) which introduce
  a large scattering. For details see the text.
  }
  \label{fig:BinSin:Separately}
\end{figure*}

The ratio $R_{B/S}$ computed for all channels together adds complexity. Thus, we
 decided to split the ratio for the evolutionary and dynamical BSs.
Fig.~\ref{fig:BinSin:Separately} presents the ratio $R_{B/S}$ divided into
evolutionary (top panel, $R_{B/S}^{evol}$) and dynamical BSs (bottom panel,
$R_{B/S}^{dyn}$). This plot will be helpful in the discussion on the ratios
$R_{B/S}$ in the next paragraphs. In
\citet[Sect.~4.1.5]{Hypki2013MNRAS.429.1221H} one can find detail analysis of
the changes of types of BSs for a typical simulation. In general, the
evolutionary BSs do not often change their types. Additionally,
Sect.~\ref{sec:PropBSs:IniitialConditions} shows that the population of
evolutionary BSs strongly depends on the initial distribution of semi-major
axes. In turn, the concentrations of the GCs influence the population of the
dynamical BSs. They change types if the density of GC is high
enough. Thus, it is safe to assume that EM are single stars, EMT are in
binaries, and CBS, CBB, EXBS, EXBB, DBS, DBB channels can be both in binaries
as well as single stars. This allows to split $R_{B/S}$ without worrying 
about the changes of types of BSs.

\subsubsection{Models from \textsc{mocca-dropping} group}
\label{sec:PropBSs:MoccaDropping}

Evolution of the ratio $R_{B/S}$ for the models from \textsc{mocca-dropping}
group is mainly a consequence of the initial distribution of semi-major axes of
binaries.

\subsubsection{\textsc{mocca-39}}

\begin{figure*}
	\includegraphics[width=\columnwidth]{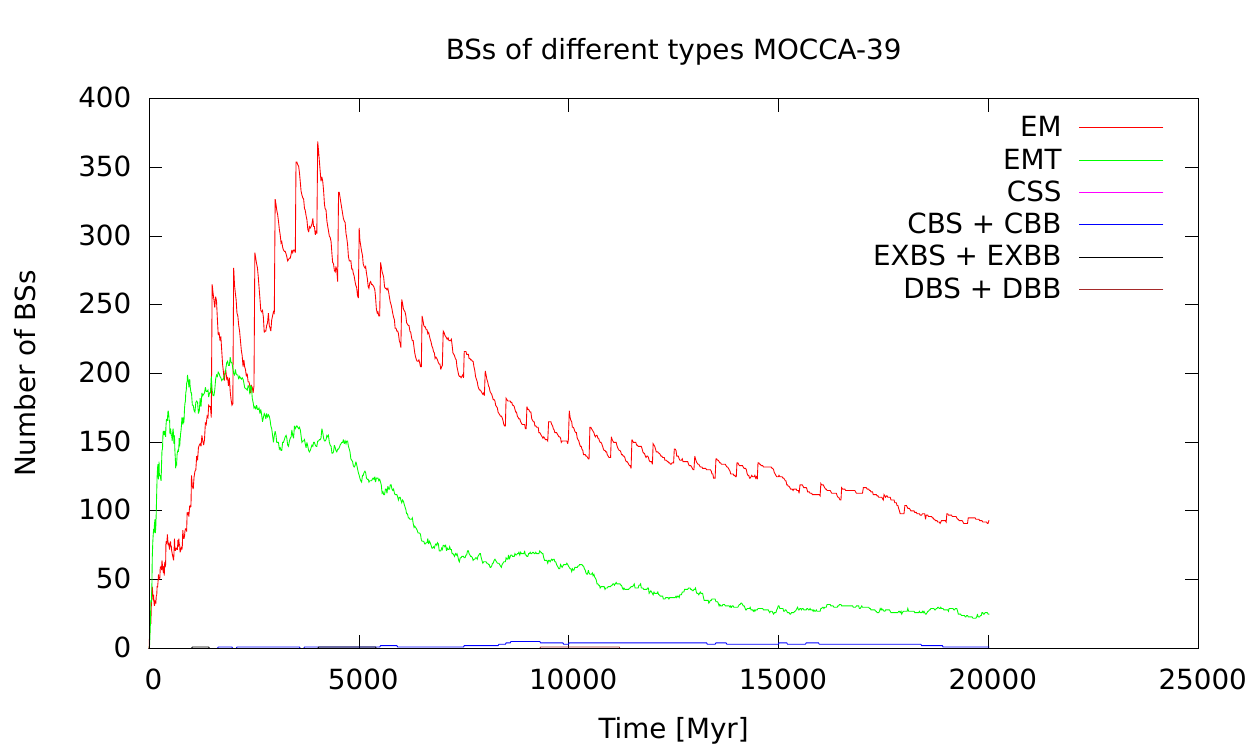}
	\includegraphics[width=\columnwidth]{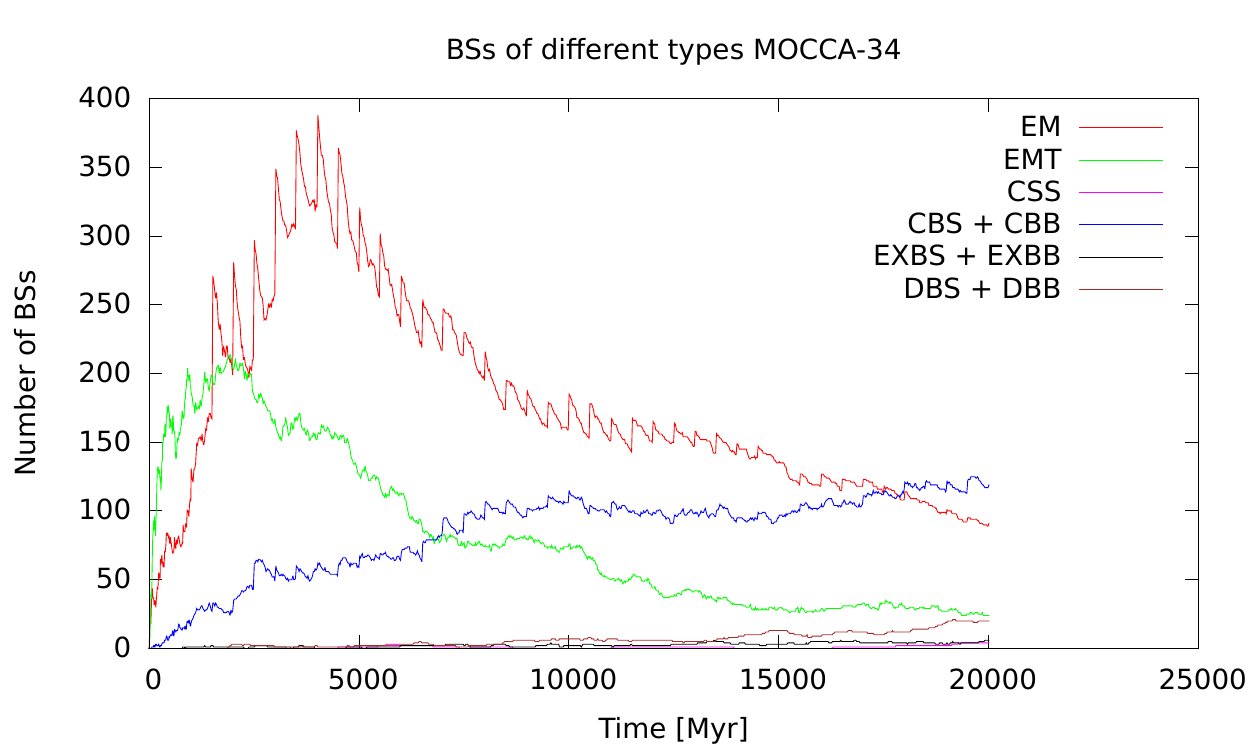}
	\includegraphics[width=\columnwidth]{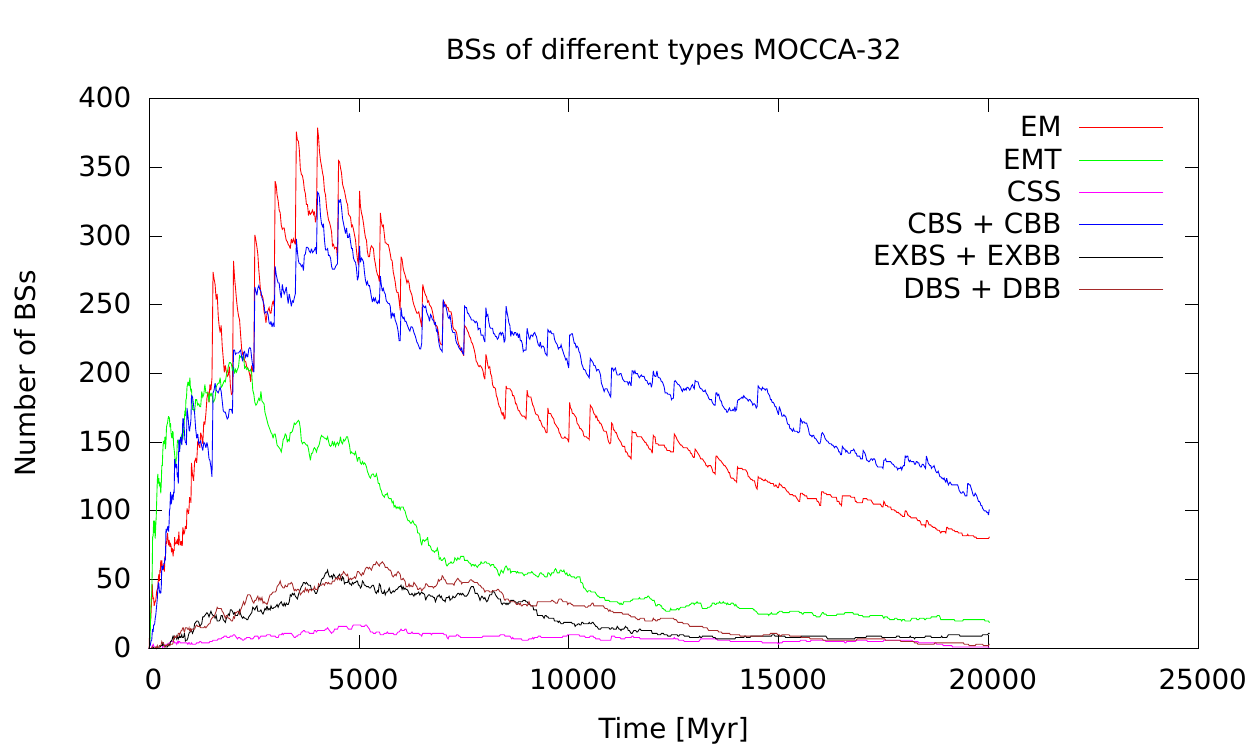}
	\includegraphics[width=\columnwidth]{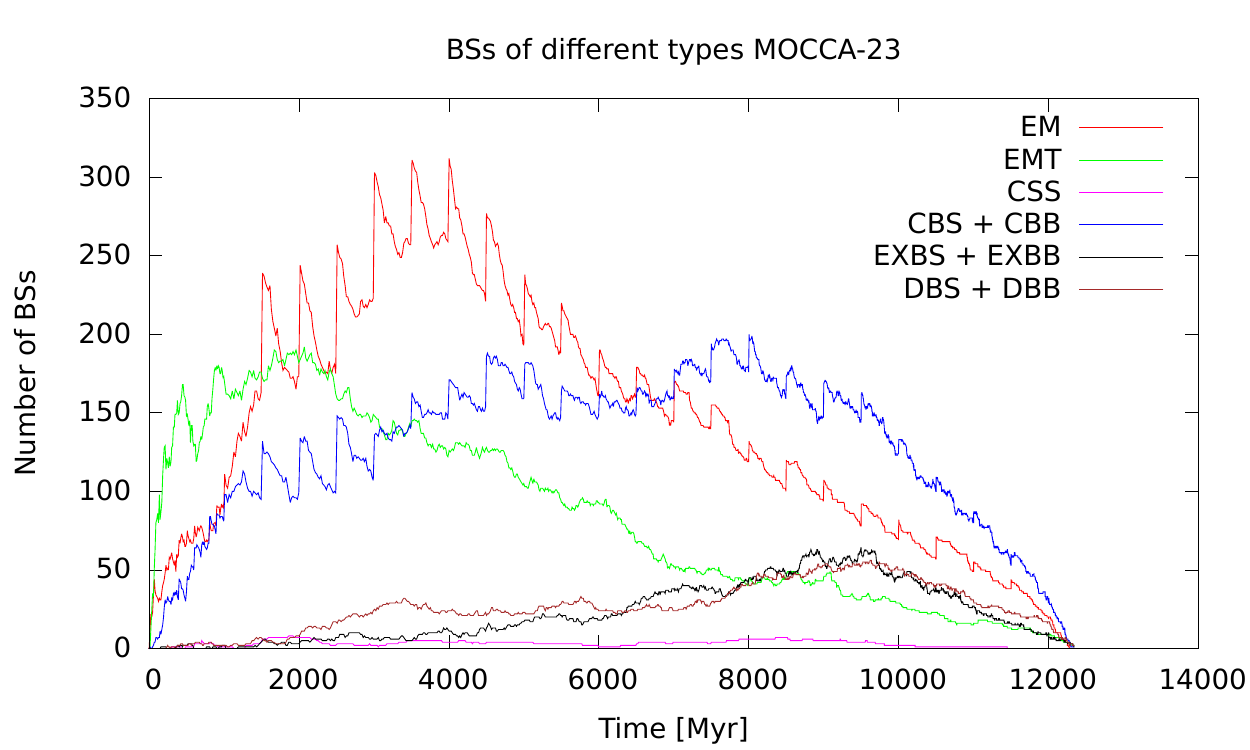}
	\includegraphics[width=\columnwidth]{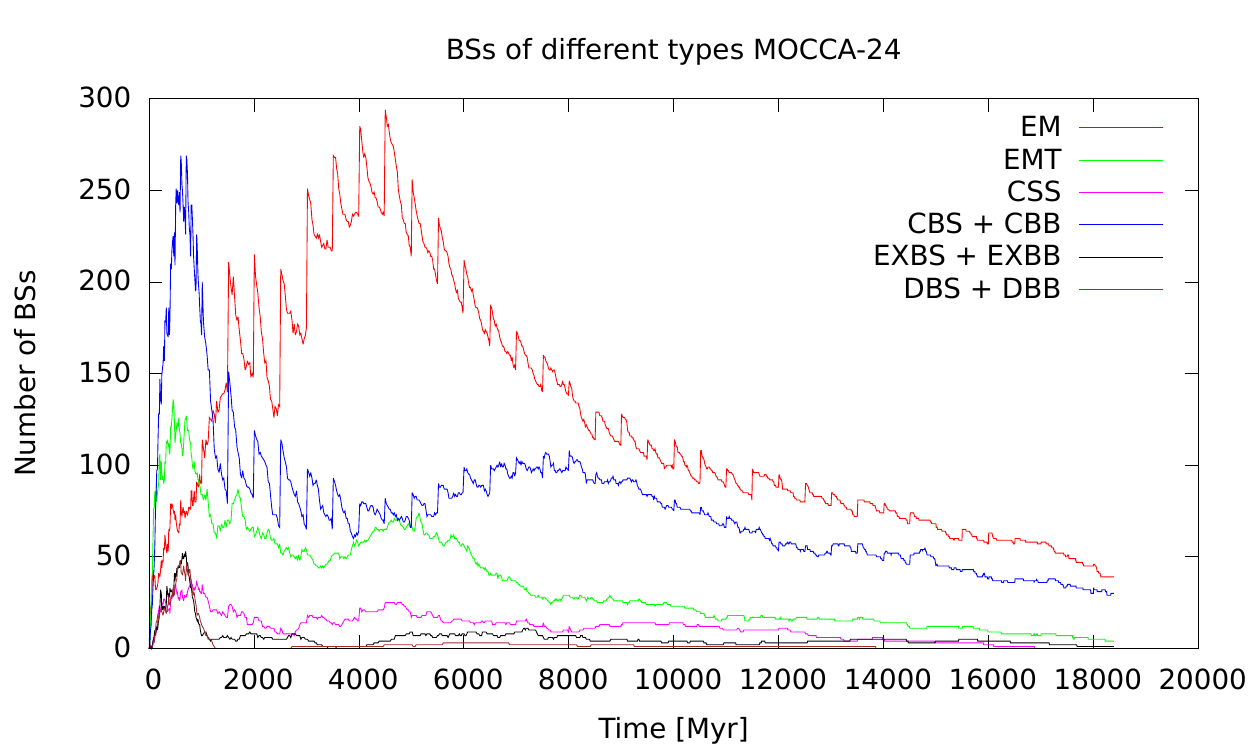}
	\includegraphics[width=\columnwidth]{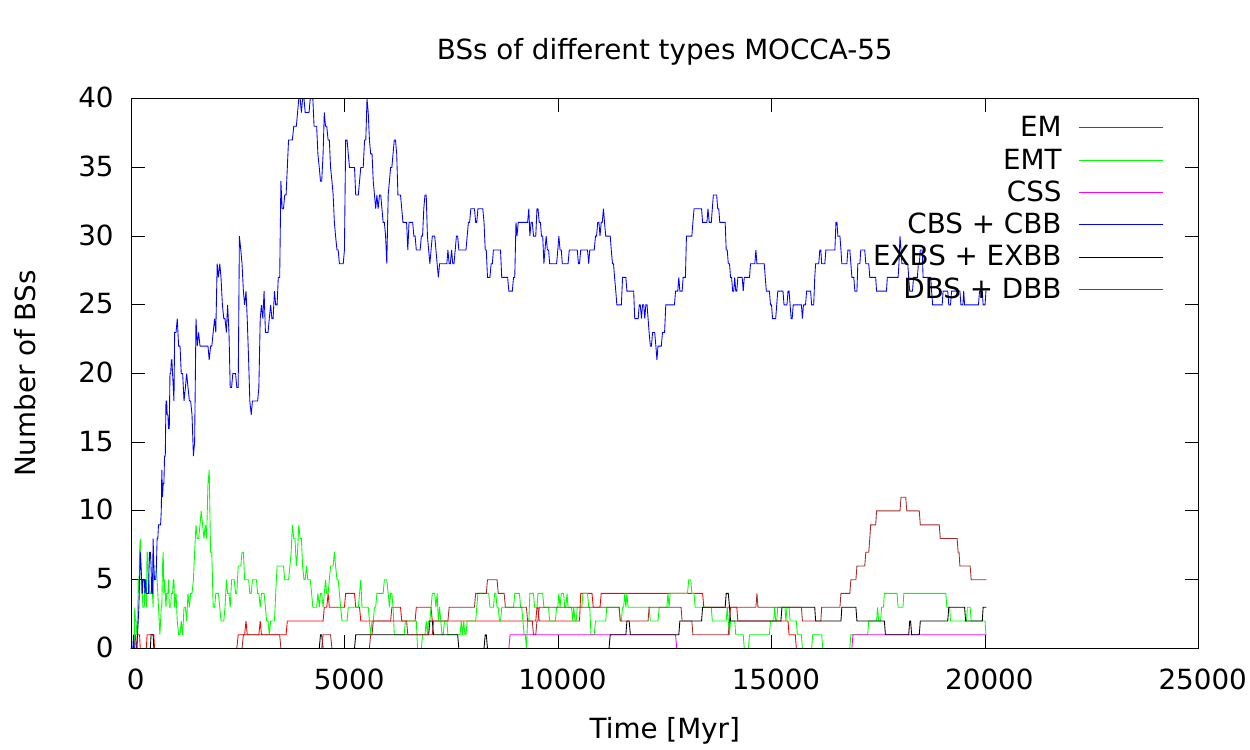}
	\includegraphics[width=\columnwidth]{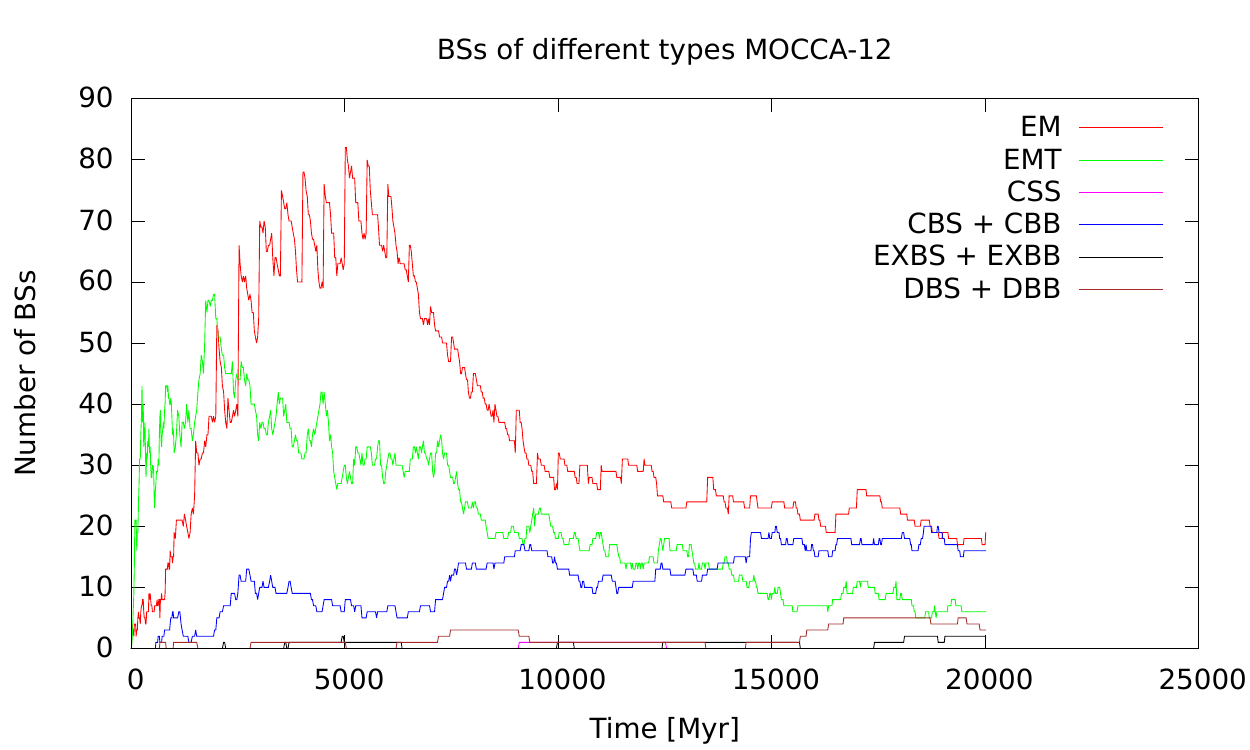}
  \caption[Population of BSs of different types for selected models]{Population of BSs of different types for selected models. Types of BSs are described in Sect.~\ref{sec:PropBSs:Channels}. The names of models are given in the titles of the plots. The parameters of models are summarized in Tab.~\ref{tab:Sim:Ini2} and Tab.~\ref{tab:Sim:Ini2}. 
  }
  \label{fig:BinSin:BSsNumbers}
\end{figure*}

A good representative of the \textsc{mocca-dropping} group is the \textsc{mocca-39}
model. It starts with 600k stars, 20\% of primordial binaries, a large
tidal radius $r_{tid} = 230$, and small concentration $c = r_{tid}/r_h = 10$
(see Tab.~\ref{tab:Sim:Ini1}). Because of the large $r_{tid}$, this model evolves
almost as an isolated GC. The number of BSs of different types for this model
are shown in Fig.~\ref{fig:BinSin:BSsNumbers} (top-left panel). The number
of EM, and EMT is a pure consequence of the initial conditions (see
Sect.~\ref{sec:PropBSs:isemi}). Because of the large $r_{tid}$ the number of BSs
created due to dynamical interactions is negligible.

The ratio $R_{B/S}$ for model \textsc{mocca-39} is represented as blue line in
Fig.~\ref{fig:BinSin:All}.
It has only EM and EMT BSs, and there is no significant contribution from 
dynamical BSs. The initial ratio $R_{B/S}$ is around $1.4$ for time 1~Gyr (the first
circle in Fig.~\ref{fig:BinSin:All}) because during that time the most active
channel is EMT (see Fig.~\ref{fig:BinSin:BSsNumbers}). After 4~Gyr the EM channel
becomes the most dominant one. Then, the ratio drops to $R_{B/S} \sim 0.4$ and
stays around this level up to 12~Gyr. The ratio does not change significantly
during that period because the number of EM and EMT decreases at almost the
same rate (see Fig.~\ref{fig:BinSin:BSsNumbers}).

The EMT channel is the most active in the model \textsc{mocca-39} (and many others)
during the first few Gyr as a result of the initial binary properties (see
Sect.~\ref{sec:PropBSs:Channels}). The number of EM BSs increases within the first
few Gyr as a result of two formation scenarios, described in
Sect.~\ref{sec:PropBSs:Channels}.

The ratio $R_{B/S}$ for the model \textsc{mocca-39} represents the simplest case.
It is equal to the ratio between the number of EMT dived by the number of
EM BSs (no dynamical BSs). The blue line in Fig.~\ref{fig:BinSin:All} ($R_{B/S}$
consists of all channels) is very much the same as in the  top panel of
Fig.~\ref{fig:BinSin:Separately} where $R_{B/S}^{evol}$ consists of only EMT and
EM. The number of  dynamical BSs is low, thus, the ratio $R_{B/S}^{dyn}$ has a
large scatter (see bottom panel in Fig.~\ref{fig:BinSin:Separately}). The model
is dominated by  evolutionary BSs. Nevertheless, the overall ratio $R_{B/S}$
stays around the value 0.3.

\subsubsection{\textsc{mocca-43}}

The ratio $R_{B/S}$  for the model \textsc{mocca-43} (light blue line in
Fig.~\ref{fig:BinSin:All}) evolves in a similar way as in \textsc{mocca-39}. The
model \textsc{mocca-43} has 50\% of primordial binaries, whereas
\textsc{mocca-39} has 20\%. The evolution of the ratio $R_{B/S}$ follows the
evolution of the ratio $R_{B/S}^{evol}$ too (see
Fig.~\ref{fig:BinSin:Separately}). For model \textsc{mocca-43} only the number
of BSs is larger, but the overall trend of the number of EM and EMT is very much
the same as for \textsc{mocca-39}. The number of  dynamical BSs is slightly
larger, but still very low. Thus, the ratio $R_{B/S}^{dyn}$ scatter a lot too
(see Fig.~\ref{fig:BinSin:Separately}).

\subsubsection{\textsc{mocca-34}, \textsc{mocca-33}}

Other examples of the \textsc{mocca-dropping} group are models
\textsc{mocca-34} (dark green) and \textsc{mocca-33} (magenta). These are the
models which are initially more concentrated than previously discussed
\textsc{mocca-39} model (dark blue). The rest of the initial conditions are the
same. The number of EM and EMT BSs is the same for these models despite the
higher concentrations (see Fig.~\ref{fig:BinSin:BSsNumbers}). This is also the
reason why the ratio $R_{B/S}^{evol}$ follows closely the ratio $R_{B/S}$  (see
Fig.~\ref{fig:BinSin:Separately}). This is another example which shows that the
concentration itself does not change the population of the evolutionary BSs
(more details on the influence of the concentration on the population of BSs of
different types one can find in Sect.~\ref{sec:PropBSs:concentrations}).

However, the higher concentrations cause that the number of dynamical BSs
becomes significant for models \textsc{mocca-34} and \textsc{mocca-33}.
Interestingly, a higher number of  dynamical BSs does not  change the
overall ratio $R_{B/S}$. Both, $R_{B/S}^{dyn}$ and $R_{B/S}^{evol}$,
oscillate around value 0.3 (see Fig.~\ref{fig:BinSin:Separately}).

\subsubsection{\textsc{mocca-32}, \textsc{mocca-23}, \textsc{mocca-24}}

The ratio $R_{B/S}$ drops below 0.3  for models which are very dense or
close to dissolution.

\textsc{mocca-32} is an example of a very concentrated model. It
starts with 600k stars and concentration $c = r_{tid}/r_h = 60$. It means that
half of the GC's mass is contained in the radius $r_h = 1.7$~[pc] (see
Tab.~\ref{tab:Sim:Ini1}). This very large initial concentration causes that the
number of  dynamical BSs is very high right from the beginning (see
Fig.~\ref{fig:BinSin:BSsNumbers}).

Another model, \textsc{mocca-23}, starts with 600k stars, but with $r_{tid}$,
thus $c = 25/10 = 2.5$~[pc]. This model dissolves completely within 12~Gyr (see
Fig.~\ref{fig:BinSin:BSsNumbers}).

The next model, \textsc{mocca-24}, is even denser. It starts with 600k stars,
$r_{tid} = 35$ and $c = 40$, thus $r_h = 0.9$~[pc]. Because of the larger
$r_{tid}$, the model takes a litte bit more time for the complete dissolution. The number of
 dynamical BSs for the \textsc{mocca-24} is dominant right from the
beginning. The high density of the GC causes that the probabilities of the
dynamical interactions are higher than for other models. Simultaneously, it
causes that within a few first Gyr many of the binaries are disrupted (half of
the binaries in the first 900~Myr). Thus, the number of  dynamical BSs drops
after 2~Gyr to a level around 100~BSs (see Fig.~\ref{fig:BinSin:BSsNumbers}).
Then, the EM channel  starts to be the dominant one, like for other models with
similar initial conditions.

It is worth to notice that for such a dense GC a smaller $r_{tid}$ and a high
concentration start to make a difference in the number of EM and EMT BSs. The
number of EM and EMT for \textsc{mocca-24} models is clearly smaller than in
\textsc{mocca-23} or \textsc{mocca-32} models despite the same initial number of
primordial binaries and the same semi-major axes distribution (for details see
Sect.~\ref{sec:PropBSs:concentrations}). For these three models
(\textsc{mocca-32}, \textsc{mocca-23}, \textsc{mocca-24}) the dynamical scales
of the GC evolution are larger than in other models. For \textsc{mocca-23},
i.e., the one which dissolves just after 12~Gyr, the dynamical times exceeds
hundreds of the half-max relaxation times ($t_{rh}$, see
Fig.~\ref{fig:BinSin:All}).

The ratio $R_{B/S}$ drops below  0.3 for models \textsc{mocca-32},
\textsc{mocca-24}, and \textsc{mocca-23}. The lower ratio $R_{B/S}$
is a consequence of the low number of EMT BSs, or increased number of DBS+DBB, or
both. Fig.~\ref{fig:BinSin:BSsNumbers} shows that for \textsc{mocca-32} model
the number of dissolved BSs (DBS+DBB) raises after the first few Gyr and stays
important until 12~Gyr. 
This is caused by the very high concentration of the GC
which highly increases  probabilities for  strong dynamical interactions
which will disrupt a binary. It causes that some of BSs become single stars.
Thus, the ratio $R_{B/S}^{dyn}$ is consequently dropping from around
$R_{B/S}^{dyn} \sim 0.4$ to $R_{B/S}^{dyn} < 0.2$ at 12~Gyr (see the bottom plot in
Fig.~\ref{fig:BinSin:Separately}). Additionally, the $R_{B/S}^{evol}$ is also
consequently dropping for later times too and goes slightly below the level 0.3
(see the top panel in Fig.~\ref{fig:BinSin:Separately}). The ratio $R_{B/S}^{evol}$
is getting lower with time because the number of EMT is getting lower. The high
concentration of the \textsc{mocca-32} model destroyed some binaries which would
otherwise create EMT BSs if their evolution was unperturbed (like e.g. in model
\textsc{mocca-39}). It concerns mainly wide EMT BSs which are formed through the
stellar winds. They are easier to disrupt than tight EMT.
However, the number of EM for \textsc{mocca-32} is similar as for the slowly
evolving \textsc{mocca-39} model (see Fig.~\ref{fig:BinSin:Separately}). They
are formed in close binaries, which are much harder to be disrupted due to 
dynamical interactions even in the dense clusters.

For the \textsc{mocca-24} model the ratios $R_{B/S}^{evol}$, $R_{B/S}^{dyn}$ and
$R_{B/S}$ drops below 0.3 due to the same reasons as for the
\textsc{mocca-32} model. Additionally, \textsc{mocca-24} has even a smaller tidal
radius, thus, the density of this cluster is even higher. Due to a  higher density
there are even more binaries destroyed because of strong dynamical
interactions. As a result, the number of EMT BSs is even smaller for this model
(see Fig.~\ref{fig:BinSin:BSsNumbers}). It causes that the ratio
$R_{B/S}^{evol}$ goes below 0.3  just after $2 t_{rh}$ (see
Fig.~\ref{fig:BinSin:Separately}). The number of DBS+DBB BSs is marginal -- for
such a dense GC, the binaries with BSs are hard and thus difficult to destroy
due to dynamical interactions. Hence, the number of dissolved binaries does
not influence the low ratio $R_{B/S}^{dyn}$. The number of CSS BSs is around
10 (formed in collisions due to dynamical interactions between two single
stars). However, their number is too small to make the ratio
$R_{B/S}^{dyn}$ significantly lower.
The only explanation is that  dynamical interactions preferentially create
BSs as single stars. It has to be connected with the high density of the
\textsc{mocca-24} model. However, the exact reason for that still has to be found.
The ratio $R_{B/S}^{dyn}$ is consequently lower than 0.3 (see
Fig.~\ref{fig:BinSin:Separately}).

The \textsc{mocca-23} is also a model for which the ratio $R_{B/S}$ drops below
0.3. It is a very fast evolving model, because it has small $r_{tid} = 25$~[pc].
For the default concentration $c = 10$  the half-mass radius is small $r_{h} =
2.5$~[pc], which makes it also a very dense GC. It needs only 12~Gyr for the
complete dissolution (see Fig.~\ref{fig:BinSin:BSsNumbers}). The ratio
$R_{B/S}^{evol}$ for this model oscillates around 0.4 and does not go below 0.3,
like for previously discussed models.
The ratio $R_{B/S}$ drops below 0.3 because there is a high number of DBS+DBB
BSs for this model (see Fig.~\ref{fig:BinSin:BSsNumbers}).

\subsubsection{\textsc{mocca-5}, \textsc{mocca-12}}

The models \textsc{mocca-5} (red), \textsc{mocca-12} (violet) are examples which
show some clear peak in the ratio $R_{B/S}$ for the dynamical times larger than
$1 t_{rh}$ (see Fig.~\ref{fig:BinSin:All}). They have a low number of  dynamical
BSs, thus the ratio $R_{B/S}^{dyn}$ scatter a lot (see
Fig.~\ref{fig:BinSin:Separately}). The number of EMT is also low ($\sim
10$~BSs), thus the ratio $R_{B/S}^{evol}$ scatter too (see
Fig.~\ref{fig:BinSin:Separately}). As a result, the ratio $R_{B/S}$ scatter when
a GC becomes dynamically old (> a few of $t_{rh}$).

\subsubsection{Models from \textsc{mocca-raising} group}
\label{sec:PropBSs:MoccaRaising}

The group \textsc{mocca-raising} consists of the models which start with low
ratios $R_{B/S}$ (see Fig.~\ref{fig:BinSin:All}). It is a result of a different
initial semi-major axes distributions of binaries in comparison to models of
the \textsc{mocca-dropping} group.

The model \textsc{mocca-55} is an example of the \textsc{mocca-raising} group
(yellow line in Fig.~\ref{fig:BinSin:All} and Fig.~\ref{fig:BinSin:Separately}).
It starts with 300k stars, with $r_{tid} = 69$~[pc] and concentration $c = 10$
(see Tab.~\ref{tab:Sim:Ini2}). The initial semi-major axes distribution is
created according to \citet{Kroupa1995aMNRAS.277.1491K}.

\begin{figure}
	\includegraphics[trim={0 1.5cm 0
	2cm},width=\columnwidth]{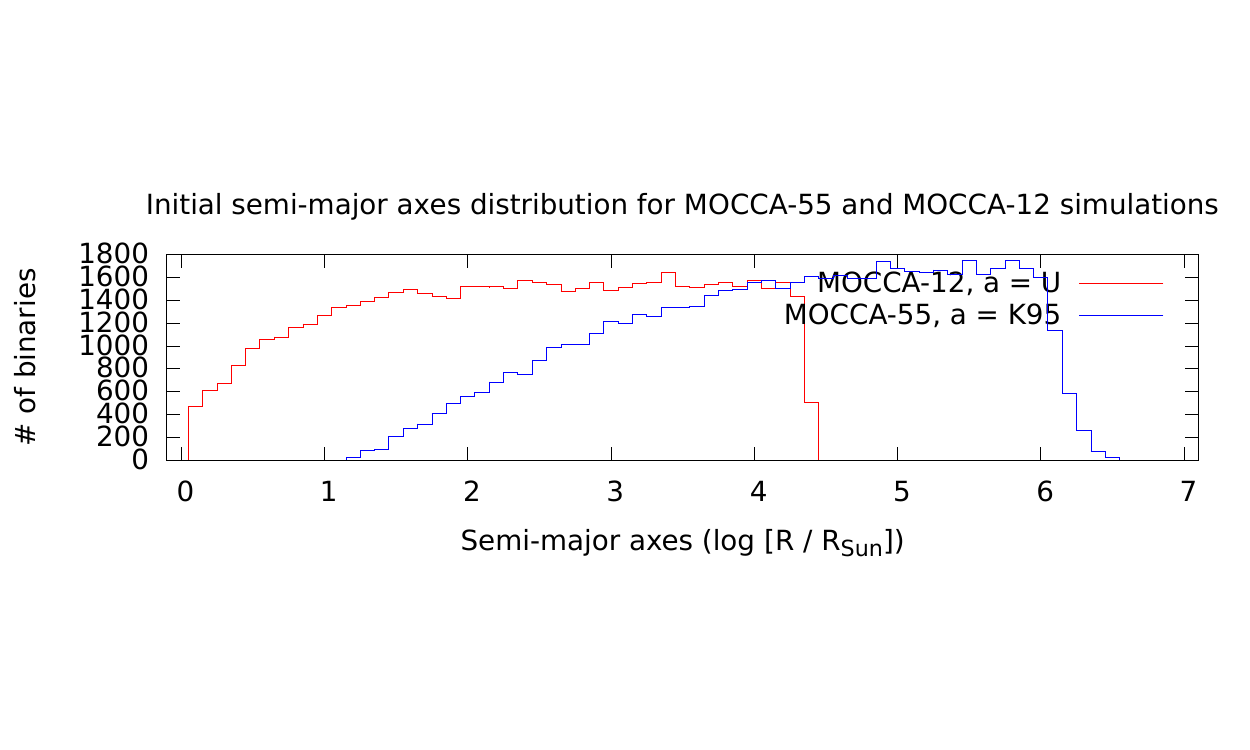}
	\includegraphics[trim={0 2cm 0
	2cm},width=\columnwidth]{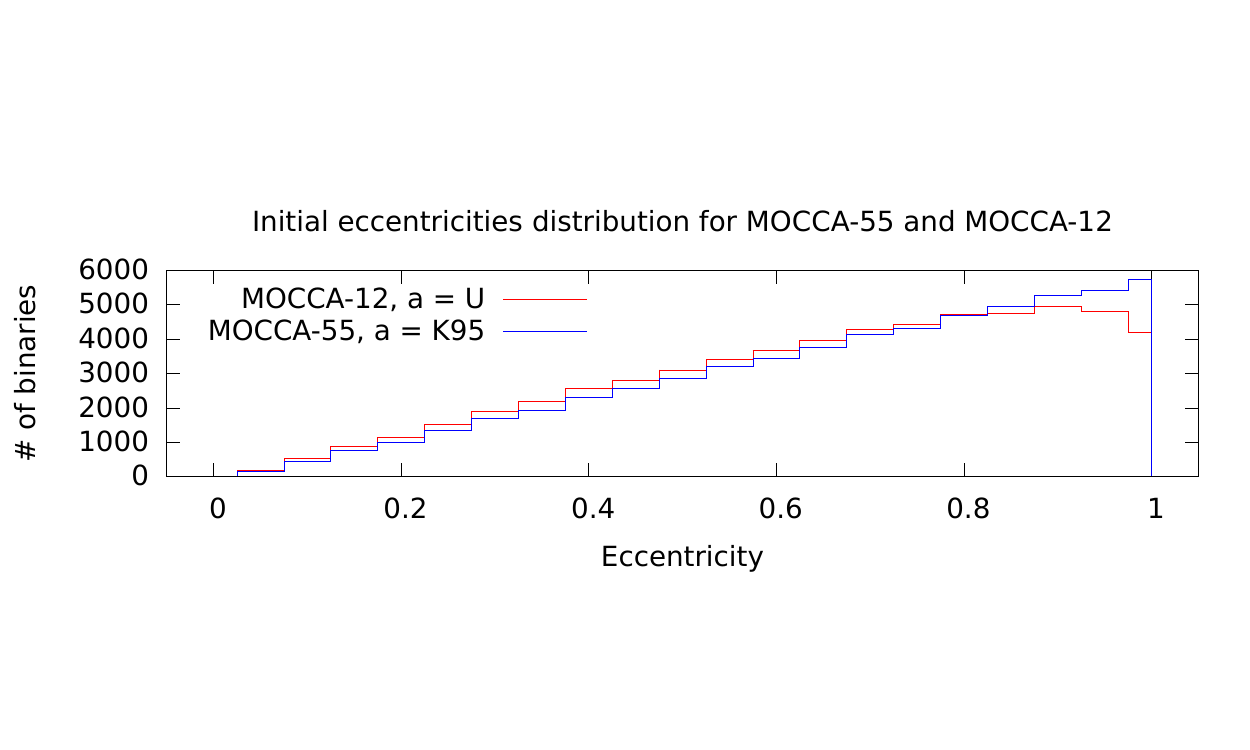}
  \caption[Semi-major axes and eccentricities distributions of models
  \textsc{mocca-55} and \textsc{mocca-12}]{Initial distributions of semi-major
  axes (top) and eccentricities (bottom) for models \textsc{mocca-55} and
  \textsc{mocca-12}. The first model is a representative of the
  \textsc{mocca-raising} group, whereas the latter one is an example of
  \textsc{mocca-dropping} group (see Sect.~\ref{sec:PropBSs:Binaries}). For the
  meaning of the symbols see Tab.~\ref{tab:Sim:Ini2}, for the number of BSs of
  these models see Fig.~\ref{fig:BinSin:BSsNumbers}, for details see text.
  }
  \label{fig:BinSin:Raising1D}
\end{figure}

Fig.~\ref{fig:BinSin:Raising1D} presents a comparison of the initial
conditions between models \textsc{mocca-55} and \textsc{mocca-12}. The latter
one is an example from the \textsc{mocca-dropping} group. They differ in only one initial parameter (see
Tab.~\ref{tab:Sim:Ini1}, Tab.~\ref{tab:Sim:Ini2}). 
The initial semi-major axes distributions are compared 
on the top panel of Fig.~\ref{fig:BinSin:Raising1D}, while
distributions of eccentricities on the bottom panel. For
\textsc{mocca-55} there are many more wide binaries, whereas the eccentricity
distribution is very much the same. However, there are small differences for 
high eccentricities (> 0.9). It is caused by the fact that the \textsc{mocca}
code, while generating the initial conditions, checks for immediate mergers. It
may happen that for a compact binary a high eccentricity will be drawn.
For such binary the periastron distance could be smaller than the sum of radii
of the system components. The binary would merge just in the first call of
the stellar evolution. In order to avoid such situation the \textsc{mocca} code
generates the eccentricity for the binary once again. For the \textsc{mocca-12}
there are many more compact binaries, thus this situation may happen more often.
As a result, there is slightly less high eccentricities for this model
(see bottom panel in Fig.~\ref{fig:BinSin:Raising1D}). All other initial
conditions for the \textsc{mocca-55} and \textsc{mocca-12} models are the same
(see Tab.~\ref{tab:Sim:Ini1} and Tab.~\ref{tab:Sim:Ini2}).

\begin{figure}
	\includegraphics[width=\columnwidth]{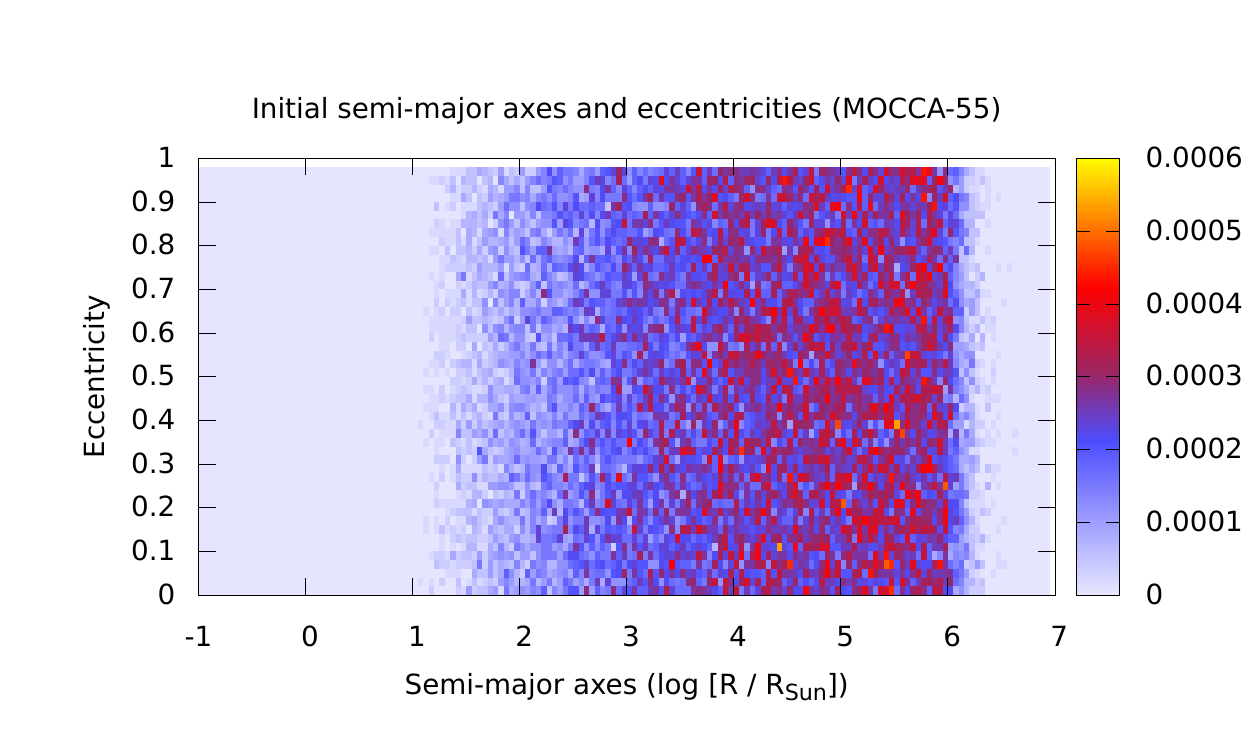}
	\includegraphics[width=\columnwidth]{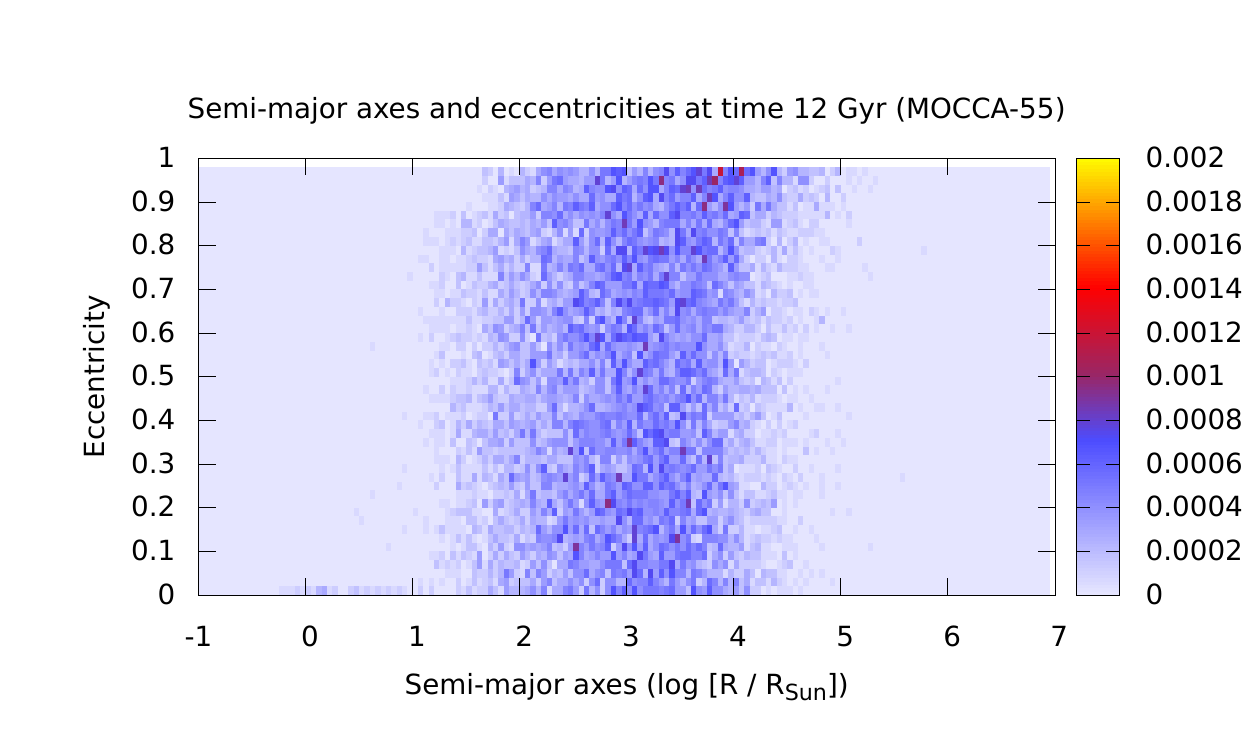}
  \caption[Density maps of the semi-major axes -- eccentricity distribution for
  $T = 0$ and $T = 12$~Gyr for the model \textsc{mocca-55}]{Density maps of the
  semi-major axes -- eccentricity distribution for $T = 0$ (top) and for time
  12~Gyr (bottom) for the model \textsc{mocca-55}. For 12~Gyr the density map is
  produced only for MS-MS binaries. For the meaning of the symbols see
  Tab.~\ref{tab:Sim:Ini2}, for the number of BSs of these models see
  Fig.~\ref{fig:BinSin:BSsNumbers}, for details see text.
  }
  \label{fig:BinSin:Raising2D}
\end{figure}

Fig.~\ref{fig:BinSin:Raising2D} shows the density maps which combines
distributions of semi-major axes with eccentricities for the model
\textsc{mocca-55} (\textsc{mocca-raising}). The top panel shows the initial
density map, whereas the bottom shows the density for 12~Gyr. The bottom plot
consists of only the binaries with main sequence stars. These are the types of
objects which still have a chance to create BSs in a physical collision (CBS,
CBB), due to mass transfer when one of the stars leaves main-sequence (EMT) or
due to a merger (EM).
Fig.~\ref{fig:BinSin:Raising2D} shows that for this model there are many wide
binaries ($a > 10^4 [R_{\odot}]$), also with high eccentricities for time $T =
0$. After 12~Gyr there are almost no binaries with semi-major axes $> 10^4
[R_{\odot}]$. However, there are still these $\lesssim 10^4 [R_{\odot}]$ with
high eccentricities ($\approx 1$).

The differences in the initial distributions of semi-major axes in binaries (see
Fig.~\ref{fig:BinSin:Raising1D}) cause the differences in the number of BSs of
various types for these models (see Fig.~\ref{fig:BinSin:BSsNumbers}). For
\textsc{mocca-12} model, which represents the group \textsc{mocca-dropping}, the
dominant channels from the beginning are EM and EMT (evolutionary BSs). The
number of dynamical BSs increases with time because GC is getting denser. In
turn, \textsc{mocca-55} has a very low number of EM and EMT and a high number of
dynamical BSs.

\begin{figure}
	\includegraphics[width=\columnwidth]{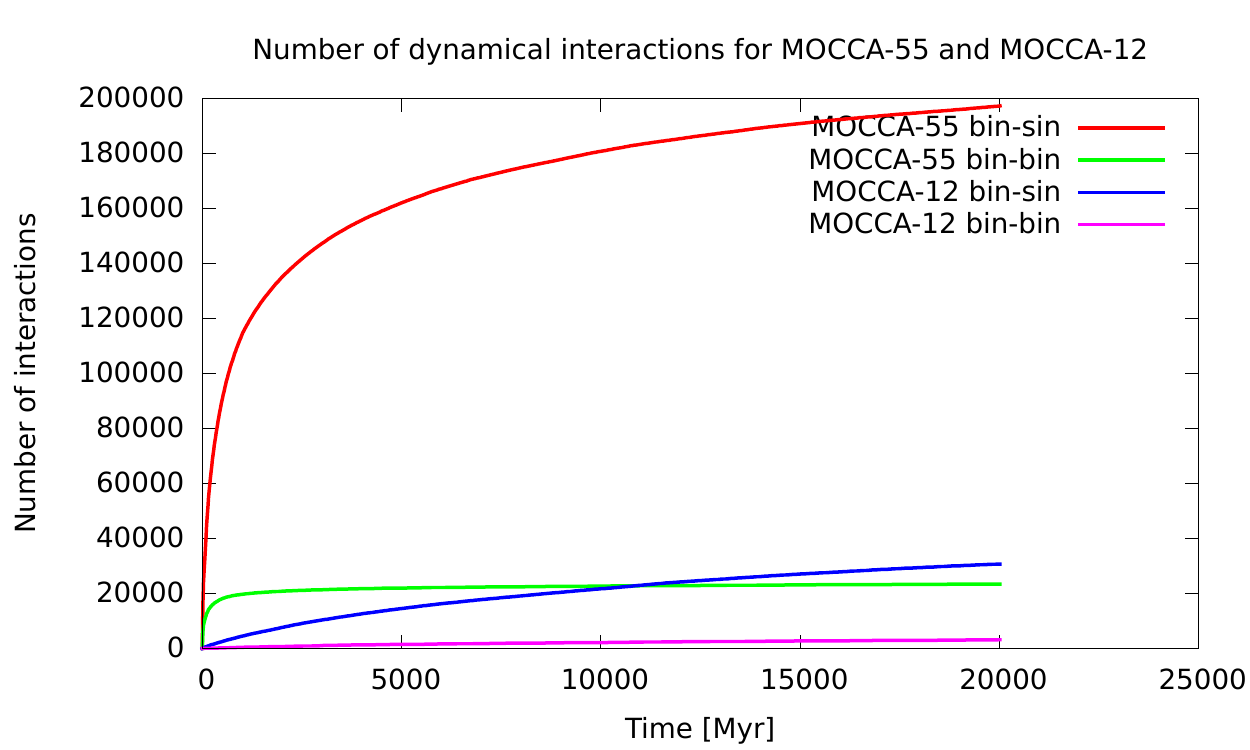}
	\includegraphics[width=\columnwidth]{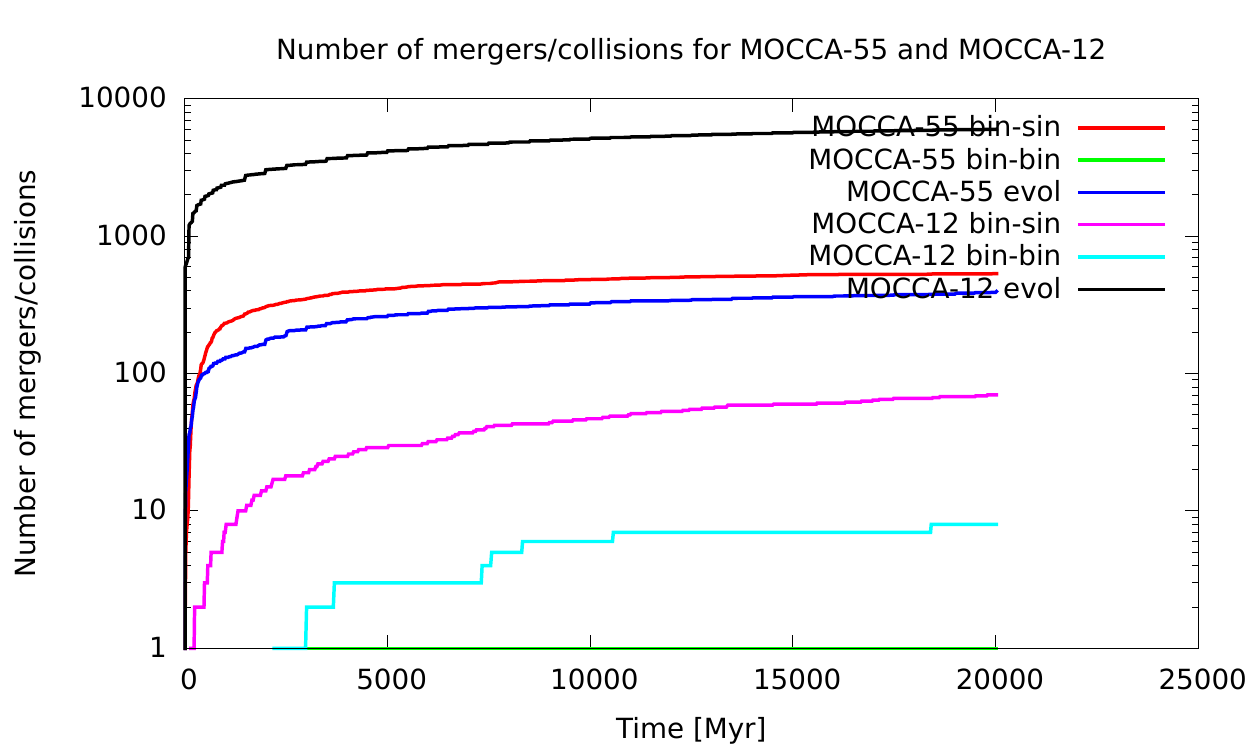}
	\caption[Number of dynamical interactions, mergers and collisions for models
	\textsc{mocca-55} and \textsc{mocca-12}]{Number of dynamical interactions (top
	panel), mergers and collisions (bottom) for models \textsc{mocca-55} and
	\textsc{mocca-12}. The term \textit{bin-sin} stands for a dynamical interaction
	between a binary and a single star, the term \textit{bin-bin} for a dynamical
	interaction between two binaries, the term \textit{merger} for coalescence
	between two components in a binary without involving any sort of a dynamical
	interaction. The \textit{merger} is a result of the stellar evolution only. The
	term \textit{collision} concerns a physical collision between two or more stars
	due to a dynamical interaction. The \textsc{mocca-55} model represents the
	group \textsc{mocca-raising}, whereas \textsc{mocca-12} the group
	\textsc{mocca-dropping}. For details see the text. 
  }
  \label{fig:BinSin:NumInterAndMergers}
\end{figure}

The differences in the number of BSs of different populations between the models
\textsc{mocca-55} and \textsc{mocca-12} is due to the different frequency of
dynamical interactions and mergers. Fig.~\ref{fig:BinSin:NumInterAndMergers}
shows on the top panel the number of the dynamical interactions between binaries
and single stars and between two binaries. The bottom panel in
Fig.~\ref{fig:BinSin:NumInterAndMergers} shows the number of evolutionary
mergers and physical collisions between stars due to dynamical interactions.

The low number of the dynamical BSs in \textsc{mocca-12} from
\textsc{mocca-dropping} group (see Fig.~\ref{fig:BinSin:BSsNumbers}) is caused
by the low number of the dynamical interactions (only $\sim 20k$ up to 12~Gyr,
see Fig.~\ref{fig:BinSin:NumInterAndMergers}). Thus, the number of
collisions due to dynamical interactions is also low. After 12~Gyr only a few
dozens of stars collided (violet and light blue line on the bottom panel in
Fig.~\ref{fig:BinSin:NumInterAndMergers}). Instead, the \textsc{mocca-12} model
has many mergers due to stellar evolution (black line). This is a result of the
initial semi-major axes distribution. For the model \textsc{mocca-12} there are many
compact binaries which create EM and EMT BSs. Dynamical
BSs become significant only later, when the density of the GC increases and the cluster
slowly goes to the core collapse.

The model \textsc{mocca-55} which represents \textsc{mocca-raising} group has
a low number of EM and EMT (see Fig.~\ref{fig:BinSin:BSsNumbers}). For this model
there is a small number of compact binaries (see Fig.~\ref{fig:BinSin:Raising1D})
and thus there are not many binaries which could possibly create evolutionary
BSs. Instead, the number of the dynamical BSs is high. The number of CBS and CBB
increases from the beginning and it is important throughout the whole GC
evolution.

The number of dynamical interactions for \textsc{mocca-55} is around 10
times larger than for the \textsc{mocca-12}. The high number of the interactions
is a result of larger semi-major axes for binaries (see
Fig.~\ref{fig:BinSin:Raising1D}). It increases probabilities of 
interactions. Many of them are just fly-by passages between objects. They do not
create large differences in semi-major axes of binaries. However, due to a large
number of them the eccentricity of some binaries increases (even to $> 0.99$).
The increased eccentricities cause that the code detects a collision when the
periastron distance is smaller than the sum of radii of stars and the binary
collides. This is not a typical collision between two stars in a dynamical
interaction. The incoming star or binary does not actually collide with the
binary, instead it changes only its properties. As a result of this interaction the
collided star is a single star. Thus, the ratio $R_{B/S}$ for the
\textsc{mocca-raising} group of models starts with small values $R_{B/S}
\lesssim 0.3$ (see Fig.~\ref{fig:BinSin:All}).

A large number of dynamical interactions disrupts many very wide binaries
(\textsc{mocca-55}). As a result, after 12~Gyr there is actually a few of them left
(see Fig.~\ref{fig:BinSin:Raising2D}). Because the number of wide binaries with
high eccentricities is getting lower with time, also the number of the dynamical
BSs created from wide binaries with high eccentricities is getting lower.
More dynamical BSs are created in physical collisions
between a star in a binary and some incoming star (or binary). As a result BSs
stay in binaries. Thus, the ratio $R_{B/S}^{dyn}$ slowly increases with time (see
Fig.~\ref{fig:BinSin:Separately}). Additionally, the $R_{B/S}$ which combines
all BSs from all channels increases too (see Fig.~\ref{fig:BinSin:All}). The
number of EM and EMT BSs is low during the whole simulation, thus the ratio
$R_{B/S}^{evol}$ scatter a lot. However, it has a small influence on the ratio
$R_{B/S}$. Dynamical BSs, after wide high-eccentricity binaries are
disrupted, are created in the same way as BSs from the
\textsc{mocca-dropping} group. Thus, the ratio $R_{B/S}$ from
\textsc{mocca-raising} group reaches the same values as the models of the
\textsc{mocca-dropping} group.

\section{Summary}
\label{sec:Summary}

Blue stragglers (BSs) are very interesting stars in terms of their formation and evolution. 
Their increased mass indicates
that they have received an additional mass during their life. The main channels of
their formation concern the stellar evolution (mass gained through the mass
transfer) and the dynamical interactions (mass gained due to physical
collisions). BSs are more numerous in globular clusters than in the field
of the Galaxy. This make the BSs in GCs extremely interesting objects to study the complex
interplay between the stellar evolution and the dynamical evolution of GCs.

Initial conditions of GCs have a fundamental influence on populations of BSs
of different types. It concerns both, the global initial parameters of a GCs, and
the initial distributions of binaries properties. Different initial
conditions can very significantly change populations of BSs.

In this context the importance of the semi-major axes distribution is crucial. The high number
of compact binaries, like for the distribution $a = UL$ (see
Fig.~\ref{fig:PropBSs:isemi0_vs_isemi1} and definitions in
Tab.~\ref{tab:Sim:Ini1}), directly relates to the high number of EM and EMT BSs.
When the initial number of compact binaries is lower, like for $a = L$
distribution, the number of EM and EMT decreases (see
Fig.~\ref{fig:PropBSs:isemi0_vs_isemi1}). It decreases even more for
distributions like K95 (see Fig.~\ref{fig:PropBSs:isemi0_vs_isemi2}) which
prefer even wider binaries.

The influence of the semi-major axes distribution on the population of EM and
EMT is a consequence of the formation scenarios of these BSs. EM are formed in
compact binaries. Thus, when the initial semi-major axis is wider, the binary
needs much more time to merge. The same applies for the first subgroup of EMT which
are also compact binaries (see Sect.~\ref{sec:PropBSs:Channels}). For them a
large semi-major axes prevent the mass transfer as well. For the second subgroup of EMT the
semi-major axes can be larger, but a high eccentricity is needed to make the
mass transfer possible. However, the number of EMT created according to this scenario is
rather low. Thus, it has a smaller impact on the overall population of EMT.

It was very unexpected to find out that a large number of wide binaries can so
significantly influence the population of dynamical BSs (CBS, CBB, see
Fig.~\ref{fig:PropBSs:isemi0_vs_isemi2}). In this mechanism wider binaries have
higher probabilities of dynamical interactions. These interactions are mostly
fly-bys but increase the eccentricities of many binaries. For a number of
them the eccentricities get so large that a collision takes place and the binary merges.
In this scenario, because of a large number of semi-major axes with wide binaries, the
dynamical BSs are the dominant ones.

The influence of the initial semi-major axes distributions on  populations of
BSs can be very valuable for narrowing down the initial conditions. 
If a good way to distinguish between evolutionary and the dynamical
BSs is found, 
it could help to give some boundaries on the initial semi-major axes
distribution -- it is a very important subject in the studies of GCs.

Higher initial concentrations of GCs have a large impact on the population of
dynamical BSs. The number of CBS and CBB BSs increase for larger
concentrations. It is a consequence of higher probabilities of strong
dynamical interactions for denser systems. If the density increased, the number
of dynamical BSs changing their types increases too. There are more binaries
which exchange their companions (EXBS, EXBB) or are dissolved (DBS, DBB)
due to strong dynamical interactions.

Surprisingly, higher initial concentrations of GCs do not have any influence
on the population of evolutionary BSs. For the \textsc{mocca-7} model the
concentration is $c = r_{tid}/r_h = 40$, which means that half of the GC's mass
is contained inside $r_h = 1.7$~[pc]. This is a very dense model and even for such
extreme conditions populations of EM and EMT BSs were not affected
noticeably. It has very important implications for observations. It 
strongly supports the theory that evolutionary BSs are results of the unperturbed
evolution of the primordial binaries. The same applies even if the concentration
is increased up to $c = 60$ -- the population of EM and EMT BSs is also not affected.

Different initial conditions have also a profound effect on the ratio between
BSs in binaries and as single stars ($R_{B/S}$). The \textsc{mocca} models
from Tab.~\ref{tab:Sim:Ini1} and Tab.~\ref{tab:Sim:Ini2} vary in many aspects.
They have various initial concentrations which implies different dynamical ages
for GCs. They have also various initial semi-major axes and eccentricities
distributions. Surprisingly, the ratio $R_{B/S}$ for all of the models gets
close to the value $R_{B/S} \sim 0.4$ within a few Gyrs.

There are two groups which have different evolutions of the ratio $R_{B/S}$. The
first group, \textsc{mocca-dropping}, concerns the models which have some
fraction of compact binaries. For them, the channels EM are EMT are important.
Their ratio $R_{B/S}$ drops from the values $\gtrsim 1.0$ to around 0.4. The
second group, \textsc{mocca-raising}, consists of models which have a large
fraction of wide binaries. For them the EM and EMT channels are almost not
important, whereas dynamical BSs are the dominant ones. The ratio $R_{B/S}$
for \textsc{mocca-raising} starts with low values and raises with time to reach
$\sim 0.4$. It is very interesting that the main reason behind this division is
the initial semi-major axes distribution. The initial concentrations of GCs do
not have a significant influence on these two groups of models.

However, an attempt to understand why the ratio $R_{B/S}$ oscillates around the
value 0.4 for such a variety of different initial conditions still has to be
determined. It only suggests that, because of some reason, $\sim 40\%$ of all
dynamical interactions lead to the creation of binaries, and the rest to the
dissolution of binaries.

\section*{Acknowledgment}

The project was supported partially by Polish National Science Center grants
DEC-2011/01/N/ST9/06000 and DEC-2012/07/B/ST9/04412.





\bibliographystyle{mn2e} 
\bibliography{biblio}

\begin{thebibliography}{}

\bibitem[\protect\citeauthoryear{{Bragaglia}, {Carretta}, {Blanco}, {Cacciari}
  \& {Kinman}}{{Bragaglia} et~al.}{2005}]{Bragaglia2005IAUS..228..243B}
{Bragaglia} A.,  {Carretta} E.,  {Blanco} A.~R.,  {Cacciari} C.,    {Kinman}
  T.~D.,  2005, in {Hill} V.,  {Fran{\c c}ois} P.,   {Primas} F.,  eds, From
  Lithium to Uranium: Elemental Tracers of Early Cosmic Evolution Vol.~228 of
  IAU Symposium, {Chemical properties of field halo and thick disk blue
  straggler stars: first results}.
pp 243--244

\bibitem[\protect\citeauthoryear{{Casertano} \& {Hut}}{{Casertano} \&
  {Hut}}{1985}]{1985ApJ...298...80C}
{Casertano} S.,  {Hut} P.,  1985, \apj, 298, 80

\bibitem[\protect\citeauthoryear{{Clarkson}, {Sahu}, {Anderson}, {Rich},
  {Smith}, {Brown}, {Bond}, {Livio}, {Minniti}, {Renzini} \&
  {Zoccali}}{{Clarkson} et~al.}{2011}]{Clarkson2011ApJ...735...37C}
{Clarkson} W.~I.,  {Sahu} K.~C.,  {Anderson} J.,  {Rich} R.~M.,  {Smith} T.~E.,
   {Brown} T.~M.,  {Bond} H.~E.,  {Livio} M.,  {Minniti} D.,  {Renzini} A.,
  {Zoccali} M.,  2011, \apj, 735, 37

\bibitem[\protect\citeauthoryear{{Davies}, {Piotto} \& {de Angeli}}{{Davies}
  et~al.}{2004}]{Davies2004MNRAS.349..129D}
{Davies} M.~B.,  {Piotto} G.,    {de Angeli} F.,  2004, \mnras, 349, 129

\bibitem[\protect\citeauthoryear{{De Marco}, {Shara}, {Zurek}, {Ouellette},
  {Lanz}, {Saffer} \& {Sepinsky}}{{De Marco}
  et~al.}{2005}]{DeMarco2005ApJ...632..894D}
{De Marco} O.,  {Shara} M.~M.,  {Zurek} D.,  {Ouellette} J.~A.,  {Lanz} T.,
  {Saffer} R.~A.,    {Sepinsky} J.~F.,  2005, \apj, 632, 894

\bibitem[\protect\citeauthoryear{{Dotter}, {Sarajedini}, {Anderson},
  {Aparicio}, {Bedin}, {Chaboyer}, {Majewski}, {Mar{\'{\i}}n-Franch}, {Milone},
  {Paust}, {Piotto}, {Reid}, {Rosenberg} \& {Siegel}}{{Dotter}
  et~al.}{2010}]{Dotter2010ApJ...708..698D}
{Dotter} A.,  {Sarajedini} A.,  {Anderson} J.,  {Aparicio} A.,  {Bedin} L.~R.,
  {Chaboyer} B.,  {Majewski} S.,  {Mar{\'{\i}}n-Franch} A.,  {Milone} A.,
  {Paust} N.,  {Piotto} G.,  {Reid} I.~N.,  {Rosenberg} A.,    {Siegel} M.,
  2010, \apj, 708, 698

\bibitem[\protect\citeauthoryear{{Ferraro}, {Beccari}, {Dalessandro},
  {Lanzoni}, {Sills}, {Rood}, {Pecci}, {Karakas}, {Miocchi} \&
  {Bovinelli}}{{Ferraro} et~al.}{2009}]{Ferraro2009Natur.462.1028F}
{Ferraro} F.~R.,  {Beccari} G.,  {Dalessandro} E.,  {Lanzoni} B.,  {Sills} A.,
  {Rood} R.~T.,  {Pecci} F.~F.,  {Karakas} A.~I.,  {Miocchi} P.,    {Bovinelli}
  S.,  2009, \nat, 462, 1028

\bibitem[\protect\citeauthoryear{{Ferraro}, {Fusi Pecci} \&
  {Bellazzini}}{{Ferraro} et~al.}{1995}]{Ferraro1995A&A...294...80F}
{Ferraro} F.~R.,  {Fusi Pecci} F.,    {Bellazzini} M.,  1995, \aap, 294, 80

\bibitem[\protect\citeauthoryear{{Ferraro} \& {Lanzoni}}{{Ferraro} \&
  {Lanzoni}}{2009}]{2009RMxAC..37...62F}
{Ferraro} F.~R.,  {Lanzoni} B.,  2009, in Revista Mexicana de Astronomia y
  Astrofisica Conference Series Vol.~37 of Revista Mexicana de Astronomia y
  Astrofisica Conference Series, {Probing the link between dynamics and stellar
  evolution: Blue Straggler Stars in Globular clusters}.
pp 62--71

\bibitem[\protect\citeauthoryear{{Ferraro}, {Messineo}, {Fusi Pecci}, {de
  Palo}, {Straniero}, {Chieffi} \& {Limongi}}{{Ferraro}
  et~al.}{1999}]{Ferraro1999AJ....118.1738F}
{Ferraro} F.~R.,  {Messineo} M.,  {Fusi Pecci} F.,  {de Palo} M.~A.,
  {Straniero} O.,  {Chieffi} A.,    {Limongi} M.,  1999, \aj, 118, 1738

\bibitem[\protect\citeauthoryear{{Ferraro}, {Paltrinieri}, {Fusi Pecci},
  {Cacciari}, {Dorman}, {Rood}, {Buonanno}, {Corsi}, {Burgarella} \&
  {Laget}}{{Ferraro} et~al.}{1997}]{1997A&A...324..915F}
{Ferraro} F.~R.,  {Paltrinieri} B.,  {Fusi Pecci} F.,  {Cacciari} C.,  {Dorman}
  B.,  {Rood} R.~T.,  {Buonanno} R.,  {Corsi} C.~E.,  {Burgarella} D.,
  {Laget} M.,  1997, \aap, 324, 915

\bibitem[\protect\citeauthoryear{{Ferraro}, {Sabbi}, {Gratton}, {Piotto},
  {Lanzoni}, {Carretta}, {Rood}, {Sills}, {Fusi Pecci}, {Moehler}, {Beccari},
  {Lucatello} \& {Compagni}}{{Ferraro}
  et~al.}{2006}]{Ferraro2006ApJ...647L..53F}
{Ferraro} F.~R.,  {Sabbi} E.,  {Gratton} R.,  {Piotto} G.,  {Lanzoni} B.,
  {Carretta} E.,  {Rood} R.~T.,  {Sills} A.,  {Fusi Pecci} F.,  {Moehler} S.,
  {Beccari} G.,  {Lucatello} S.,    {Compagni} N.,  2006, \apjl, 647, L53

\bibitem[\protect\citeauthoryear{{Fiorentino}, {Lanzoni}, {Dalessandro},
  {Ferraro}, {Bono} \& {Marconi}}{{Fiorentino}
  et~al.}{2014}]{Fiorentino2014ApJ...783...34F}
{Fiorentino} G.,  {Lanzoni} B.,  {Dalessandro} E.,  {Ferraro} F.~R.,  {Bono}
  G.,    {Marconi} M.,  2014, \apj, 783, 34

\bibitem[\protect\citeauthoryear{{Fossati}, {Mochnacki}, {Landstreet} \&
  {Weiss}}{{Fossati} et~al.}{2010}]{Fossati2010A&A...510A...8F}
{Fossati} L.,  {Mochnacki} S.,  {Landstreet} J.,    {Weiss} W.,  2010, \aap,
  510, A8

\bibitem[\protect\citeauthoryear{{Fregeau} \& {Rasio}}{{Fregeau} \&
  {Rasio}}{2007}]{Fregeau2007ApJ...658.1047F}
{Fregeau} J.~M.,  {Rasio} F.~A.,  2007, \apj, 658, 1047

\bibitem[\protect\citeauthoryear{{Fuhrmann}, {Chini}, {Hoffmeister} \&
  {Stahl}}{{Fuhrmann} et~al.}{2011}]{Fuhrmann2011MNRAS.416..391F}
{Fuhrmann} K.,  {Chini} R.,  {Hoffmeister} V.~H.,    {Stahl} O.,  2011, \mnras,
  416, 391

\bibitem[\protect\citeauthoryear{{Fusi Pecci}, {Ferraro}, {Corsi}, {Cacciari}
  \& {Buonanno}}{{Fusi Pecci} et~al.}{1992}]{1992AJ....104.1831F}
{Fusi Pecci} F.,  {Ferraro} F.~R.,  {Corsi} C.~E.,  {Cacciari} C.,
  {Buonanno} R.,  1992, \aj, 104, 1831

\bibitem[\protect\citeauthoryear{{Geller} \& {Mathieu}}{{Geller} \&
  {Mathieu}}{2011}]{Geller2011Natur.478..356G}
{Geller} A.~M.,  {Mathieu} R.~D.,  2011, \nat, 478, 356

\bibitem[\protect\citeauthoryear{{Geller} \& {Mathieu}}{{Geller} \&
  {Mathieu}}{2012}]{Geller2012AJ....144...54G}
{Geller} A.~M.,  {Mathieu} R.~D.,  2012, \aj, 144, 54

\bibitem[\protect\citeauthoryear{{Giersz}}{{Giersz}}{1998}]{Giersz1998MNRAS.298.1239G}
{Giersz} M.,  1998, \mnras, 298, 1239

\bibitem[\protect\citeauthoryear{{Giersz}}{{Giersz}}{2001}]{Giersz2001MNRAS.324..218G}
{Giersz} M.,  2001, \mnras, 324, 218

\bibitem[\protect\citeauthoryear{{Giersz}}{{Giersz}}{2006}]{Giersz2006MNRAS.371..484G}
{Giersz} M.,  2006, \mnras, 371, 484

\bibitem[\protect\citeauthoryear{{Giersz} \& {Heggie}}{{Giersz} \&
  {Heggie}}{1994a}]{Giersz1994MNRAS.268..257G}
{Giersz} M.,  {Heggie} D.~C.,  1994a, \mnras, 268, 257

\bibitem[\protect\citeauthoryear{{Giersz} \& {Heggie}}{{Giersz} \&
  {Heggie}}{1994b}]{Giersz1994MNRAS.270..298G}
{Giersz} M.,  {Heggie} D.~C.,  1994b, \mnras, 270, 298

\bibitem[\protect\citeauthoryear{{Giersz} \& {Heggie}}{{Giersz} \&
  {Heggie}}{1996}]{Giersz1996MNRAS.279.1037G}
{Giersz} M.,  {Heggie} D.~C.,  1996, \mnras, 279, 1037

\bibitem[\protect\citeauthoryear{{Giersz} \& {Heggie}}{{Giersz} \&
  {Heggie}}{1997}]{Giersz1997MNRAS.286..709G}
{Giersz} M.,  {Heggie} D.~C.,  1997, \mnras, 286, 709

\bibitem[\protect\citeauthoryear{{Giersz}, {Heggie}, {Hurley} \&
  {Hypki}}{{Giersz} et~al.}{2011}]{Giersz2011arXiv1112.6246G}
{Giersz} M.,  {Heggie} D.~C.,  {Hurley} J.,    {Hypki} A.,  2011, ArXiv
  e-prints

\bibitem[\protect\citeauthoryear{{Giersz}, {Heggie}, {Hurley} \&
  {Hypki}}{{Giersz} et~al.}{2013}]{Giersz2013MNRAS.431.2184G}
{Giersz} M.,  {Heggie} D.~C.,  {Hurley} J.~R.,    {Hypki} A.,  2013, \mnras,
  431, 2184

\bibitem[\protect\citeauthoryear{{Giersz} \& {Spurzem}}{{Giersz} \&
  {Spurzem}}{1994}]{Giersz1994MNRAS.269..241G}
{Giersz} M.,  {Spurzem} R.,  1994, \mnras, 269, 241

\bibitem[\protect\citeauthoryear{{Harris}}{{Harris}}{1996}]{Harris1996AJ....112.1487H}
{Harris} W.~E.,  1996, \aj, 112, 1487

\bibitem[\protect\citeauthoryear{{Heggie}}{{Heggie}}{1975}]{Heggie1975MNRAS.173..729H}
{Heggie} D.~C.,  1975, \mnras, 173, 729

\bibitem[\protect\citeauthoryear{{H{\'e}non}}{{H{\'e}non}}{1971}]{Henon1971Ap&SS..14..151H}
{H{\'e}non} M.~H.,  1971, \apss, 14, 151

\bibitem[\protect\citeauthoryear{{Hills} \& {Day}}{{Hills} \&
  {Day}}{1976}]{1976ApL....17...87H}
{Hills} J.~G.,  {Day} C.~A.,  1976, \aplett, 17, 87

\bibitem[\protect\citeauthoryear{{Hurley}, {Pols} \& {Tout}}{{Hurley}
  et~al.}{2000}]{Hurley2000MNRAS.315..543H}
{Hurley} J.~R.,  {Pols} O.~R.,    {Tout} C.~A.,  2000, \mnras, 315, 543

\bibitem[\protect\citeauthoryear{{Hurley}, {Tout} \& {Pols}}{{Hurley}
  et~al.}{2002}]{Hurley2002MNRAS.329..897H}
{Hurley} J.~R.,  {Tout} C.~A.,    {Pols} O.~R.,  2002, \mnras, 329, 897

\bibitem[\protect\citeauthoryear{{Hypki} \& {Giersz}}{{Hypki} \&
  {Giersz}}{2013}]{Hypki2013MNRAS.429.1221H}
{Hypki} A.,  {Giersz} M.,  2013, \mnras, 429, 1221

\bibitem[\protect\citeauthoryear{{Kaluzny}, {Rozyczka}, {Thompson} \&
  {Zloczewski}}{{Kaluzny} et~al.}{2009}]{Kaluzny2009AcA....59..371K}
{Kaluzny} J.,  {Rozyczka} M.,  {Thompson} I.~B.,    {Zloczewski} K.,  2009,
  \actaa, 59, 371

\bibitem[\protect\citeauthoryear{{Kaluzny}, {Rucinski}, {Thompson}, {Pych} \&
  {Krzeminski}}{{Kaluzny} et~al.}{2007}]{Kaluzny2007AJ....133.2457K}
{Kaluzny} J.,  {Rucinski} S.~M.,  {Thompson} I.~B.,  {Pych} W.,    {Krzeminski}
  W.,  2007, \aj, 133, 2457

\bibitem[\protect\citeauthoryear{{Kaluzny}, {Thompson}, {Rucinski}, {Pych},
  {Stachowski}, {Krzeminski} \& {Burley}}{{Kaluzny}
  et~al.}{2007}]{Kaluzny2007AJ....134..541K}
{Kaluzny} J.,  {Thompson} I.~B.,  {Rucinski} S.~M.,  {Pych} W.,  {Stachowski}
  G.,  {Krzeminski} W.,    {Burley} G.~S.,  2007, \aj, 134, 541

\bibitem[\protect\citeauthoryear{{Knigge}, {Leigh} \& {Sills}}{{Knigge}
  et~al.}{2009}]{Knigge2009Natur.457..288K}
{Knigge} C.,  {Leigh} N.,    {Sills} A.,  2009, \nat, 457, 288

\bibitem[\protect\citeauthoryear{{Kroupa}}{{Kroupa}}{1995a}]{Kroupa1995aMNRAS.277.1491K}
{Kroupa} P.,  1995a, \mnras, 277, 1491

\bibitem[\protect\citeauthoryear{{Kroupa}}{{Kroupa}}{1995b}]{Kroupa1995bMNRAS.277.1507K}
{Kroupa} P.,  1995b, \mnras, 277, 1507

\bibitem[\protect\citeauthoryear{{Kroupa}, {Gilmore} \& {Tout}}{{Kroupa}
  et~al.}{1991}]{Kroupa1991MNRAS.251..293K}
{Kroupa} P.,  {Gilmore} G.,    {Tout} C.~A.,  1991, \mnras, 251, 293

\bibitem[\protect\citeauthoryear{{Kroupa}, {Tout} \& {Gilmore}}{{Kroupa}
  et~al.}{1993}]{Kroupa1993MNRAS.262..545K}
{Kroupa} P.,  {Tout} C.~A.,    {Gilmore} G.,  1993, \mnras, 262, 545

\bibitem[\protect\citeauthoryear{{Kroupa}, {Weidner}, {Pflamm-Altenburg},
  {Thies}, {Dabringhausen}, {Marks} \& {Maschberger}}{{Kroupa}
  et~al.}{2013}]{Kroupa2013pss5.book..115K}
{Kroupa} P.,  {Weidner} C.,  {Pflamm-Altenburg} J.,  {Thies} I.,
  {Dabringhausen} J.,  {Marks} M.,    {Maschberger} T.,  2013, {The Stellar and
  Sub-Stellar Initial Mass Function of Simple and Composite Populations}.
p.~115

\bibitem[\protect\citeauthoryear{{Lanzoni}, {Sanna}, {Ferraro}, {Valenti},
  {Beccari}, {Schiavon}, {Rood}, {Mapelli} \& {Sigurdsson}}{{Lanzoni}
  et~al.}{2007}]{Lanzoni2007ApJ...663.1040L}
{Lanzoni} B.,  {Sanna} N.,  {Ferraro} F.~R.,  {Valenti} E.,  {Beccari} G.,
  {Schiavon} R.~P.,  {Rood} R.~T.,  {Mapelli} M.,    {Sigurdsson} S.,  2007,
  \apj, 663, 1040

\bibitem[\protect\citeauthoryear{{Lee} \& {Carney}}{{Lee} \&
  {Carney}}{2002}]{Lee2002AJ....124.1511L}
{Lee} J.-W.,  {Carney} B.~W.,  2002, \aj, 124, 1511

\bibitem[\protect\citeauthoryear{{Leigh}, {Sills} \& {Knigge}}{{Leigh}
  et~al.}{2007}]{Leigh2007ApJ...661..210L}
{Leigh} N.,  {Sills} A.,    {Knigge} C.,  2007, \apj, 661, 210

\bibitem[\protect\citeauthoryear{{Leigh}, {Sills} \& {Knigge}}{{Leigh}
  et~al.}{2008}]{2008IAUS..246..331L}
{Leigh} N.,  {Sills} A.,    {Knigge} C.,  2008, in {E.~Vesperini, M.~Giersz, \&
  A.~Sills} ed., IAU Symposium Vol.~246 of IAU Symposium, {Where the Blue
  Stragglers Roam: Searching for a Link Between Formation and Environment}.
pp 331--335

\bibitem[\protect\citeauthoryear{{Mapelli}, {Ripamonti}, {Tolstoy},
  {Sigurdsson}, {Irwin} \& {Battaglia}}{{Mapelli}
  et~al.}{2007}]{Mapelli2007MNRAS.380.1127M}
{Mapelli} M.,  {Ripamonti} E.,  {Tolstoy} E.,  {Sigurdsson} S.,  {Irwin} M.~J.,
     {Battaglia} G.,  2007, \mnras, 380, 1127

\bibitem[\protect\citeauthoryear{{Mapelli}, {Sigurdsson}, {Ferraro}, {Colpi},
  {Possenti} \& {Lanzoni}}{{Mapelli} et~al.}{2006}]{2006MNRAS.373..361M}
{Mapelli} M.,  {Sigurdsson} S.,  {Ferraro} F.~R.,  {Colpi} M.,  {Possenti} A.,
    {Lanzoni} B.,  2006, \mnras, 373, 361

\bibitem[\protect\citeauthoryear{{Mardling} \& {Aarseth}}{{Mardling} \&
  {Aarseth}}{2001}]{Mardling2001MNRAS.321..398M}
{Mardling} R.~A.,  {Aarseth} S.~J.,  2001, \mnras, 321, 398

\bibitem[\protect\citeauthoryear{{Mateo}, {Fischer} \& {Krzeminski}}{{Mateo}
  et~al.}{1995}]{Mateo1995AJ....110.2166M}
{Mateo} M.,  {Fischer} P.,    {Krzeminski} W.,  1995, \aj, 110, 2166

\bibitem[\protect\citeauthoryear{{Mateo}, {Harris}, {Nemec} \&
  {Olszewski}}{{Mateo} et~al.}{1990}]{Mateo1990AJ....100..469M}
{Mateo} M.,  {Harris} H.~C.,  {Nemec} J.,    {Olszewski} E.~W.,  1990, \aj,
  100, 469

\bibitem[\protect\citeauthoryear{{Mathieu} \& {Geller}}{{Mathieu} \&
  {Geller}}{2009}]{Mathieu2009Natur.462.1032M}
{Mathieu} R.~D.,  {Geller} A.~M.,  2009, \nat, 462, 1032

\bibitem[\protect\citeauthoryear{{McCrea}}{{McCrea}}{1964}]{1964MNRAS.128..147M}
{McCrea} W.~H.,  1964, \mnras, 128, 147

\bibitem[\protect\citeauthoryear{{Milone}, {Piotto}, {Bedin} \&
  {Sarajedini}}{{Milone} et~al.}{2008}]{2008MmSAI..79..623M}
{Milone} A.~P.,  {Piotto} G.,  {Bedin} L.~R.,    {Sarajedini} A.,  2008,
  \memsai, 79, 623

\bibitem[\protect\citeauthoryear{{Monelli}, {Cassisi}, {Mapelli}, {Bernard},
  {Aparicio}, {Skillman}, {Stetson}, {Gallart}, {Hidalgo}, {Mayer} \&
  {Tolstoy}}{{Monelli} et~al.}{2012}]{Monelli2012ApJ...744..157M}
{Monelli} M.,  {Cassisi} S.,  {Mapelli} M.,  {Bernard} E.~J.,  {Aparicio} A.,
  {Skillman} E.~D.,  {Stetson} P.~B.,  {Gallart} C.,  {Hidalgo} S.~L.,  {Mayer}
  L.,    {Tolstoy} E.,  2012, \apj, 744, 157

\bibitem[\protect\citeauthoryear{{Otulakowska}, {Olech}, {Pych}, {Pamyatnykh},
  {Zdravkov} \& {Rucinski}}{{Otulakowska}
  et~al.}{2011}]{Otulakowska2011AcA....61..161O}
{Otulakowska} M.,  {Olech} A.,  {Pych} W.,  {Pamyatnykh} A.~A.,  {Zdravkov} T.,
     {Rucinski} S.~M.,  2011, \actaa, 61, 161

\bibitem[\protect\citeauthoryear{{Piotto}, {De Angeli}, {King}, {Djorgovski},
  {Bono}, {Cassisi}, {Meylan}, {Recio-Blanco}, {Rich} \& {Davies}}{{Piotto}
  et~al.}{2004}]{2004ApJ...604L.109P}
{Piotto} G.,  {De Angeli} F.,  {King} I.~R.,  {Djorgovski} S.~G.,  {Bono} G.,
  {Cassisi} S.,  {Meylan} G.,  {Recio-Blanco} A.,  {Rich} R.~M.,    {Davies}
  M.~B.,  2004, \apjl, 604, L109

\bibitem[\protect\citeauthoryear{{Pritchet} \& {Glaspey}}{{Pritchet} \&
  {Glaspey}}{1991}]{Pritchet1991ApJ...373..105P}
{Pritchet} C.~J.,  {Glaspey} J.~W.,  1991, \apj, 373, 105

\bibitem[\protect\citeauthoryear{{Pych}, {Kaluzny}, {Krzeminski},
  {Schwarzenberg-Czerny} \& {Thompson}}{{Pych}
  et~al.}{2001}]{Pych2001A&A...367..148P}
{Pych} W.,  {Kaluzny} J.,  {Krzeminski} W.,  {Schwarzenberg-Czerny} A.,
  {Thompson} I.~B.,  2001, \aap, 367, 148

\bibitem[\protect\citeauthoryear{{Sandage}}{{Sandage}}{1953}]{1953AJ.....58...61S}
{Sandage} A.~R.,  1953, \aj, 58, 61

\bibitem[\protect\citeauthoryear{{Santolamazza}, {Marconi}, {Bono}, {Caputo},
  {Cassisi} \& {Gilliland}}{{Santolamazza}
  et~al.}{2001}]{Santolamazza2001ApJ...554.1124S}
{Santolamazza} P.,  {Marconi} M.,  {Bono} G.,  {Caputo} F.,  {Cassisi} S.,
  {Gilliland} R.~L.,  2001, \apj, 554, 1124

\bibitem[\protect\citeauthoryear{{Shara}, {Saffer} \& {Livio}}{{Shara}
  et~al.}{1997}]{Shara1997ApJ...489L..59S}
{Shara} M.~M.,  {Saffer} R.~A.,    {Livio} M.,  1997, \apjl, 489, L59

\bibitem[\protect\citeauthoryear{{Sollima}, {Lanzoni}, {Beccari}, {Ferraro} \&
  {Fusi Pecci}}{{Sollima} et~al.}{2008}]{2008A&A...481..701S}
{Sollima} A.,  {Lanzoni} B.,  {Beccari} G.,  {Ferraro} F.~R.,    {Fusi Pecci}
  F.,  2008, \aap, 481, 701

\bibitem[\protect\citeauthoryear{{Stodolkiewicz}}{{Stodolkiewicz}}{1986}]{Stodolkiewicz1986AcA....36...19S}
{Stodolkiewicz} J.~S.,  1986, \actaa, 36, 19

\bibitem[\protect\citeauthoryear{{Thompson}, {Kaluzny}, {Rucinski},
  {Krzeminski}, {Pych}, {Dotter} \& {Burley}}{{Thompson}
  et~al.}{2010}]{Thompson2010AJ....139..329T}
{Thompson} I.~B.,  {Kaluzny} J.,  {Rucinski} S.~M.,  {Krzeminski} W.,  {Pych}
  W.,  {Dotter} A.,    {Burley} G.~S.,  2010, \aj, 139, 329

\bibitem[\protect\citeauthoryear{{Tillich}, {Przybilla}, {Scholz} \&
  {Heber}}{{Tillich} et~al.}{2010}]{Tillich2010A&A...517A..36T}
{Tillich} A.,  {Przybilla} N.,  {Scholz} R.-D.,    {Heber} U.,  2010, \aap,
  517, A36

\bibitem[\protect\citeauthoryear{{Wang}, {Spurzem}, {Aarseth}, {Giersz},
  {Askar}, {Berczik}, {Naab}, {Schadow} \& {Kouwenhoven}}{{Wang}
  et~al.}{2016}]{Wang2016MNRAS.458.1450W}
{Wang} L.,  {Spurzem} R.,  {Aarseth} S.,  {Giersz} M.,  {Askar} A.,  {Berczik}
  P.,  {Naab} T.,  {Schadow} R.,    {Kouwenhoven} M.~B.~N.,  2016, \mnras, 458,
  1450

\bibitem[\protect\citeauthoryear{{Zinn} \& {Searle}}{{Zinn} \&
  {Searle}}{1976}]{1976ApJ...209..734Z}
{Zinn} R.,  {Searle} L.,  1976, \apj, 209, 734

\end{thebibliography}



\appendix



\bsp	
\label{lastpage}
\end{document}